\documentclass[journal, twoside]{IEEEtran}
\pdfoutput=1
\usepackage{graphics, amsmath}
\usepackage{epsfig}
\usepackage{subfigure}

\usepackage{enumitem}

\usepackage{amsthm}
\renewcommand{\QED}{{\QEDopen}}
\theoremstyle{definition}

\begin{document}

\newtheorem{theorem}{Theorem}
\newtheorem{lemma}[theorem]{Lemma}
\newtheorem{proposition}[theorem]{Proposition}
\newtheorem{corollary}[theorem]{Corollary}
\newtheorem{conjecture}[theorem]{Conjecture}
\newtheorem{condition_set}[theorem]{Condition Set}
\newtheorem{definition}[theorem]{Definition}
\newtheorem{algorithm}[theorem]{Algorithm}
\newtheorem{remark}[theorem]{Remark}
\newcommand{\bre}{\begin{equation}}
\newcommand{\ere}{\end{equation}}
\newcommand{\be}{{\bf {e}}}
\newcommand{\ee}\]
\newcommand{\bra}{\begin{eqnarray}}
\newcommand{\era}{\end{eqnarray}}
\newcommand{\bfg}{\begin{figure}[hbtp]}
\newcommand{\efg}{\end{figure}}
\newcommand{\bver}{\begin{verbatim}}
\newcommand{\ever}{\end{verbatim}}
\newcommand{\bit}{\begin{itemize}}
\newcommand{\eit}{\end{itemize}}
\newcommand{\ben}{\begin{enumerate}}
\newcommand{\een}{\end{enumerate}}
\newcommand{\ett}{\mbox{$\eta$} }
\newcommand{\coeff}[1]{\lfloor #1\rfloor}
\newcommand{\ceil}[1]{\lceil #1\rceil}
\newcommand{\floor}[1]{[ #1 ]}
\newcommand{\bfloor}[1]{\left[ #1 \right]}
\newcommand{\dgr}[1]{\mbox{$#1^{\circ}$}}
\newcommand{\cu}{\mbox{cosmic $\mu$ }}
\newcommand{\csa}[2]{\mbox{$\cos^2(#1 - #2)$}}
\newcommand{\csb}[2]{\mbox{$\cos2(#1 - #2)$}}
\newcommand{\balpha}{\mbox{\boldmath $\alpha$}}
\newcommand{\bbeta}{\mbox{\boldmath $\beta$}}
\newcommand{\blambda}{\mbox{\boldmath $\lambda$}}
\newcommand{\bkappa}{\mbox{\boldmath $\kappa$}}
\newcommand{\bXi}{\mbox{\boldmath $\Xi$}}
\newcommand{\brho}{\mbox{\boldmath $\rho$}}
\newcommand{\bchi}{\mbox{\bf x}}
\newcommand{\bal}{\mbox{\boldmath $\alpha$}_0}
\newcommand{\Exp}{\mbox{E}}
\newcommand{\given}{\: | \:}
\newcommand{\Ker}{\mbox{Ker}\,}
\newcommand{\tildepss}{\tilde \epsilon_{s}}
\newcommand{\tildepssl}{\tilde \epsilon_{s_l}}
\newcommand{\epss}{\epsilon_s}
\newcommand{\epssl}{\epsilon_{s_l}}
\newcommand{\bphi}{{\mathbf \Phi}}
\newcommand{\bepsilon}{\mbox{\boldmath $\epsilon$}}
\newcommand{\bOmega}{{\mathbf \Omega}}
\newcommand{\btOmega}{\tilde{\mathbf \Omega}}
\newcommand{\bSigma}{{\mathbf \Sigma}}
\newcommand{\btheta}{\mbox{\boldmath $\theta$}}
\newcommand{\bzeta}{\mbox{\boldmath $\zeta$}}
\newcommand{\hbs}{{\hat{\bf s}}}
\newcommand{\hbb}{{\hat{\bf b}}}
\newcommand{\hs}{{\hat{s}}}
\newcommand{\hf}{{\hat{f}}}
\newcommand{\hb}{{\hat{b}}}
\newcommand{\br}{{\bf r}}
\newcommand{\bR}{{\bf R}}
\newcommand{\hS}{{\hat{S}}}
\newcommand{\ts}{{\tilde{s}}}
\newcommand{\botheta}{\mbox{\boldmath ${\bar\theta}$}}
\newcommand{\otheta}{{\bar\theta}}
\newcommand{\tlambda}{\tilde{\lambda}}
\newcommand{\trho}{\tilde{\rho}}
\newcommand{\bsigma}{\mbox{\boldmath $\sigma$}}
\newcommand{\btsigma}{\mbox{\boldmath $\tilde\sigma$}}
\newcommand{\Bernoulli}{\textrm{Bernoulli}}
\newcommand{\mmse}{\textrm{mmse}}
\newcommand{\Good}{\textrm{Good}}
\newcommand{\design}{\textrm{design}}
\newcommand{\naive}{\textrm{naive}}

\newcommand{\we}{\overrightarrow{e}}
\newcommand{\bA}{{\bf A}}
\newcommand{\bB}{{\bf B}}
\newcommand{\bC}{{\bf C}}
\newcommand{\bI}{{\bf I}}
\newcommand{\bg}{{\bf{g}}}
\newcommand{\bG}{{\bf{G}}}
\newcommand{\bE}{{\bf E}}
\newcommand{\bF}{{\bf F}}
\newcommand{\bm}{{\bf m}}
\newcommand{\bk}{{\bf k}}
\newcommand{\bof}{{\bf f}}
\newcommand{\rmf}{{\rm f}}
\newcommand{\norm}[1]{\|#1\|}
\newcommand{\oQ}{\overline Q}
\newcommand{\tQ}{\tilde Q}
\newcommand{\tD}{\tilde D}
\newcommand{\oD}{\bar D}
\newcommand{\oV}{\bar V}
\newcommand{\ox}{\bar x}
\newcommand{\tP}{\tilde P}
\newcommand{\tN}{\tilde N}
\newcommand{\tA}{\tilde A}
\newcommand{\tM}{\tilde M}
\newcommand{\tX}{\tilde X}
\newcommand{\tY}{\tilde Y}
\newcommand{\tW}{\tilde W}
\newcommand{\tm}{\tilde m}
\newcommand{\tc}{\tilde c}
\newcommand{\tx}{\tilde x}
\newcommand{\tv}{\tilde v}
\newcommand{\tH}{\tilde H}
\newcommand{\btW}{\tilde {\bf W}}
\newcommand{\btA}{\tilde {\bf A}}
\newcommand{\btB}{\tilde {\bf B}}
\newcommand{\btV}{\tilde {\bf V}}
\newcommand{\btH}{\tilde {\bf H}}
\newcommand{\btZ}{\tilde {\bf Z}}
\newcommand{\btY}{\tilde {\bf Y}}
\newcommand{\btM}{\tilde {\bf M}}
\newcommand{\btX}{\tilde {\bf X}}
\newcommand{\btx}{\tilde {\bf x}}
\newcommand{\btm}{\tilde {\bf m}}
\newcommand{\bFDPC}{\bF_{DPC}}
\newcommand{\tdet}{\tilde {\textrm{det}}}
\newcommand{\btT}{\tilde {\bf T}}
\newcommand{\btS}{\tilde {\bf S}}
\newcommand{\btn}{\tilde {\bf n}}
\newcommand{\btv}{\tilde {\bf v}}
\newcommand{\bty}{\tilde {\bf y}}
\newcommand{\btp}{\tilde {\bf p}}
\newcommand{\bta}{\tilde {\bf a}}
\newcommand{\PrEras}{\hat{\Pr}}
\newcommand{\ErasEpsilon}{\hat{\epsilon}}
\newcommand{\ErasCY}{\hat{\cY}}
\newcommand{\tab}{\ \ \ \ }
\newcommand{\bigtab}{\tab \tab \tab}
\newcommand{\bc}{{\bf c}}
\newcommand{\type}{{\mathrm type}}
\newcommand{\QEC}{{\mathrm{QEC}}}
\newcommand{\scnd}{{\mathrm {scnd}}}
\newcommand{\APP}{{\mathrm {APP}}}
\newcommand{\LLR}{{\mathrm {LLR}}}
\newcommand{\Cov}{{\mathrm{Cov}}}
\newcommand{\cov}{{\mathrm{cov}}}
\newcommand{\GF}{{\mathrm {GF}}}
\newcommand{\mymod}{{\mathrm{mod}\:}}
\newcommand{\rank}{{\mathrm{rank\:}}}
\newcommand{\spn}{{\mathrm{span\:}}}
\newcommand{\E}{{\mathrm E}}
\newcommand{\bmu}{\mbox{\boldmath $\mu$}}
\newcommand{\bxi}{\mbox{\boldmath $\xi$}}
\newcommand{\cL}{{\cal L}}
\newcommand{\bN}{{\bf N}}
\newcommand{\cB}{{\cal B}}
\newcommand{\cH}{{\cal H}}
\newcommand{\cU}{{\cal U}}
\newcommand{\cW}{{\cal W}}
\newcommand{\baa}{\begin{eqnarray*}}
\newcommand{\eaa}{\end{eqnarray*}}

\newcommand{\ds}{{\:d\bf s}}
\newcommand{\du}{{\:d\bf u}}
\newcommand{\dy}{{\:d\bf y}}
\newcommand{\dH}{{\:d\bf H}}
\newcommand{\dHyus}{\dH\dy\du\ds}
\newcommand{\diag}{\textrm{diag}}
\newcommand{\BPSK}{\textrm{BPSK}}
\newcommand{\eras}{\textrm{Erasure}}
\newcommand{\BEC}{\textrm{BEC}}
\newcommand{\EEC}{\textrm{EEC}}
\newcommand{\unrevealed}{\textrm{unrevealed}}
\newcommand{\Verdu}{Verd\'{u}}
\newcommand{\Measson}{M\'{e}asson}
\newcommand{\MAP}{\textrm{MAP}}
\newcommand{\DF}{\textrm{DF}}
\newcommand{\CF}{\textrm{CF}}
\newcommand{\out}{\textrm{out}}

\newcommand{\EE}{\mbox{\boldlarge E}}

\newcommand{\bs}{{\bf s}}
\newcommand{\mR}{{\mathrm R}}
\newcommand{\mL}{{\mathrm L}}
\newcommand{\ba}{{\bf a}}
\newcommand{\bb}{{\bf b}}
\newcommand{\bq}{{\bf q}}
\newcommand{\qstar}{{q^{\star}}}
\newcommand{\Qstar}{{Q^{\star}}}
\newcommand{\Pe}{{P_{\mathrm{e}}}}
\newcommand{\xstar}{{x^{\star}}}
\newcommand{\Xstar}{{X^{\star}}}
\newcommand{\Ystar}{{Y^{\star}}}
\newcommand{\ystar}{{y^{\star}}}
\newcommand{\rhostar}{{\rho^{\star}}}
\newcommand{\bp}{{\bf p}}
\newcommand{\bX}{{\bf X}}
\newcommand{\obX}{{\bar {\bf X}}}
\newcommand{\obx}{{\bar {\bf x}}}
\newcommand{\obm}{{\bar {\bf m}}}
\newcommand{\obY}{{\bar {\bf Y}}}
\newcommand{\oby}{{\bar {\bf y}}}
\newcommand{\obZ}{{\bar {\bf Z}}}
\newcommand{\obU}{{\bar {\bf U}}}
\newcommand{\obW}{{\bar {\bf W}}}
\newcommand{\obu}{{\bar {\bf u}}}
\newcommand{\obw}{{\bar {\bf w}}}
\newcommand{\obN}{{\bar {\bf N}}}
\newcommand{\obM}{{\bar {\bf M}}}
\newcommand{\obB}{{\bar {\bf B}}}
\newcommand{\oX}{{\bar {X}}}
\newcommand{\oY}{{\bar {Y}}}
\newcommand{\oU}{{\bar {U}}}
\newcommand{\oW}{{\bar {W}}}
\newcommand{\ou}{{\bar {u}}}
\newcommand{\ow}{{\bar {w}}}
\newcommand{\oR}{{\bar {R}}}
\newcommand{\oM}{{\bar {M}}}
\newcommand{\oB}{{\bar {B}}}
\newcommand{\bU}{{\bf U}}
\newcommand{\bW}{{\bf W}}
\newcommand{\bY}{{\bf Y}}
\newcommand{\bV}{{\bf V}}
\newcommand{\bZ}{{\bf Z}}
\newcommand{\bT}{{\bf T}}
\newcommand{\bS}{{\bf S}}
\newcommand{\bM}{{\bf M}}
\newcommand{\bH}{{\bf H}}
\newcommand{\DFT}{{\mathrm{DFT}}}
\newcommand{\IDFT}{{\mathrm{IDFT}}}
\newcommand{\ldeg}{{\mathrm ldeg}}
\newcommand{\Real}{{\mathrm Re}}
\newcommand{\weight}{{\mathrm weight}}
\newcommand{\xor}{\oplus}
\newcommand{\bu}{{\bf u}}
\newcommand{\bv}{{\bf v}}
\newcommand{\bt}{{\bf t}}
\newcommand{\bd}{{\bf d}}
\newcommand{\bD}{{\bf D}}
\newcommand{\bw}{{\bf w}}
\newcommand{\bn}{{\bf n}}
\newcommand{\bx}{{\bf x}}
\newcommand{\bxstar}[1]{{\bf x}_{#1}^\star}
\newcommand{\by}{{\bf y}}
\newcommand{\bz}{{\bf z}}
\newcommand{\bone}{{\bf 1}}
\newcommand{\bzr}{{\bf 0}}
\newcommand{\cA}{{\cal A}}
\newcommand{\cP}{{\cal P}}
\newcommand{\cE}{{\cal E}}
\newcommand{\cF}{{\cal F}}
\newcommand{\cR}{{\cal R}}
\newcommand{\cS}{{\cal S}}
\newcommand{\cT}{{\cal T}}
\newcommand{\cX}{{\cal X}}
\newcommand{\cY}{{\cal Y}}
\newcommand{\bare}{{\bar{e}}}
\newcommand{\oS}{\overline{S}}
\newcommand{\oN}{\overline{N}}
\newcommand{\oA}{\overline{A}}
\newcommand{\oP}{\overline{P}}
\newcommand{\op}{\overline{p}}
\newcommand{\hpi}{\hat{\pi}}
\newcommand{\hP}{\hat{P}}
\newcommand{\hdelta}{\hat{\delta}}
\newcommand{\hvarepsilon}{\hat{\varepsilon}}
\newcommand{\hR}{\hat{R}}
\newcommand{\hE}{\hat{E}}
\newcommand{\hbe}{{\bf \hat{e}}}
\newcommand{\tR}{\tilde{R}}
\newcommand{\hbF}{\hat{\bf F}}
\newcommand{\hbE}{\hat{\bf E}}
\newcommand{\hF}{\hat{F}}
\newcommand{\hbU}{{\bf{\hat{U}}}}
\newcommand{\tbU}{{\bf{\tilde{U}}}}
\newcommand{\hbS}{{\bf{\hat{S}}}}
\newcommand{\hw}{\hat{w}}
\newcommand{\cC}{{\mathcal{C}}}
\newcommand{\cCstar}{{\mathcal{C^\star}}}
\newcommand{\cN}{{\mathcal{N}}}
\newcommand{\cI}{{\mathcal{I}}}
\newcommand{\cIbar}{{\bar{\mathcal{I}}}}
\newcommand{\cG}{\mathcal{G}}
\newcommand{\Erasure}{\diamond}
\newcommand{\hcC}{\hat{\mathcal{C}}}
\newcommand{\cD}{\mathcal{D}}
\newcommand{\hcD}{\hat{\mathcal{D}}}
\newcommand{\hD}{\hat{D}}
\newcommand{\tcD}{\tilde{\mathcal{D}}}
\newcommand{\tcC}{\tilde{\cal C}}
\newcommand{\tC}{\tilde{C}}
\newcommand{\hA}{\hat{A}}
\newcommand{\hB}{\hat{B}}
\newcommand{\hC}{\hat{C}}
\newcommand{\he}{\hat{e}}
\newcommand{\btc}{\tilde{\bf c}}
\newcommand{\hbr}{\hat{\bf r}}
\newcommand{\hbw}{\hat{\bf w}}
\newcommand{\hbv}{\hat{\bf v}}
\newcommand{\hbc}{\hat{\bf c}}
\newcommand{\hby}{\hat{\bf y}}
\newcommand{\hX}{\hat{X}}
\newcommand{\hx}{\hat{x}}
\newcommand{\hY}{\hat{Y}}
\newcommand{\hU}{\hat{U}}
\newcommand{\hW}{\hat{W}}
\newcommand{\hV}{\hat{V}}
\newcommand{\hv}{\hat{v}}
\newcommand{\hy}{\hat{y}}
\newcommand{\hm}{\mathrm{{\hat m}}}
\newcommand{\hbx}{\hat{\bf x}}
\newcommand{\hbm}{\hat{\bf m}}
\newcommand{\hbX}{\hat{\bf X}}
\newcommand{\hbY}{\hat{\bf Y}}
\newcommand{\beginproof}{\noindent \textbf{Proof: }  }
\newcommand{\finproof}{\noindent $\Box$\\}
\newcommand{\ve}{\varepsilon}
\newcommand{\emptyline}{$\:\\ $}
\def\refeq#1{\: {\stackrel{ (#1)}{=}} \: }
\def\defined{\: {\stackrel{\scriptscriptstyle \Delta}{=}} \: }
\def\psdir#1{#1}
\def\MSE{{\rm MSE}}
\def\argmax{\mathop{\rm argmax}}
\def\CB{\mathop{\rm BS}}
\def\defined{\: {\stackrel{\scriptscriptstyle \Delta}{=}} \: }
\def\leqa{\buildrel \rm {\scriptscriptstyle (1)} \over \leq}
\def\leqb{\buildrel \rm {\scriptscriptstyle (2)} \over \leq}
\def\leqc{\buildrel \rm {\scriptscriptstyle (3)} \over \leq}
\def\eqa{\buildrel \rm {\scriptscriptstyle (1)} \over =}
\def\eqb{\buildrel \rm {\scriptscriptstyle (2)} \over =}
\def\eqc{\buildrel \rm {\scriptscriptstyle (3)} \over =}
\newfont{\boldlarge}{msbm10 scaled 1100}
\newcommand{\RR}{\mbox{\boldlarge R}}
\newcommand{\tr}{\mathrm{tr}}

\newcommand{\myeqref}[2]{(\ref{#1}#2)}
\newcommand{\labeleq}[1]{\label{eq:#1}}
\newcommand{\refsec}[1]{(\ref{section:#1})}
\newcommand{\labelsec}[1]{\label{section:#1}}

\newcommand{\m}{{\mathrm{m}}}
\newcommand{\e}{{\mathrm{e}}}
\newcommand{\rb}[2]{{{r}^{(#1)}_{#2}}}
\newcommand{\Rb}[2]{{{R}^{(#1)}_{#2}}}
\newcommand{\rbt}[2]{{{r}^{(#1)}_{#2}}'}
\newcommand{\hrb}[2]{{\hat{{r}}^{(#1)}_{#2}}}
\newcommand{\hrbt}[2]{{{\hat{{r}}}^{(#1)}_{#2}}'}
\newcommand{\orb}[2]{{\overline{{r}}^{(#1)}_{#2}}}
\newcommand{\lb}[2]{{{l}^{(#1)}_{#2}}}
\newcommand{\Lb}[2]{{{L}^{(#1)}_{#2}}}
\newcommand{\lbt}[2]{{{l}^{(#1)}_{#2}}'}
\newcommand{\bYBP}{{\bf Y}^{\textrm{\tiny{BP}}}}
\newcommand{\byBP}{{\bf y}^{\textrm{\tiny{BP}}}}
\newcommand{\beBP}{{\bf e}^{\textrm{\tiny{BP}}}}

\newcommand{\varepsilono}{{\varepsilon_\mathrm{o}}}

\newcommand{\YBP}[1]{{Y}^{\textrm{\tiny{BP}}}_{#1}}
\newcommand{\yBP}[1]{{y}^{\textrm{\tiny{BP}}}_{#1}}
\newcommand{\eBP}[1]{{e}^{\textrm{\tiny{BP}}}_{#1}}

\newcommand{\SRC}{{\textrm{s}}}
\newcommand{\RLY}{{\textrm{r}}}
\newcommand{\StR}{{\textrm{sr}}}
\newcommand{\StD}{{\textrm{sd}}}
\newcommand{\RtD}{{\textrm{rd}}}
\newcommand{\BPR}{{\textrm{r}}}
\newcommand{\BPD}{{\textrm{d}}}
\newcommand{\mxx}{{\textrm{max}}}
\newcommand{\bYBPrelay}{{\bf Y}^{\textrm{\tiny{BP}}}_\BPR}
\newcommand{\cbYBPrelay}{{\mbox{\boldmath $\mathcal{Y}$}}^{\textrm{\tiny{BP}}}_\BPR}
\newcommand{\byBPrelay}{{\bf y}^{\textrm{\tiny{BP}}}_\BPR}
\newcommand{\yBPrelay}[1]{{y}^{\textrm{\tiny{BP}}}_{\BPR,i}}
\newcommand{\yBPdest}[1]{{y}^{\textrm{\tiny{BP}}}_{\BPD,i}}
\newcommand{\hbyBPrelay}{\hat{\bf y}^{\textrm{\tiny{BP}}}_\BPR}
\newcommand{\bEBPrelay}{{\bf E}^{\textrm{\tiny{BP}}}_\BPR}
\newcommand{\beBPrelay}{{\bf e}^{\textrm{\tiny{BP}}}_\BPR}
\newcommand{\bYBPdest}{{\bf Y}^{\textrm{\tiny{BP}}}_\BPD}
\newcommand{\byBPdest}{{\bf y}^{\textrm{\tiny{BP}}}_\BPD}
\newcommand{\bYBPdestTag}{{\bf Y'}^{\textrm{\tiny{BP}}}_\BPD}
\newcommand{\Final}{{\mathrm{fin}}}
\newcommand{\byBPdestTag}{{ \bf y_{\rm{\BPD}}^{\textrm{\tiny{BP}}}}'}
\newcommand{\yBPdestTag}[1]{{ y_{\textrm{\BPD},#1}^{\textrm{\tiny{BP}}}}'}
\newcommand{\YBPdest}[1]{{Y}^{\textrm{\tiny{BP}}}_{\BPD,#1}}
\newcommand{\YBPdestTag}[1]{{Y'}^{\textrm{\tiny{BP}}}_{\BPD,#1}}
\newcommand{\hbYBPrelay}{{\bf \hat Y}^{\textrm{\tiny{BP}}}_\BPR}
\newcommand{\hbYBPrelayTag}{{\bf \hat Y}^{\textrm{\tiny{BP}}'}_\BPR}
\newcommand{\hbYBPrelayDoubleTag}{{\bf \hat Y}^{\textrm{\tiny{BP}}''}_\BPR}
\newcommand{\hYBPrelayTag}[1]{{\hat Y}^{\textrm{\tiny{BP}}'}_{\BPR,#1}}
\newcommand{\hYBPrelayDoubleTag}[1]{{\hat Y}^{\textrm{\tiny{BP}}''}_{\BPR,#1}}
\newcommand{\hbEBPrelay}{{\bf \hat E}^{\textrm{\tiny{BP}}}_\BPR}
\newcommand{\hYBPrelay}[1]{{\hat Y}^{\textrm{\tiny{BP}}}_{\BPR,#1}}
\newcommand{\hyBPrelay}[1]{{\hat y}^{\textrm{\tiny{BP}}}_{\BPR,#1}}
\newcommand{\oYBPrelay}[1]{{\overline Y}^{\textrm{\tiny{BP}}}_{\BPR,#1}}
\newcommand{\hEBPrelay}[1]{{\hat E}^{\textrm{\tiny{BP}}}_{\BPR,#1}}
\newcommand{\EBPrelay}[1]{{E}^{\textrm{\tiny{BP}}}_{\BPR,#1}}
\newcommand{\hbYBPrelaySub}[1]{{\bf \hat Y}^{\textrm{\tiny{BP}}}_{\BPR,#1}}
\newcommand{\hYBPrelaySub}[1]{{\hat Y}^{\textrm{\tiny{BP}}}_{\BPR,#1}}
\newcommand{\YBPrelay}[1]{{Y}^{\textrm{\tiny{BP}}}_{\BPR,#1}}
\newcommand{\cYBPrelay}[1]{\mathcal{Y}^{\textrm{\tiny{BP}}}_{\BPR,#1}}
\newcommand{\bYBPrelaySub}[1]{{\bf Y}^{\textrm{\tiny{BP}}}_{\BPR,#1}}
\newcommand{\YBPrelaySub}[1]{{Y}^{\textrm{\tiny{BP}}}_{\BPR,#1}}
\newcommand{\deltaRelayBP}{{\delta}^{\textrm{\tiny{BP}}}_\BPR}
\newcommand{\xRelayBP}{{x}^{\textrm{\tiny{BP}}}_\BPR}
\newcommand{\hdeltaRelayBP}{{\hat{\delta}}^{\textrm{\tiny{BP}}}_\BPR}
\newcommand{\DRelayBP}{{D}^{\textrm{\tiny{BP}}}_\BPR}
\newcommand{\cDRelayBP}{{\mathcal{D}}^{\textrm{\tiny{BP}}}_\BPR}
\newcommand{\hDRelayBP}{{\hat{D}}^{\textrm{\tiny{BP}}}_\BPR}
\newcommand{\hdeltaRelayBPInd}[1]{{\hat{\delta}}^{\textrm{\tiny{BP}}}_{\BPR,#1}}
\newcommand{\hDeltaRelayBPInd}[1]{{\hat{\Delta}}^{\textrm{\tiny{BP}}}_{\BPR,#1}}
\newcommand{\DeltaRelayBPInd}[1]{{{\Delta}}^{\textrm{\tiny{BP}}}_{\BPR,#1}}
\newcommand{\bcErelay}{\mbox{\boldmath $\mathcal{E}_\StR$}}
\newcommand{\cErelay}[1]{\mbox{$\mathcal{E}_{\StR,#1}$}}
\newcommand{\bcErelayBP}{\mbox{\boldmath $\mathcal{E}^{\textrm{\tiny{BP}}}_\BPR$}}
\newcommand{\cErelayBP}[1]{\mbox{$\mathcal{E}^{\textrm{\tiny{BP}}}_{\BPR,#1}$}}
\newcommand{\hbcErelayBP}{\mbox{\boldmath $\hat{\mathcal{E}}^{\textrm{\tiny{BP}}}_\BPR$}}
\newcommand{\hcErelayBP}[1]{\mbox{$\hat{\mathcal{E}}^{\textrm{\tiny{BP}}}_{\BPR,#1}$}}
\newcommand{\mmm}{}       
\newcommand{\varepsilonstar}{\varepsilon^\star}
\newcommand{\fstar}{f^\star}
\newcommand{\bYCt}{{\bf Y}^{+}_3}
\newcommand{\bYCti}{{\bf Y}^{+(i)}_3}
\newcommand{\YCt}[1]{{Y}^{+}_{3,#1}}
\newcommand{\bEt}{{\bf E}^{+}_3}
\newcommand{\bEti}{{\bf E}^{+(i)}_3}
\newcommand{\Eti}[1]{{\bf E}^{+(i)}_{3,#1}}
\newcommand{\bet}{{\bf e}^{+}_3}
\newcommand{\Et}[1]{{E}^{+}_{3,#1}}
\newcommand{\SNRstar}{{\textrm{SNR}^\star}}
\newcommand{\SNR}{{\textrm{SNR}}}
\newcommand{\snr}{{\textrm{snr}}}
\newcommand{\SUD}{{\textrm{SUD}}}
\newcommand{\MUD}{{\textrm{MUD}}}
\newcommand{\UB}{{\textrm{UB}}}
\newcommand{\WZ}{{\textrm{WZ}}}
\newcommand{\comment}[1]{}
\newcommand{\topc}[1]{$\stackrel{\circ}{\rm #1}$}
\newcommand{\bbb}[1]{$<$ \textbf{To do: #1} $>$}
\newcommand{\ccc}[1]{$<$ \textbf{Amir: #1} $>$}
\newcommand\addabove[2]{ \:{\stackrel{\scriptscriptstyle \mathrm{(#2)}}{#1}} \:}
\newcommand{\etal}{{\it et al.}}
\renewcommand{\thesection}{\Roman{section}}

\newlength{\tmpbigbar}

\newcommand{\bigbar}[1]{
 \setlength{\tmpbigbar}{\unitlength}
 \settowidth{\unitlength}{\mbox{$#1$}}
  \stackrel{\barpic}{#1}
 \setlength{\unitlength}{\tmpbigbar}
}
\newcommand{\barpic}{\begin{picture}(1,0.01)(0,0)
\put(0.15,0){\line(1R,0){0.7}}
\end{picture}
}

\newcommand{\myfigure}[3]{\begin{figure}[htp]
\begin{center}
\epsfig{#1}
\end{center}
\caption{#2}\label{#3}
\end{figure}}

\newcommand{\myfigurestar}[3]{\begin{figure*}[htp]
\begin{center}
\epsfig{#1}
\end{center}
\caption{#2}\label{#3}
\end{figure*}} 

\title{Soft-Decoding-Based Strategies for Relay and Interference Channels:  Analysis and Achievable Rates Using LDPC Codes}

\author{Amir~Bennatan, Shlomo~Shamai~(Shitz) and A.~Robert~Calderbank\thanks{The work of S.~Shamai (Shitz) is supported by the Israel Science Foundation
(ISF), NEWCOM++ and NEWCOM\#, EU 7th Framework Program.  The work of R.~Calderbank is supported in part by NSF under grant DMS 0701226, by ONR under grant N00173-06-1-G006, and by AFOSR under grant FA9550-05-1-0443.}\thanks{ A.~Bennatan and R.~Calderbank were with the Program in Applied
and Computational Mathematics (PACM), Princeton University, Princeton, NJ 08540 USA.  A.~Bennatan is now with Samsung Israel R\&D Center (SIRC), Ramat Gan, 52522, Israel (e-mail: amir.b@samsung.com).  R. Calderbank is with the Department of Computer Science, Duke University,
Durham, NC 27708 USA (e-mail: robert.calderbank@duke.edu).
S. Shamai (Shitz) is with the Department of Electrical Engineering, Technion
Israel Institute of Technology, Technion City, Haifa 32000, Israel (e-mail:
sshlomo@ee.technion.ac.il).}}

\markboth{Submitted to the IEEE Transactions on Information Theory}
{Bennatan~\etal: Soft-Decoding-Based Strategies for Relay and Interference Channels}

\maketitle \setcounter{page}{1}

\begin{abstract}
We provide a rigorous mathematical analysis of two communication strategies:
soft decode-and-forward (soft-DF) for relay channels, and soft partial interference-cancelation (soft-IC) for interference channels.  Both strategies involve soft estimation, which assists the decoding process.  We consider LDPC codes, not because of their practical benefits, but because of their analytic tractability, which enables an asymptotic analysis similar to random coding methods of information theory.  Unlike some works on the closely-related demodulate-and-forward, we assume non-memoryless, code-structure-aware estimation.   With soft-DF, we develop {\it simultaneous density evolution} to  bound the decoding error probability at the destination.  This result applies to erasure relay channels.  In one variant of soft-DF, the relay applies Wyner-Ziv coding to enhance its communication with the destination, borrowing from compress-and-forward.   To analyze soft-IC, we adapt existing techniques for iterative multiuser detection, and focus on binary-input additive white Gaussian noise (BIAWGN) interference channels.  We prove that optimal point-to-point codes are unsuitable for soft-IC, as well as for all strategies that apply partial decoding to improve upon single-user detection (SUD) and multiuser detection (MUD), including Han-Kobayashi (HK).
\end{abstract}
\section{Introduction}\label{sec:Introduction}
Relay and interference channels capture two of the fundamental phenomena that characterize wireless networks:  The potential of cooperation to achieve better performance and interference between nodes (e.g., resulting from the shared wireless medium).

\begin{figure}
\centering
\subfigure[A relay channel.]{%
\begin{minipage}[b]{0.5\textwidth}
\centering
\epsfig{file=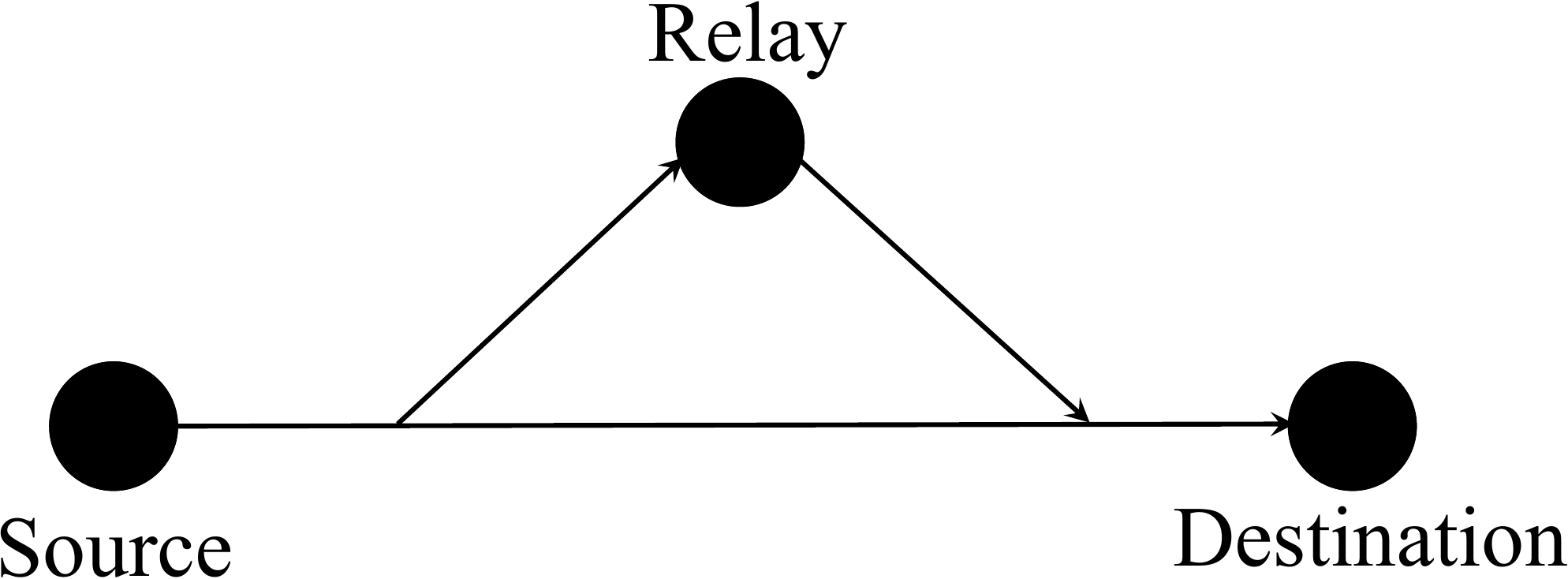, width = 6cm}
\end{minipage}}%
\\
\subfigure[An interference channel.]{%
\begin{minipage}[b]{0.5\textwidth}
\centering
\epsfig{file=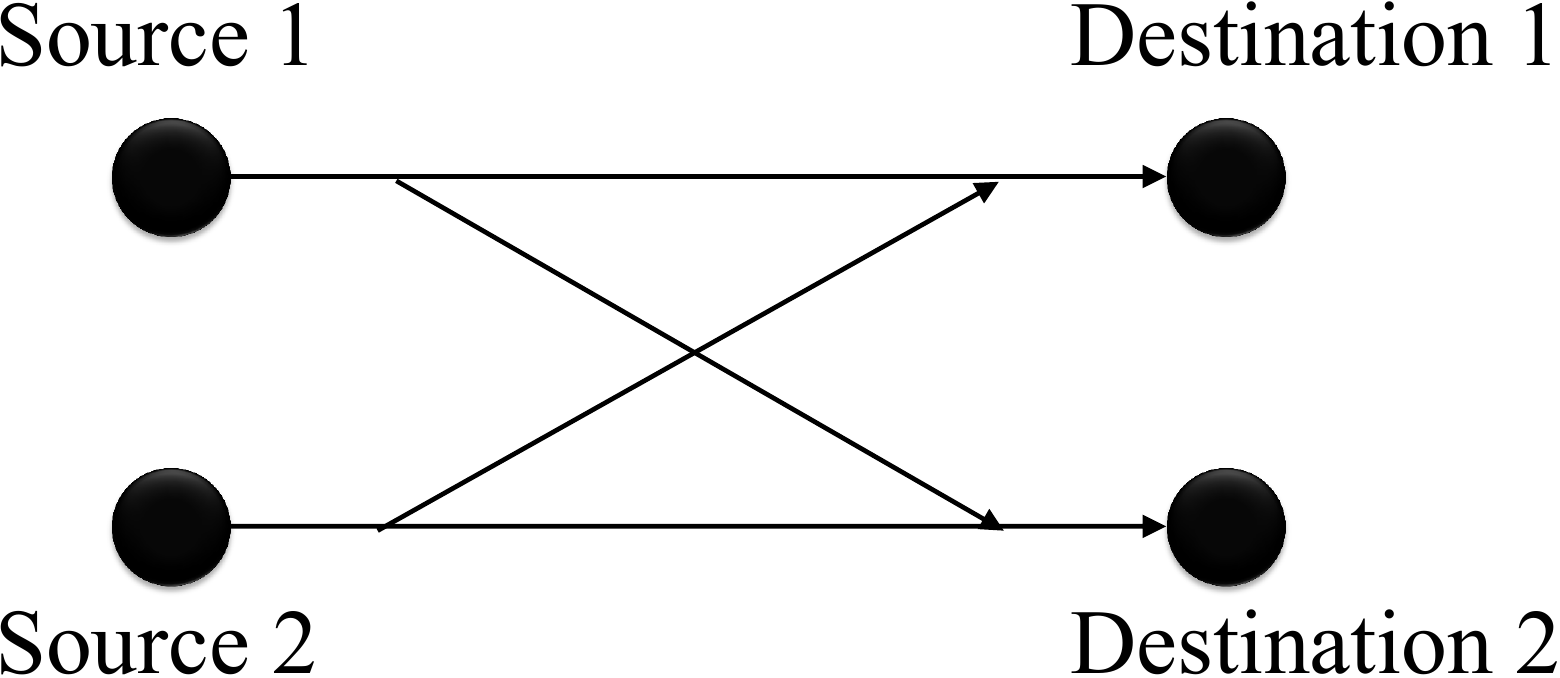, width = 6cm}
\end{minipage}}%
\caption{Two multi-terminal channel models.} \label{fig:MultiTerminal_Channels}
\end{figure}

Both channels are illustrated in Fig.~\ref{fig:MultiTerminal_Channels}.  In a relay channel (introduced by van der Meulen~\cite{VanderMeulen}), a single pair of source and destination nodes wish to communicate, and are aided by a {\it relay} node, which lends its resources to support their communications.   An interference channel (introduced by Shannon~\cite{ShannonInterference}) is characterized by two pairs of nodes, each pair consisting of a source and destination that wish to communicate.  Unlike point-to-point (P2P) channels, each destination experiences interference resulting from the signal produced by the source of the other pair.  The capacities of both channels are in general still unknown.

In this paper, we focus on two classes of communication strategies.  The first applies to relay channels and the second to interference channels.
\vspace{0.2cm}
\begin{itemize}[leftmargin=1em]
\item {\bf Soft Decode-and-Forward (Soft-DF).} The relay computes an estimate of the signal transmitted by the source, and forwards it to the destination over the channel between them.  The destination obtains from the channel a noisy version of the estimate, and uses it to help decode the source message.  Estimation reduces the level of noise in the forwarded signal.
\item {\bf Soft Partial Interference Cancelation (Soft-IC).}  Each destination in an interference channel computes an estimate of the interfering signal (codeword), in addition to decoding the message from its respective source. Estimation helps the destination overcome the interference.  \vspace{0.1cm}
\end{itemize}
\vspace{0.2cm}
With both strategies, estimation can be perceived as a form of {\it partial decoding}.  In both cases, we would ideally have liked to {\it decode} rather than estimate (asymptotically zero-error estimation).  With soft-DF, this would have enabled the relay to better cooperate with the source, for example by functioning as a virtual transmit antenna.  With soft-IC, this would have enabled each destination to eliminate the interference, for example by subtracting the interfering signal from the received channel output (over additive interference channels).  In some cases, however, the relevant signal is too weak to enable reliable decoding.  In such cases, estimation is a useful substitute, enabling partial cooperation with the source and partial cancelation of the interference, respectively.

The term soft-DF was coined by Sneessens and Vandendorpe~\cite{SoftDF1}, but related concepts were also examined extensively in several other works, e.g.,~\cite{SoftDF_DTC,SoftDF3,SoftDF2,SoftDF7,Laneman_New,Reliability_Exchange,Jafar}.  While an overwhelming body of research exists on soft-IC, the majority of it (e.g.,~\cite{SoftInterferenceCancellation,boutros-caire,Giuseppe_II,SoftIC2,Sanderovich}) focuses on the approach's application as a component of larger, iterative schemes designed to achieve complete (rather than partial) decoding of multiple signals.   The concepts of soft-IC for the purpose of partial decoding of interference, as considered in this paper, were proposed by Divsalar~\etal~\cite{SoftICPartial_4}  as well as~\cite{SoftICPartial_2,SoftICPartial_3,SoftICPartial_1}.

The significance of partial decoding has long been recognized in the information-theoretic literature.  However, it typically takes a different form than estimation.
The best known strategies for relay and interference channels (in terms of achievable communication rates) are (respectively) partial decode-and-forward\footnote{In~\cite{Gerhard_Maric_Cooperative} it is known as {\it multipath} decode-and-forward.} (partial-DF)~\cite[Theorem~7]{CoverElGamal}, \cite{Gerhard_Maric_Cooperative} and Han-Kobayashi (HK)~\cite{HanKobayashi}.  Both strategies achieve partial decoding by {\it rate-splitting}.\footnote{This term was coined by Rimoldi and Urbanke~\cite{Urbanke_Rate_Splitting}, in the context of coding for multiple-access channels.}  With this approach, codes are constructed by combining two (generally, two or more) other, auxiliary codes.  With the partial-DF strategy, for example, this gives the relay the option of decoding {\it one} of the two auxiliary codewords that constitute the signal from the source, amounting to a partial decoding of the signal.\footnote{An interesting related strategy is {\it compute-and-forward}~\cite{Nazer}, which applies to a channel characterized by multiple sources and relays.  With this strategy, each relay decodes part of a collection of linear equations of the source messages, amounting to partial decoding of its received signal. }

The above references on partial-DF and HK include mathematical analyses of the strategies in an information-theoretic context, which focuses on achievable rates at asymptotically large block lengths.   An interesting question that arises is how soft-DF and soft-IC perform in a similar theoretical context, under the same measure of performance.  Existing results in the literature (e.g.,~\cite{SoftICPartial_4,SoftDF1}), however, rely on extensive simulations and heuristic evidence (e.g., EXIT charts), focusing on practical implementation rather than a rigorous analysis.  One exception~\cite{Laneman_New} will be discussed shortly.\footnote{In~\cite{Dabora}, the authors provide an analysis of {\it estimate-and-forward}.  In their context, however, it does {\it not} involve estimation at the relay, and is synonymous with compress-and-forward~\cite{Gerhard_Gaussian_Relay,Junshan}.}

In this paper we develop rigorous bounds on the achievable rates with soft-DF and soft-IC, similar to the ones on the partial-DF and HK strategies.  One important difference
involves our codes.   The results for the latter two strategies~\cite{Gerhard_Maric_Cooperative,HanKobayashi} focus on randomly generated\footnote{With memoryless, identically distributed code symbols.} component codes (within the rate-splitting framework).  In our analysis of soft-DF and soft-IC, we focus on structured codes, namely LDPC codes~\cite{Gallager_PhD}.

Our interest in LDPC codes, however, is motivated not by their practical benefits.  Kramer~\cite{Gerhard_Turbo_Symp} argued that research of such codes may provide insight into fundamental communication limits of multi-terminal channels.  Random codes  were traditionally used in the information-theoretic literature because of their analytic tractability, and also because they were the first to be proven asymptotically optimal for point-to-point channels.  However, such codes are not guaranteed to be optimal in multi-terminal settings.  Recent results (e.g., Philosof and
Zamir~\cite{PhilosofRandom}, Nazer and Gastpar~\cite{Nazer} and Narayanan~\etal~\cite{Narayanan}) point to several advantages of non-random, structured codes in various settings.\footnote{Koetter~\etal~\cite{Koetter_Networks1,Koetter_Networks2} have taken the opposing view, and argue in favor of concatenating randomly generated (or any other) P2P-optimal codes with combinatorial network codes.  They prove optimality, however, only for networks of decoupled point-to-point channels (P2P), leaving out many interesting scenarios, including the ones of this paper.}  As the capacity regions of the relay and interference channels are in general unknown, our results are of similar theoretical interest.

LDPC codes have several advantages which make them well suited for a theoretical analysis.  Most importantly, they possess an analytically tractable soft-estimation algorithm.  LDPC codes' belief propagation (BP) algorithm, which is typically used for decoding, actually computes bitwise estimates.  The algorithm is often applied in scenarios where the level of channel
noise is low enough for the estimation error to be negligibly small, essentially amounting to complete (and not partial) decoding.  In this paper we apply it in other scenarios as well.\footnote{A similar approach was taken by
Barak~\etal~\cite{Ohad}, in the context of communication over
erasure channels with unknown erasure probabilities.}  BP can be analyzed using the {\it density evolution} paradigm~\cite{Luby_Irregular,Urbanke_Message_Passing}, which provides tight bounds at asymptotically large block lengths, similar to the random-coding analysis of information theory.  Finally, LDPC codes are a broad family, which includes a variety of codes.  By manipulating the codes' {\it degree distributions}, they can be tailored to a diverse range of channels and applications.

While our main interest is theoretical, our analysis may also benefit practical applications.  Density evolution (mentioned above), despite its focus on asymptotically long block lengths, plays a valuable role in the design of practical codes for P2P communications (in specifying the codes' degree distributions).  Our work can similarly be applied to codes for soft-DF and soft-IC.

In~\cite{Laneman_New,Jafar}, the authors provide an analysis of {\it demodulate-and-forward}\footnote{In~\cite{Jafar} the strategy is called {\it estimate-and-forward}, not to be confused with a different interpretation of the same term in~\cite{Dabora}.} (DmF), which is closely related to soft-DF.  Unlike soft-DF, however, the DmF relay applies bit-wise memoryless estimation, i.e., the estimation of each codebit's value relies only on the corresponding channel output symbol.   Such estimation is in general suboptimal.  With soft-DF (e.g.,~\cite{SoftDF1}), estimation is aware of the code structure, and exploits the statistical dependencies between codebits, that the structure implies.

Such code-structure-aware estimation presents a challenge to the analysis of soft-DF.  The non-memorylessness of the relay's estimation implies that the components of the estimation error vector, unlike white noise, are statistically dependent (in general), and the correlation patterns are complex.  Furthermore, the error cannot be argued to be independent of the codebook, because the estimation process relies on its structure.  This complicates the analysis of the destination decoder, which (as noted above) uses a noisy version of the relay's estimate, obtained via the channel between them.

In this paper, we develop a technique called {\it simultaneous density evolution} (sim-DE), which overcomes this problem for erasure relay channels.  As with standard density evolution, sim-DE provides rigorous bounds on the decoding error at asymptotically large block lengths.  The technique applies to soft-DF-BP, an instance of soft-DF which uses LDPC codes and invokes BP both for estimation and decoding.

To further increase the achievable rates, we also develop soft-DF-BP2, which improves upon soft-DF-BP by using Wyner-Ziv coding~\cite{WynerZiv} to enhance the communication between the relay and destination.  This idea borrows from compress-and-foward (CF)~\cite{CoverElGamal,Gerhard_Gaussian_Relay,Junshan}.  We develop rigorous bounds on the achievable compression rates, and show how the structure of LDPC codes can be exploited to improve them.  Our bounds make use of some standard analysis tools for LDPC codes, including {\it stopping sets}~\cite{Urbanke_Book}.

Unlike soft-DF, analysis of soft-IC is straightforward using existing techniques in the literature.  We define soft-IC-BP as
an instance of {\it iterative multiuser detection} (e.g.,~\cite{boutros-caire,AmraouiUrbanke}), and show how the strategy can be used to achieve partial decoding of the interference, when complete decoding is not possible.    We focus on binary-input additive white Gaussian noise (BIAWGN) interference channels, and use density evolution to obtain a rigorous analysis.

We are also interested in deeper insight into properties of codes that work well with soft-DF and soft-IC.  Interestingly, both strategies appear to explicitly disfavor optimal P2P codes in many contexts.  Optimal codes for BIAWGN P2P channels, for example, are defined by their reliable (asymptotically zero-error) decoding at SNRs above the Shannon limit.\footnote{Rigorous definitions, which consider code {\it sequences}, will be provided in Sections~\ref{sec:Limitations_Good_Codes} and~\ref{sec:Good-Interference}.}  Such codes, however, exhibit a {\it threshold effect} (see Fig.~\ref{fig:1}), which makes them less suitable to partial decoding: At SNRs {\it below} the Shannon limit, where complete decoding is not possible, their estimation error ``explodes,'' and coincides with that of an uncoded stream of bits.  P2P-suboptimal codes, by contrast, typically exhibit a {\it lower} estimation error at such SNRs.  This fact has long been known in the design of turbo codes~\cite[Sec.~II.B]{Turbo_Codes} and was proven by Peleg~\etal~\cite{PelegExtrinsicGood} to hold for {\it any} P2P-optimal code sequence, over BIAWGN P2P channels.\footnote{In related work,
Bustin and Shamai~\cite{RonitShlomo}  focused on the MMSE curve.  They showed that Gaussian superposition codes are optimal in terms of shaping the curve at one SNR, subject to minimal constraints on the decoding SNR threshold and rate.  They also showed that such codes are characterized by {\it two} thresholds, analogous to the threshold displayed in Fig.~\ref{fig:1} (see also Merhav~\etal~\cite{NeriGuoShlomo}).}

P2P-optimal codes' threshold effect, however, does not establish their unsuitability when partial decoding is only one component of a larger strategy.  For example, with both the soft-IC and HK strategies, each interference-channel destination completely decodes the signal from the corresponding source, in addition to partially decoding the interference.
In this paper, we establish limitations on the codes' performance over BIAWGN interference channels.  Namely, we prove that with any P2P-optimal codes and regardless of the decoding strategy, reliable communication over such channels is bounded by the rates achievable with multi-user detection (MUD) and single-user detection (SUD).\footnote{A similar result was developed independently in~\cite[Theorem~2]{Shlomo_Rem2}, but in the context of {\it non}-binary-input AWGN interference channels.}  The latter two strategies correspond to complete decoding of the interference, and no decoding of it, respectively.  P2P-suboptimal codes, by contrast, can often be designed to achieve better rates.

\addtocounter{footnote}{+1}

\myfigure{file=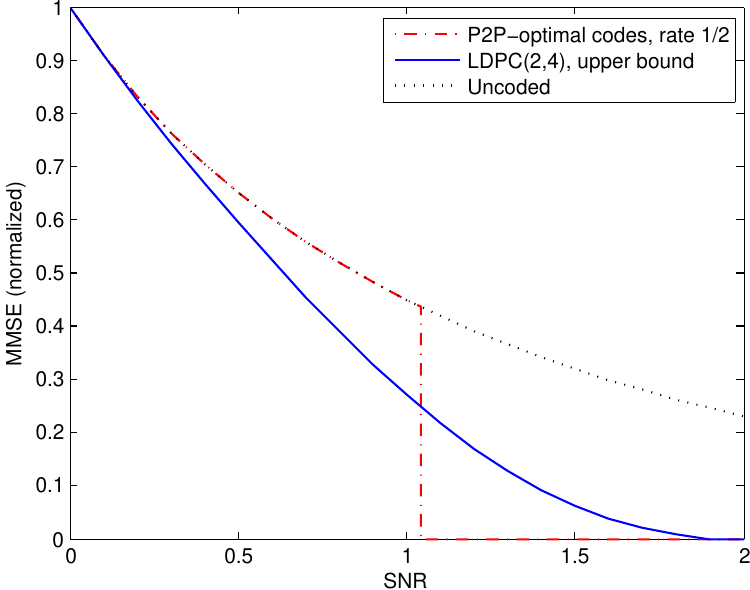, width = 8cm}{The threshold effect of P2P-optimal codes over a BIAWGN P2P channel.   The LDPC (2,4) curve corresponds to a specific ensemble of LDPC codes known to be P2P-suboptimal.  At SNRs above the Shannon limit (SNR~$>$~1.044), the P2P-optimal code's MMSE (which is zero) outperforms the suboptimal LDPC code's upper bound.$^{12}$  Below this limit, the optimal code's MMSE abruptly ``explodes" and coincides with the MMSE of uncoded communications, while the LDPC code achieves substantially better performance.  See Appendix~\ref{apdx:Rigorous_Fig_Bad_Codes} for a rigorous discussion.}{fig:1}

By this discussion, LDPC codes may appear unsuitable for soft-DF and soft-IC, as they are known primarily for their near-optimal performance over many P2P channels (see e.g., Miller and Burshtein~\cite{GadiDavid_MLD}).  However, as noted above, by manipulating their degree distributions, a variety of LDPC codes can be constructed, many of which (including the LDPC (2,4) codes examined in Fig.~\ref{fig:1}) are P2P-suboptimal.

This paper is organized as follows.  In
Sec.~\ref{sec:Preliminaries} we introduce preliminary
notations, definitions and relevant background on LDPC codes.  In
Sec.~\ref{sec:BEC_relay} we define erasure relay channels, soft-DF-BP and develop sim-DE.  We briefly discuss the limitations of P2P-optimal codes.  In Sec.~\ref{sec:Soft-DF-BP2} we define soft-DF-BP2 and develop bounds on its performance.  In Sec.~\ref{sec:Interference} we define BIAWGN interference channels and soft-IC-BP.  We also prove the limitations of P2P-optimal codes in this context. In Sec.~\ref{sec:Numerical} we present numerical examples
of our strategies, and compare them to other strategies.  In
Sec.~\ref{sec:Conclusion} we discuss our results and  conclude the paper.
\footnotetext{While relying on the LDPC code's upper bound may seem unfair at SNRs above the Shannon limit, we conjecture that the true MMSE is not much better.  More importantly, our main interest is in SNRs below the limit.}

\section{Preliminaries}\label{sec:Preliminaries}

\subsection{General Notation}\label{sec:Notation}
Vector values will be denoted by boldface (e.g., $\bx$) and scalars by normalface (e.g., $x$).  Random variables will be upper-cased ($X$) and their instantiations lower-cased ($x$).  The expectation operator will be denoted by $\EE$. The exponential function will be denoted by $\exp(x)$ and $\e^x$ (we will use both notations interchangeably). The natural logarithm (to the base $\e$) will be denoted by $\ln$ and the base 2 logarithm by $\log$.  Correspondingly, all communication rates are given in bits per channel use.
The interval  $\{x\in\RR:a \leq x \leq b\}$ will be denoted by $[a,b]$, and the interval $\{x\in\RR:a < x < b\}$ by $(a,b)$.

Given a node $i$ in a graph,  the set of nodes that are adjacent to $i$ will be denoted by $\cN(i)$.  The binary entropy function will be denoted by $h(x)$, i.e.,
\begin{eqnarray*}
h(x) = -x\cdot \log x - (1-x)\log(1-x).
\end{eqnarray*}

The block length of a code will be denoted by $n$, unless stated otherwise.  We will let $o(1)$ denote a term that approaches zero as $n\rightarrow\infty$.
\subsection{BIAWGN and Binary Erasure Channels}\label{sec:BIAWGN and BEC}

We now define two point-to-point (P2P) channels which we will use throughout the paper to classify codes as P2P-optimal or P2P-suboptimal.  The binary-input additive white Gaussian noise (BIAWGN) channel is characterized by the equation
\begin{eqnarray}\label{eq:BIAWGN}
Y = X + Z,
\end{eqnarray}
where $Y$ is the channel output, $X$ (the transmitted signal) is taken from $\{\pm1\}$, and $Z$ is a zero-mean real-valued Gaussian random variable with variance $\sigma^2$, whose realizations at different time instances are statistically independent.  $\sigma$ is a positive constant.

The binary erasure channel (BEC) is characterized by
\begin{eqnarray}\label{eq:43}
Y = \left\{
      \begin{array}{ll}
        \Erasure, & \hbox{with probability $\varepsilon$} \\
        X, & \hbox{otherwise,}
      \end{array}
    \right.
\end{eqnarray}
where $Y$ is the channel output,  $X$ is the channel input, and is taken from $\{0,1\}$, $\varepsilon \in [0,1]$ is a constant.  The symbol $\Erasure$ indicates an ``erasure'' event.  We assume that the channel transitions at different time instances are independent.  We let BEC($\varepsilon$) denote a BEC with erasure probability $\varepsilon$.

\subsection{Notations for Analysis of Erasures}\label{sec:Notation_Erasures}
The following notations will be useful in our analysis of erasure channels.  For simplicity, we rewrite~\eqref{eq:43} as
\begin{eqnarray}\label{eq:BEC_Erasure_Noise}
Y = X + E,
\end{eqnarray}
where $E$ is an {\it erasure noise} random variable, denoted $E \sim \eras(\varepsilon)$.  $E$ equals $\Erasure$ with probability $\varepsilon$ and $0$ otherwise.  Addition  of two values $x_1, x_2 \in \{0,1,\Erasure\}$ is defined as
\begin{eqnarray}
x_1 + x_2 \defined \left\{
                                                            \begin{array}{ll}
                                                              \Erasure, & \hbox{$x_1 = \Erasure$ or $x_2 = \Erasure$} \\
                                                              x_1\xor x_2, & \hbox{otherwise,}
                                                            \end{array}
                                                          \right. \label{eq:Erasure_Addition}
\end{eqnarray}
where $\xor$ denotes modulo-2 addition.  Note that the sum $E_1 + E_2$ of two independent erasure noise variables $E_1 \sim \eras(\varepsilon_1)$ and $E_2 \sim \eras(\varepsilon_2)$ is also an erasure noise, distributed as $\eras(\varepsilon_1 \circ \varepsilon_2)$ where
\begin{eqnarray}\label{eq:circ}
\varepsilon_1 \circ \varepsilon_2 \defined \varepsilon_1 + \varepsilon_2\cdot(1 - \varepsilon_1).
\end{eqnarray}
We also define the product of two values $x_1, x_2 \in \{0,1,\Erasure\}$ as follows.
\begin{eqnarray}
x_1\cdot x_2 \defined \left\{
                                                            \begin{array}{ll}
                                                              \Erasure, & \hbox{$x_1 = x_2 = \Erasure$} \\
                                                              x, & \hbox{$x \in \{0,1\}$ and:\:\:\:\:$x_1 = x, x_2 = \Erasure$, or}\\                                                              & \hbox{\quad\quad\quad\quad\quad\quad\quad\:$x_1 = \Erasure, x_2 = x$, or}\\
                                                              \: & \hbox{\quad\quad\quad\quad\quad\quad\quad\:$x_1 = x_2 = x$}\\
\textrm{0}, & \hbox{$x_1, x_2 \in  \{0,1\},\:x_1\neq x_2$.}
                                                            \end{array}
                                                          \right.\label{eq:Erasure_Multiplication}
\end{eqnarray}
Note that although the product $x_1\cdot x_2$ is defined in the last case ($x_1, x_2 \in  \{0,1\},\:x_1\neq x_2$) we will not encounter this case in practice.  We define the product between two vectors $\bx_1,\bx_2 \in \{0,1,\Erasure\}^n$ as the vector obtained by multiplying their respective components.

We define the {\it erasure rate} of a vector $\bx \in \{0,1,\Erasure\}^n$, denoted $P(\Erasure \given \bx)$, as the empirical fraction of erasures in the vector, i.e.,
\begin{eqnarray}\label{eq:erasure_rate}
P(\Erasure \given \bx) = \frac{1}{n}\#\{i\::\: x_i = \Erasure \},
\end{eqnarray}
where the $\#$ symbol denotes the cardinality of the set.  We define the {\it erasure indicator vector} of $\bx$, denoted $I_\Erasure (\bx)$, as the following vector.
\begin{eqnarray}\label{eq:erasure_indicator}
\be = I_\Erasure (\bx)\quad \Leftrightarrow \quad
e_i = \left\{
              \begin{array}{ll}
                \Erasure, & \hbox{$x_i = \Erasure$} \\
                0, & \hbox{$x_i \neq \Erasure$,}
              \end{array}
            \right.\quad i = 1,\ldots,n.
\end{eqnarray}

Finally, we will also be interested in {\it erasure-erasure channels} (EEC), which resemble BEC channels, but accept erasures at their inputs.  That is, the channel equation of an EEC($\varepsilon$) is given by~\eqref{eq:BEC_Erasure_Noise}, where $E\sim\eras(\varepsilon)$ but $X$ is defined over the set  $\{0,1,\Erasure\}$, rather than just $\{0,1\}$.  The capacity $C$ of an EEC($\varepsilon$) can be shown to equal\footnote{This can easily be verified by observing that by symmetry, the EEC capacity-achieving input distribution gives equal probabilities to the symbols 0 and 1.  The right-hand side of~\eqref{eq:C_EEC} is the mutual information $I(X;Y)$ for a given probability $\theta$ of $\Erasure$, and~\eqref{eq:C_EEC2} is the maximizing $\theta$, evaluated by differentiating~\eqref{eq:C_EEC}.}
\begin{eqnarray}\label{eq:C_EEC}
C = h(\theta + (1-\theta)\varepsilon) + (1-\theta)\Big{(}1-\varepsilon -h(\varepsilon)\Big{)},
\end{eqnarray}
where
\newcommand{\AuxK}{{2^{1-h(\varepsilon)/(1-\varepsilon)}}}
\begin{eqnarray}\label{eq:C_EEC2}
\theta = 1 - \frac{\AuxK}{(1+\AuxK)(1-\varepsilon)}.
\end{eqnarray}

\subsection{LDPC Codes}\label{sec:LDPC}
We now briefly describe the essential features of LDPC codes which we will use in our analysis.  See e.g.,~\cite{Urbanke_Book}, for a comprehensive review.

We define LDPC codes in the standard way, using a bipartite {\it Tanner graph}~\cite{Tanner}, as in Fig.~\ref{fig:Tanner}.  The nodes on its left are called
\textit{variable nodes}, and each corresponds to a transmitted codebit.  The nodes on the right are \textit{check nodes}, and each corresponds to
a parity-check.  The codewords of the LDPC code are defined by the condition that at each check node, the set of codebits corresponding to adjacent variable nodes, must sum to zero (modulo-2).

\myfigure{file=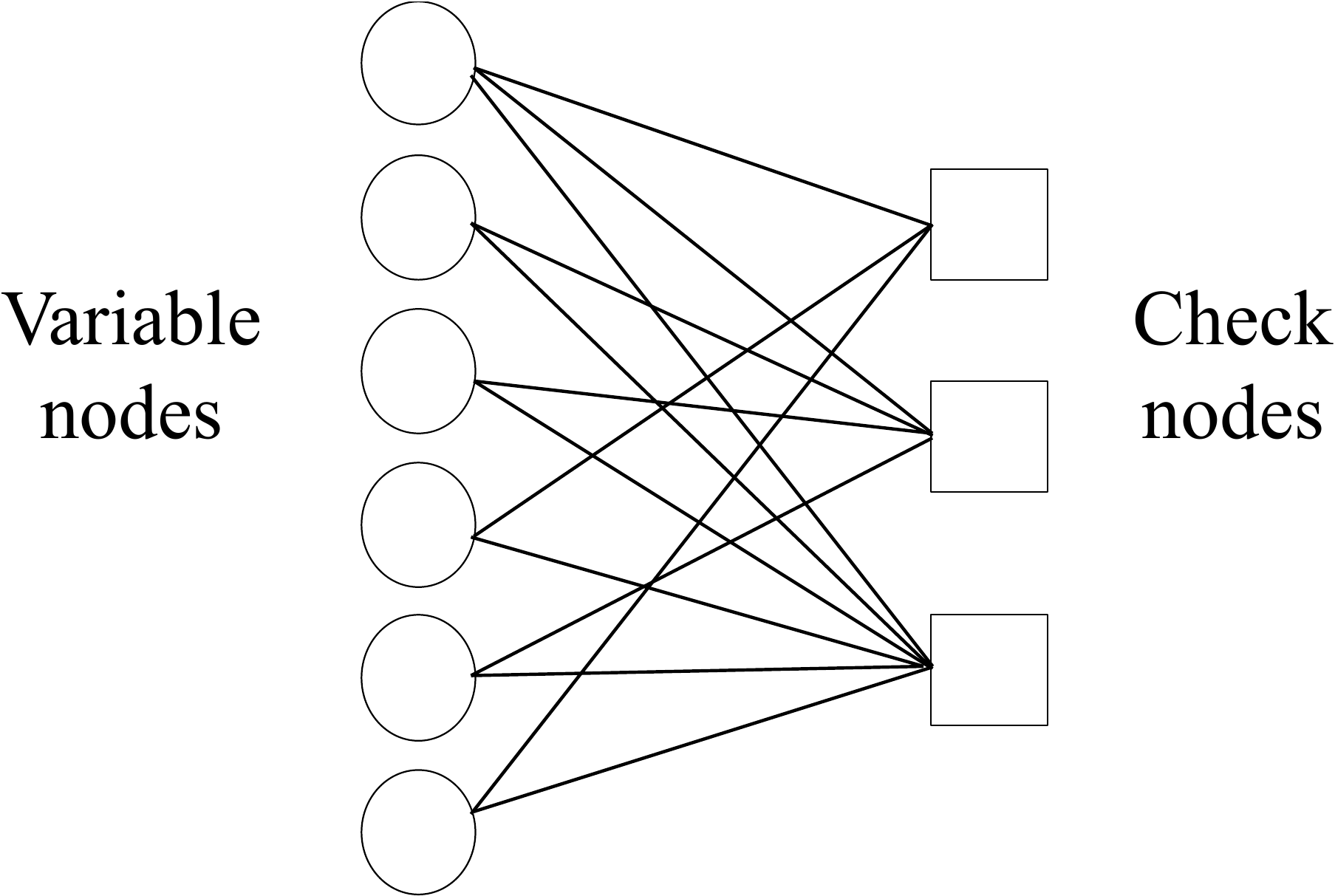, width = 6cm}{An example of the Tanner graph of an LDPC code.  }{fig:Tanner}

Following Luby~\etal~\cite{Luby_Irregular}, we characterize Tanner graphs by their {\it degree distributions} (from an edge perspective) ${\lambda,\rho}$ where $\lambda = (\lambda_1,\ldots,\lambda_{i_\mxx})$ and $\rho =
(\rho_1,\ldots,\rho_{j_\mxx})$.  $\lambda_i, i = 1,\ldots,i_\mxx$, denotes the fraction of graph edges which have left degree $i$, i.e., are connected to a variable node of degree $i$, and $\rho_j, j = 1,\ldots,j_\mxx$, denotes the fraction of edges having right degree $j$, i.e., are connected to a check node of degree $j$.  While $\lambda_i$ refers to the fraction of edges, the fraction of variable nodes of degree $i$ can be shown to equal\footnote{The fraction $\trho_j$ of check nodes of degree $j$ can similarly be obtained from $\rho$, but is not required in our analysis.}
\begin{eqnarray}\label{eq:tlambda}
\tlambda_i = \frac{\lambda_i/i}{\sum_{k=1}^{i_\mxx} (\lambda_k/k)}.
\end{eqnarray}

A Tanner graph is said to be $(c,d)$-regular if it is characterized by $(\lambda,\rho)$ where $\lambda_c = 1$  and $\rho_d = 1$, i.e., all variable nodes have degree $c$ and all check nodes have degree $d$.   A graph is called $(\lambda,d)$-{\it right-regular} if it is characterized by $(\lambda,\rho)$ where $\rho_d = 1$.

The LDPC $(\lambda, \rho)$ code {\it ensemble} is the set of codes whose Tanner graphs are characterized by $(\lambda, \rho)$.  As often encountered in information theory, analysis of LDPC codes is greatly simplified by focusing on the average performance of a code selected at random from such an ensemble, rather than on the performance of an individual code.  We use the procedure of Luby~\etal~\cite[Sec.~III.A]{Luby_Irregular} to randomly generate Tanner graphs that correspond to a given $(\lambda, \rho)$.  Different pairs $(\lambda,\rho)$ may correspond to substantially different performance, and so much of the analysis of LDPC codes focuses on finding effective pairs.  We use the shorthand notation LDPC ($c,d$) to denote the code ensemble that corresponds to ($c,d$)-regular Tanner graphs.

The rate of a $(\lambda, \rho)$ LDPC code is lower bounded by its {\it design rate}, defined as follows.
\begin{eqnarray}\label{eq:design_rate}
R_\design = 1 - \frac{\sum_j \rho_j / j}{\sum_i \lambda_i / i}.
\end{eqnarray}
\subsection{Belief Propagation (BP) over the BEC}\label{sec:BP}
We now provide the details of LDPC codes' BP algorithm,  specialized for case of transmission over the BEC~\cite{Luby_Irregular}.  The input to the algorithm is the channel output vector $\by \in \{0,1,\Erasure\}^n$ and the algorithm's output is a vector $\byBP \in \{0,1,\Erasure\}^n$ of decisions (estimates) for the various bits.  Our analysis also includes cases where many of the components $\{\yBP{i}\}$ remain erasures, indicating that the algorithm was unable to decode the corresponding transmitted bits.  Such scenarios correspond to partial decodings of the transmitted codeword.

The algorithm relies on the Tanner graph, as defined above.  We make use of notation which was introduced in Sec.~\ref{sec:Notation_Erasures}.
\begin{algorithm}[Belief-propagation (BP) over the BEC]\label{alg:BP}$\:$
\begin{enumerate}
\item {\bf Iterations:} Perform the following steps, alternately.
\begin{itemize}
\item {\it Variable-to-check iteration number $\ell \ge 0$}: At all edges $(i,j)$ compute the variable-to-check (or right-bound) messages $\rb{\ell}{i j}$ as follows.
\begin{eqnarray}\label{eq:Variable-to-check}
\rb{\ell}{i j} = \left\{
                      \begin{array}{ll}
                        y_i, & \hbox{$\ell = 0$,} \\
                       y_i\cdot \prod_{j'\in \cN(i)\backslash j}\lb{\ell}{j'i}  , & \hbox{$\ell > 0$,}
                      \end{array}
                    \right.
\end{eqnarray}
where $\lb{\ell}{j' i}$ is a check-to-variable message computed in the preceding check-to-variable iteration, and multiplication is defined as in~\eqref{eq:Erasure_Multiplication}.

\item {\it Check-to-variable iteration number $\ell \ge 1$}: At all edges $(j,i)$ compute the check-to-variable (or left-bound) messages $\lb{\ell}{j i}$ as follows.
\begin{eqnarray}\label{eq:Check-to-variable}
\lb{\ell}{j i} = \sum_{i'\in \cN(j)\backslash i}  \rb{\ell-1}{i' j},
\end{eqnarray}
where addition is defined as in~\eqref{eq:Erasure_Addition}.
\end{itemize}
\item {\bf Stopping criterion:} The number of variable-to-check messages whose value equals erasure, is computed at the end of each iteration.  This number is guaranteed to decrease (i.e., improve) or to remain the same, from one iteration to the next.  Decoding stops at the first iteration when it has not strictly decreased.
\item {\bf Final decisions:} For each $i = 1,\ldots,n$ compute.
\begin{eqnarray}\label{eq:Final_Iteration}
\yBP{i} = y_i\cdot \prod_{j\in \cN(i)}  \lb{t}{j i},
\end{eqnarray}
where $t$ denotes the number of the last iteration.
\end{enumerate}
\end{algorithm}

We define the estimation error vector, denoted $\beBP$, as the erasure indicator vector $\beBP = I_\Erasure(\byBP)$ (see~\eqref{eq:erasure_indicator}).
By the definition of BP above and the transition probabilities of the BEC (Sec.~\ref{sec:BIAWGN and BEC}), whenever BP outputs a non-erasure symbol (0 or 1), it is guaranteed to be correct.    Thus, the following holds.
\begin{eqnarray}\label{eq:error_vector}
\byBP = \bx + \beBP.
\end{eqnarray}
\section{Soft-DF-BP}\label{sec:BEC_relay}
\subsection{Erasure Relay Channel Model}\label{sec:BEC_relay_model}

While soft-DF-BP can be defined for arbitrary relay channels, in this paper we focus on the following {\it erasure relay} model.  This model is simple enough for rigorous analysis, but retains the essential challenges facing the design of relay communication strategies.

\myfigure{file=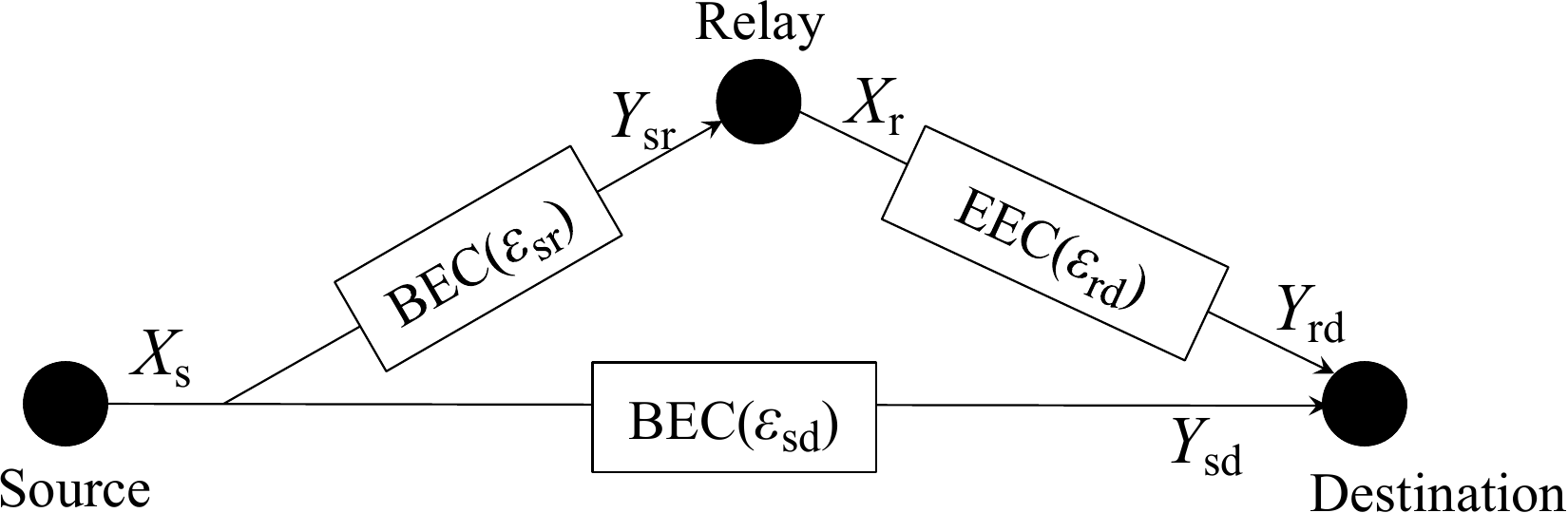, width = 8cm}{The binary erasure relay channel.}{fig:BECRelay}
Fig.~\ref{fig:BECRelay} depicts an $(\varepsilon_\StD,\varepsilon_\StR,\varepsilon_\RtD)$ erasure relay channel. It is a variation of the model suggested by Kramer~\cite{Gerhard_Turbo_Symp} and is a special case of the models of~\cite{Primitive_Relay, CoverElGamal}. It consists of three nodes (source, relay and destination), and three erasure channels, linking the nodes.

Like~\cite{Primitive_Relay}, we assume that the channels to the destination from the source and from the relay are decoupled.  This means that the destination receives two independent channel observations from its links with these nodes.  We continue, however, to assume that the channels from the source to the relay and destination, are coupled, meaning that their channel outputs are dependent on a single channel input broadcasted by the source.  Using the notation of Sec.~\ref{sec:Notation_Erasures}, the channel equations are given by
\begin{eqnarray}
Y_\StD &=& X_\SRC + E_\StD \label{eq:Source_Dest_Noise}\\
Y_\StR&=& X_\SRC + E_\StR \label{eq:Relay_Erasure_Noise}\\
Y_\RtD &=& X_\RLY + E_\RtD\:, \label{eq:Dest_Erasure}
\end{eqnarray}
where $X_\SRC$ and $X_\RLY$ are the channel inputs at the source and the relay, respectively,  $Y_\StR$,  $Y_\StD$ and $Y_\RtD$ are the outputs of the source--relay, source--destination and relay--destination channels, respectively.  $E_\StD$, $E_\StR$ and $E_\RtD$ are independent erasure noise variables, distributed as $\eras(\varepsilon_\StD)$, $\eras(\varepsilon_\StR)$ and $\eras(\varepsilon_\RtD)$, respectively.

While the channels from the source are binary-erasure channels (BECs), the relay-destination channel is an EEC (see Sec.~\ref{sec:Notation_Erasures}), meaning that its input  $X_\RLY$ is defined over the set $\{0,1,\Erasure\}$, rather than just $\{0,1\}$.

Following~\cite{CoverElGamal}, we assume that the relay is full-duplex, meaning that it can listen and transmit simultaneously.  We define communication strategies and achievable rates in the standard way, see~\cite{CoverElGamal}.  Specifically, the signal transmitted by the relay at time $i$ may depend only on the channel outputs it observed at times $j = 1,\ldots,i-1$.

\subsection{Definition of Soft-DF-BP}\label{sec:Soft-DF-BP}

Soft-DF is based on amplify-and-forward (AF)~\cite{Laneman_PhD}. With AF, the relay forwards its channel output vector to the destination, via its channel to that node. The destination thus obtains a noisy version of this vector.  It combines it with its observation of the source-destination channel output, and attempts to decode using both. The vector received from the relay experiences the accumulation of the noises along the source-relay and the relay-destination links.

With soft-DF, the relay first attempts to estimate the source's codeword from its received channel output.  It forwards this estimate to the destination, rather than the raw channel output.  Estimation reduces the level of noise in the forwarded signal, thus improving its quality and the performance of the destination.

\myfigure{file=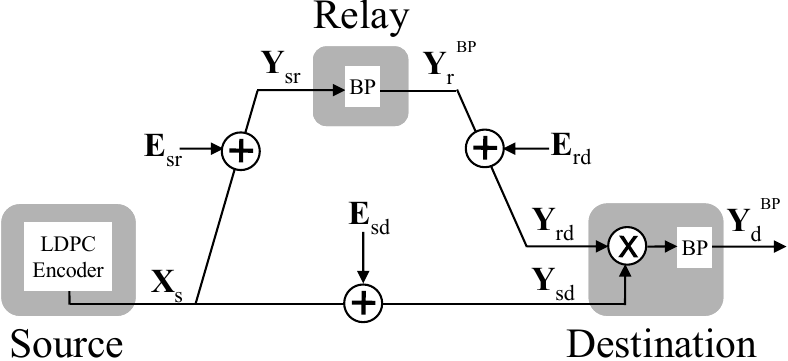, width = 8cm}{Soft-DF-BP. We have replaced the channels of Fig.~\ref{fig:BECRelay} with the actual channel equations,~\eqref{eq:Source_Dest_Noise}, \eqref{eq:Relay_Erasure_Noise}, \eqref{eq:Dest_Erasure}.  Our use of uppercase (unlike our description in Algorithm~\ref{alg:Soft-DF-BP}) relates to our analysis, where the respective values are random variables.}{fig:soft-DF-BP}

Soft-DF-BP is illustrated in Fig.~\ref{fig:soft-DF-BP}.  It is a realization of soft-DF which uses an LDPC code $\cC$ at the source, and applies BP (Algorithm~\ref{alg:BP}) at both the relay and the destination, to achieve soft estimation and decoding, respectively.  Formally, the strategy is provided below:
\begin{algorithm}[Soft-DF-BP] \label{alg:Soft-DF-BP}$\:$
\begin{itemize}
\item {\bf Source:}  Select a codeword $\bx_\SRC$ from the code $\cC$, and transmit it over the channel.
\item {\bf Relay:} Apply BP with respect to $\cC$ to compute the estimate of $\bx_\SRC$ from the channel output $\by_\StR$.  The estimate is denoted $\byBPrelay$, and is transmitted to the destination.
\item {\bf Destination:}
Apply BP with respect to $\cC$ to decode the vector $\by_\RtD\cdot\by_\StD$ (i.e., provide this vector at the input to the algorithm), where multiplication is defined as in Sec.~\ref{sec:Notation_Erasures}.  The output of the algorithm is denoted $\byBPdest$.
\end{itemize}
\end{algorithm}

Note that the vector $\by_\RtD\cdot\by_\StD$, which is used above at the destination, can be shown to be a sufficient statistic for $(\by_\RtD, \by_\StD)$ with respect to $\bx_\SRC$.

Recall from Sec.~\ref{sec:BP} that the output of BP includes erasures at indices whose corresponding transmitted codebits were not discovered.  Our analysis of soft-DF-BP will thus focus on the erasure rate $P(\Erasure \given \byBPdest)$, defined as in~\eqref{eq:erasure_rate}.
\subsection{Simultaneous Density Evolution}\label{sec:Simultaneous_DE}
Over P2P channels, analysis of BP is typically achieved using the density evolution paradigm~\cite{Luby_Erasure}, \cite{Urbanke_Message_Passing}, \cite{Urbanke_Book}.  This paradigm includes a numerical algorithm for approximating the erasure rate at the output of BP.  The approximation is asymptotically precise, in the sense that the realized erasure rate can be proven to approach it in probability, exponentially with the LDPC code's block length.   This is known as the {\it concentration theorem}~\cite[Theorem~2]{Urbanke_Message_Passing},~\cite{Luby_Erasure}.

In our context, we would like to apply the paradigm to an analysis of BP at the relay and at the destination.  Namely, we would like to bound the erasure rates of their outputs  $\byBPrelay$ and (more importantly) $\byBPdest$.  The application, however, is complicated by the distribution of the relay's estimation error, which, as mentioned in Sec.~\ref{sec:Introduction}, differs from additive white noise.  This, in turn, means that the input $\by_\RtD\cdot \by_\StD$ to BP at the destination, cannot (in general) be modeled as the output of a memoryless channel, violating an assumption of the above-mentioned
 concentration theorem.  This is best seen from the following derivation.
\begin{eqnarray}
\by_\RtD\cdot \by_\StD  &\addabove{=}{a}& (\bx_\RLY + \be_\RtD)\cdot (\bx_\SRC + \be_\StD) \nonumber \\
&\addabove{=}{b}& (\byBPrelay + \be_\RtD)\cdot (\bx_\SRC + \be_\StD) \label{eq:n1}\\
&\addabove{=}{c}& (\bx_\SRC +  \beBPrelay + \be_\RtD)\cdot (\bx_\SRC + \be_\StD).\nonumber
\end{eqnarray}
(a) follows by the channel equations~\eqref{eq:Source_Dest_Noise} and~\eqref{eq:Dest_Erasure}.  (b) follows by the operation of the soft-DF-BP relay, which sends its estimate ($\byBPrelay$) over the channel (Algorithm~\ref{alg:Soft-DF-BP}).  In (c), we have defined $\beBPrelay = I_\Erasure(\byBPrelay)$ to be the relay's estimation error ($I_\Erasure(\cdot)$ is defined as in~\eqref{eq:erasure_indicator}), and the equality follows as in~\eqref{eq:error_vector}.

By the nature of our estimation, which exploits the code structure and is {\it not} symbol-by-symbol, the components of $\beBPrelay$ are strongly correlated (in general), unlike memoryless noise.  Consequently, the channel from $\bx_\SRC$ to $\by_\RtD\cdot \by_\StD$ is not memoryless.

To overcome this problem, our analysis of soft-DF-BP focuses on the performance of a different algorithm, which we call {\it simultaneous BP} (sim-BP).  The input to sim-BP is a triplet of vectors, $(\by_\StR, \by_\StD, \be_\RtD)$, where  $\by_\StR$ and $\by_\StD$ correspond to the observations of the relay and destination over their respective channels from the source, and  $\be_\RtD$ is the realization of the noise vector along the relay-destination channel.  The output of sim-BP is an estimate of the codeword $\bx_\SRC$ that was transmitted by the source.\footnote{More precisely, a {\it pair} of estimates is outputted, as will be clarified shortly.}

Sim-BP is intended only as a theoretical tool for analysis.  The algorithm cannot be realized, because the relay and destination are physically separated, and so combined access to both their channel observations ($\by_\StR$ and $\by_\StD$) is not possible.  Access to the noise $\be_\RtD$ is also not physically possible.  However, we will design the algorithm so that its output is degraded (in a sense that will be defined) with respect to the output of soft-DF-BP, and thus its performance could be used to bound that
 of soft-DF-BP.  Most importantly, its structure will enable rigorous analysis using an instance of density evolution, which we will call {\it simultaneous density evolution} (sim-DE).

Before describing sim-BP, we first introduce the following notation for the messages of soft-DF-BP's two decoders:  $\rb{\BPR,\ell}{i j}$ and $\lb{\BPR,\ell}{j i}$ denote variable-to-check and  check-to-variable messages (respectively) at iteration $\ell$ of the relay BP decoder, and $\rb{\BPD,\ell}{i j}$ and $\lb{\BPD,\ell}{j i}$ denote the equivalent values at the destination.

Sim-BP is obtained by merging soft-DF-BP's two BP decoders  (at the relay and the destination), and by replacing the use of  $\by_\RtD\cdot \by_\StD$ (discussed above) with different values.   Namely, it is a message-passing algorithm, and alternates between
variable-to-check and check-to-variable  iterations. Its messages are {\it pairs}, denoted $(\rb{\BPR,\ell}{i j},\rb{\BPD,\ell}{i j})$ and $(\lb{\BPR,\ell}{j i},\lb{\BPD,\ell}{j i})$.  Components $\rb{\BPR,\ell}{i j}$ and $\lb{\BPR,\ell}{j i}$ are identical to the messages exchanged with soft-DF-BP by the relay's BP decoder, at iteration $\ell$.  Components $\rb{\BPD,\ell}{i j}$ and $\lb{\BPD,\ell}{j i}$ replace the messages of the destination's BP decoder, and are slightly different, as explained below.

With soft-DF-BP, the components of $\by_\RtD\cdot \by_\StD$ appear in the expression for the variable-to-check messages of the destination's BP decoder~\eqref{eq:Variable-to-check}.  We rewrite this expression below, explicitly as it is used at that node.
\begin{eqnarray}\label{eq:Variable-to-check_Dest}
\rb{\BPD,\ell}{i j} = \left\{
                      \begin{array}{ll}
                        y_{\RtD,i}\cdot y_{\StD,i}, & \hbox{$\ell = 0$,} \\
                       y_{\RtD,i}\cdot y_{\StD,i}\cdot \prod_{j'\in \cN(i)\backslash j}\lb{\BPD,\ell}{j'i}  , & \hbox{$\ell > 0$.}
                      \end{array}
                    \right.
\end{eqnarray}
The components of $\by_{\RtD}$ satisfy (see~\eqref{eq:n1})
\begin{eqnarray}\label{eq:46}
y_{\RtD,i} = \yBPrelay{i} + e_{\RtD,i}
\end{eqnarray}
To overcome the difficulties involving  $\by_\RtD\cdot \by_\StD$, with sim-BP we replace~\eqref{eq:Variable-to-check_Dest} with
\begin{eqnarray}\label{eq:Variable-to-check_Dest_Simultaneous}
\rb{\BPD,\ell}{i j} = \left\{
                      \begin{array}{ll}
                        \hrb{\BPR,\ell}{i j}\cdot y_{\StD,i}, & \hbox{$\ell = 0$,} \\
                       \hrb{\BPR,\ell}{i j}\cdot y_{\StD,i}\cdot \prod_{j'\in \cN(i)\backslash j}\lb{\BPD,\ell}{j'i}  , & \hbox{$\ell > 0$,}
                      \end{array}
                    \right.
\end{eqnarray}
where
\begin{eqnarray} \label{eq:hrb}
\hrb{\BPR,\ell}{i j} \defined \rb{\BPR,\ell}{i j} + e_{\RtD,i},
\end{eqnarray}
and where $e_{\RtD,i}$ is the same realization of the erasure noise as in~\eqref{eq:46} (which, as mentioned above, is included in the input to the algorithm).  Note that while we did not change the expressions for computing messages $\{\lb{\BPD,\ell}{i j}\}$, their dependence on $\{\rb{\BPD,\ell}{i j}\}$ (see below) means that they too differ from the equivalent soft-DF-BP messages.

We summarize this discussion below.  This description applies to the same Tanner graph of the code $\cC$ as used by soft-DF-BP.
\begin{algorithm}[Simultaneous Belief Propagation (sim-BP)] \label{alg:Simultaneous_BP}$\:$
\begin{enumerate}
\item {\bf Iterations:} Perform the following steps, alternately.
\begin{itemize}
\item {\it Variable-to-check iteration number $\ell \ge 0$}: At all edges $(i,j)$ compute the variable-to-check pair $(\rb{\BPR,\ell}{i j},\rb{\BPD,\ell}{i j})$ as follows: $\rb{\BPR,\ell}{i j}$ is computed by the following expression, which was obtained by making appropriate substitutions to~\eqref{eq:Variable-to-check}.
\begin{eqnarray}\label{eq:SimBP_RB_Relay}
\rb{\BPR,\ell}{i j} = \left\{
                      \begin{array}{ll}
                        y_{\StR,i}, & \hbox{$\ell = 0$,} \\
                       y_{\StR,i}\cdot \prod_{j'\in \cN(i)\backslash j}\lb{\BPR,\ell}{j'i}  , & \hbox{$\ell > 0$.}
                      \end{array}
                    \right.
\end{eqnarray}
$\rb{\BPD,\ell}{i j}$ is computed by~\eqref{eq:Variable-to-check_Dest_Simultaneous}.
\item {\it Check-to-variable iteration number $\ell \ge 1$}: At all edges $(j,i)$ compute the check-to-variable pair $(\lb{\BPR,\ell}{j i},\lb{\BPD,\ell}{j i})$ using the following expressions, which were obtained by making appropriate substitutions to~\eqref{eq:Check-to-variable}.
\begin{eqnarray}\label{eq:Sim_BP_Check-to-variable}
\lb{\BPR,\ell}{j i} &=& \sum_{i'\in \cN(j)\backslash i}  \rb{\BPR,\ell-1}{i' j}, \nonumber \\
\lb{\BPD,\ell}{j i} &=& \sum_{i'\in \cN(j)\backslash i}  \rb{\BPD,\ell-1}{i' j}.
\end{eqnarray}
\end{itemize}
\item {\bf Stopping Criterion:} Decoding stops after a pre-determined number $t$ of iterations.\footnote{This is different from our criterion for BP (Algorithm~\ref{alg:BP}), where decoding was allowed to continue until it could no longer improve.  As our goal here is an upper bound on the performance of soft-DF-BP, this is a valid simplification.  Asymptotically, the difference is meaningless.  With BP, our motivation not to limit the number of iterations was to enable its analysis using the stopping set mechanism, see Sec.~\ref{sec:Quantization_Noise}.}
\item {\bf Final decisions:} For each $i = 1,\ldots,n$ compute the pair $(\yBPrelay{i},\yBPdest{i})$ using the following expressions, which were obtained by making appropriate substitutions to~\eqref{eq:Final_Iteration}.
\begin{eqnarray}\label{eq:Sim_BP_Final_Iteration}
\yBPrelay{i} &=& y_{\StR,i}\cdot \prod_{j\in \cN(i)}  \lb{\BPR,t}{j i}, \nonumber \\
\yBPdest{i} &=& y_{\RtD,i}\cdot y_{\StD,i} \cdot \prod_{j\in \cN(i)}  \lb{\BPD,t}{j i},
\end{eqnarray}
where $y_{\RtD,i}$ is computed by~\eqref{eq:46}.
\end{enumerate}

\end{algorithm}
Note that while sim-BP outputs a pair of vectors $(\byBPrelay,\byBPdest)$, in practice we are only interested in $\byBPdest$, which is our final estimate of the transmitted source codeword.  We now relate it to the output of the destination's BP decoder with soft-DF-BP. We begin with the following definition.

\begin{definition}\label{def:Degraded}
Let $\bx \in \{0,1,\Erasure\}^n$ and $\by \in \{0,1,\Erasure\}^n$.  We say that  $\by$ is {\it degraded} with respect to $\bx$ if the set of indices $\{i\::\: y_i = \Erasure\}$ contains the set $\{i\::\: x_i = \Erasure\}$.
\end{definition}
\begin{theorem}\label{theorem:Degrade}
Consider an instance of communication using soft-DF-BP.  Let $\by_\StR$ and $\by_\StD$ denote the outputs of the source-relay and source-destination channels, respectively, and let $\be_\RtD$ denote the noise along the relay-destination channel.   Let $\byBPdest$ denote the strategy's output at the destination.  Let $\byBPdestTag$
denote the output of sim-BP when provided with the above vectors at its input; i.e., with the triplet $(\by_\StR, \by_\StD, \be_\RtD)$.
Then $\byBPdestTag$ is degraded with respect to $\byBPdest$ in the sense of Definition~\ref{def:Degraded}.
\end{theorem}
By this theorem, we can use sim-BP to upper bound the erasure rate at the output of soft-DF-BP at the destination.  The theorem makes intuitive sense, because the messages $\{\rb{\BPR,\ell}{i j}\}$, which sim-BP uses in~\eqref{eq:Variable-to-check_Dest_Simultaneous}, are intermediate values, computed in the process of BP, while the components $\yBPrelay{i}$, which the destination's BP decoder uses in~\eqref{eq:Variable-to-check_Dest}, are final decisions, whose quality is expected to be better.
This argument will be made rigorous in Appendix~\ref{apdx:Proof_of_Lemma_Degrade}.

Sim-BP preserves the essential features of BP which make its analysis using density evolution possible.  Namely, it is a message-passing algorithm, and its inputs $(\by_\StR, \by_\StD, \be_\RtD)$ are outputs of memoryless channels.

Simultaneous density evolution (sim-DE) tracks the quantities $P_\mR^{(\ell)}(x_\BPR, x_\BPD)$ and $P_\mL^{(\ell)}(x_\BPR,x_\BPD)$, corresponding to the joint probability functions of message pairs $(\Rb{\BPR,\ell}{i j},\Rb{\BPD,\ell}{i j})$ and $(\Lb{\BPR,\ell}{j i},\Lb{\BPD,\ell}{j i})$ (upper-case denotes random variables), respectively.

The inputs to sim-DE  are a triplet $(\varepsilon_\StR,\varepsilon_\StD,\varepsilon_\RtD)$ and a pair $(\lambda,\rho)$, that characterize the relay channel and the LDPC code (Sec.~\ref{sec:LDPC}), respectively.  The details of the algorithm are provided in Appendix~\ref{apdx:Details_DE}.  The algorithm concludes by outputting  $P^{(\Final)}(x_\BPR, x_\BPD)$, corresponding to the distribution of a final decision pair $(\YBPrelay{i},\YBPdest{i})$.  We let $P^{(\Final)}_\BPD(\Erasure)$ denote the probability that $\YBPdest{i} = \Erasure$, obtained from the appropriate marginal distribution of $P^{(\Final)}(x_\BPR, x_\BPD)$.  The following theorem relates this value to the performance of sim-BP.

\begin{theorem}\label{theorem:Validity_Simlutaneous_DE}
Consider an application of sim-BP to the outputs and noise of an erasure relay channel. That is, let $(\bY_\StR, \bY_\StD,\bE_\RtD)$ be the random outputs of the source-relay and source-destination channels, and the noise along the relay-destination channel, respectively.  Assume the code $\cC$ used was selected at random from a $(\lambda,\rho)$ LDPC code ensemble of block length $n$ (see Sec.~\ref{sec:LDPC}).
Let $(\bYBPrelay, \bYBPdest)$ denote the output of sim-BP, when provided with $(\bY_\StD, \bY_\StR,\bE_\RtD)$ as inputs, and
applied to the Tanner graph of $\cC$.  Then for any $\xi > 0$ and large enough $n$, the following holds.
\begin{eqnarray*}
\Pr\left[ \left| P(\Erasure \given \bYBPdest) - P^{(\Final)}_\BPD(\Erasure)\right| > \xi \right] < \e^{-\eta \xi^2 n},
\end{eqnarray*}
where $P(\Erasure \given \bYBPdest)$ is the empirical erasure rate (defined as in~\eqref{eq:erasure_rate}), $P^{(\Final)}_\BPD(\Erasure)$ is as defined above, and  $\eta > 0$ is some constant, which is independent of $n$ and $\xi$.
\end{theorem}
The proof of the theorem is an application of~\cite[Theorems~4.94 and~4.96]{Urbanke_Book}\footnote{Its validity follows from the fact that sim-BP is a symmetric message-passing algorithm (where symmetry is defined as in~\cite[Definition~4.83]{Urbanke_Book}) and the channel between components of $\bX_\SRC$ and $(\bY_\StD, \bY_\StR,\bE_\RtD)$ is binary-input, memoryless and output-symmetric, defined as in~\cite[Definition~4.8]{Urbanke_Book}. } and is omitted.

The theorem implies that the erasure rate at the output of sim-BP approaches sim-DE's prediction in probability, exponentially as $n\rightarrow\infty$.   Combined with Theorem~\ref{theorem:Degrade}, the theorem gives us an upper bound on the erasure rate at the output of soft-DF-BP.

\subsection{Limitations of P2P-Optimal Codes}\label{sec:Limitations_Good_Codes}

To demonstrate the unsuitability of P2P-optimal codes to soft-DF-BP, we now extend the results of Peleg~\etal~\cite{PelegExtrinsicGood} from BIAWGN P2P channels to binary erasure P2P channels.  Specifically, we show that P2P-optimal codes exhibit a threshold effect similar to the one mentioned in Sec.~\ref{sec:Introduction}, making them suitable to either complete decoding, or none, but not to partial decoding.  Our results, however, are weaker than our equivalent ones for interference channels (later, Sec.~\ref{sec:Good-Interference}).  Specifically, they are confined to soft-DF, unlike our results in Sec.~\ref{sec:Good-Interference}, which apply to all interference channel strategies.  Our results focus only on soft decoding at the relay, which is only one component of soft-DF.  Lastly, our generalization of the results of Peleg~\etal~\cite{PelegExtrinsicGood} applies only to linear codes.

Our definition of P2P-optimality focuses on {\it sequences} of codes $\cC = \{\cC_n\}_{n=1}^\infty$.  Given such a sequence, we define its {\it rate} as the limit of the rates of the individual codes, if the limit exists.
\begin{definition}\label{def:Good codes}
Let $\cC = \{\cC_n\}_{n=1}^\infty$ be a sequence of codes of rate $R$. We say that $\cC$ is {\it P2P-optimal for the BEC} if the following holds.
    \begin{eqnarray*}
    \lim_{n \rightarrow \infty} \Pe(\cC_n; \varepsilon) = 0,\quad\quad \forall \varepsilon < \varepsilonstar,
    \end{eqnarray*}
where $\varepsilonstar = 1-R$ is the BEC Shannon limit for rate $R$ and $\Pe(\cC_n; \varepsilon)$ is the probability of error under ML decoding, when the code $\cC_n$ is used over a BEC with an erasure probability of $\varepsilon$.
\end{definition}
In the above, we define the Shannon limit in the usual way, as the inverse of the Shannon capacity function.  That is, the BEC Shannon limit for rate $R \in [0,1]$ is the maximal erasure probability $\varepsilonstar$ such that reliable communication is possible at rate $R$.

The following theorem extends the results of Peleg~\etal~\cite{PelegExtrinsicGood} to BEC channels.  The theorem focuses on $P_{\MAP}(\cC;\:\varepsilon)$, the expected erasure rate (defined as in~\eqref{eq:erasure_rate}) at the output of a maximum {\it a posteriori} (MAP) decoder for a linear code\footnote{MAP decoding of linear codes over the BEC produces vectors over the alphabet $\{0,1,\Erasure\}$ (see e.g.,~\cite[Sec.~3.2.1]{Urbanke_Book}).  More precisely, the bitwise {\it a-posteriori} probability $P_{X_i|\bY}(1\given \by)$ on which it relies can be shown to belong to the set $\{0,1,1/2\}$, indicating complete confidence (0 or 1) in the decoded bit or complete lack of it (1/2).  In some formulations of the algorithm, a random decision is made when $P_{X_i|\bY}(1\given \by) = 1/2$.  In this paper, we assume 1/2 is mapped to $\Erasure$, producing the desired alphabet.  } $\cC$, when used over a BEC($\delta$).
\begin{theorem}\label{theorem:Good_over_BEC}
Let $\{\cC_n\}_{n=1}^\infty$ be a sequence of linear codes, of rate $R$, which is P2P-optimal for the BEC.  Then the following holds.
\begin{eqnarray}\label{eq:PMAP_Good}
\lim_{n\rightarrow\infty} P_{\MAP}(\cC_n;\:\varepsilon) = \left\{
                                                       \begin{array}{ll}
                                                         \varepsilon, & \hbox{$\varepsilon > \varepsilonstar$;} \\
                                                         0, & \hbox{$\varepsilon < \varepsilonstar$.}
                                                       \end{array}
                                                     \right.\quad \forall \varepsilon\in[0,1]\backslash\varepsilonstar,
\end{eqnarray}
where $\varepsilonstar = 1-R$  is the BEC Shannon limit for rate $R$.
\end{theorem}
The proof of the theorem is a variation of the proof of~\cite[Eq.~(14)]{PelegExtrinsicGood} and is provided in Appendix~\ref{apdx:Proof_Theorem_Good_over_BEC}.  It relies on the relationship between mutual information and input estimates, which was recently discovered in several contexts (see Ashikhmin~\etal~\cite{AlexeiEXIT}, Palomar and \Verdu~\cite{PalomarMutInf} and \Measson~\etal~\cite{CyrilGeneralizedArea}).

The results of this theorem resemble the ones that were presented in Fig.~\ref{fig:1}.  Specifically, P2P-optimal codes are excellent at low values of $\varepsilon$ (paralleling {\it high} SNRs in Fig.~\ref{fig:1}), and their MAP decoding provides an asymptotically zero average erasure rate.  At high values of $\varepsilon$, however, they abruptly ``explode," and their MAP decoding output closely resembles the raw channel signal at its input.  P2P-suboptimal codes, by comparison, typically exhibit superior performance.  In our work, we are explicitly interested in high communication rates $R > 1-\varepsilon$, where suboptimal codes have an advantage in terms of the estimation error at the relay.

\section{Soft-DF-BP2}\label{sec:Soft-DF-BP2}
\subsection{Definition of Soft-DF-BP2}
Recall that with soft-DF-BP (Algorithm~\ref{alg:Soft-DF-BP}), the relay sends the estimate $\byBPrelay$ over its channel to the destination, which subsequently receives a noisy version of the vector.  With soft-DF-BP2, the two nodes (relay and destination) use Wyner-Ziv coding to improve the communication between them and reduce the level of this noise.   This technique borrows from compress-and-forward (CF)~\cite{Gerhard_Gaussian_Relay}.

\myfigurestar{file=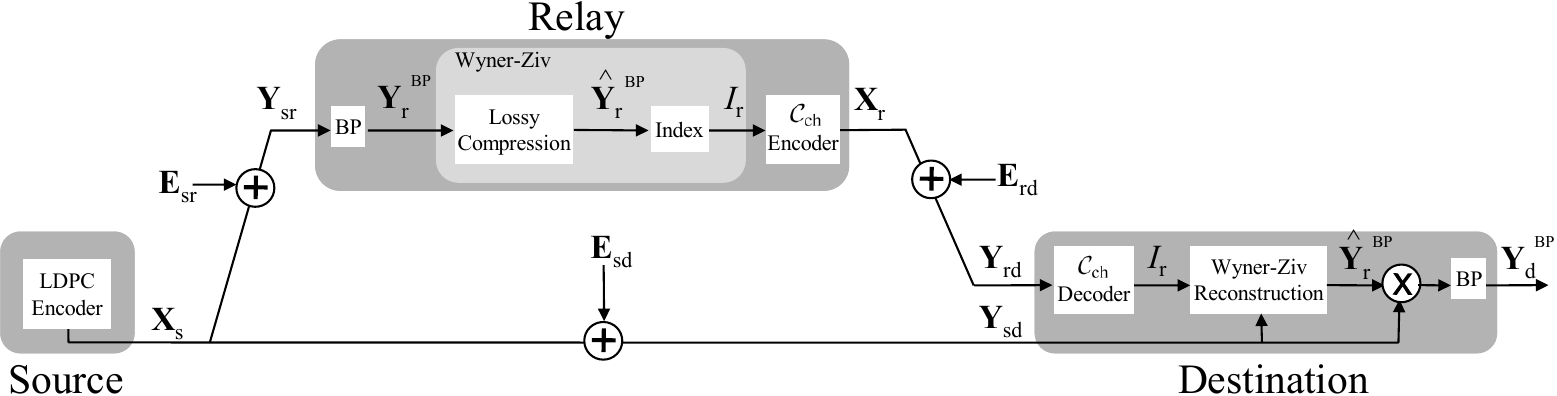, width = 18cm}{Soft-DF-BP2 (non-analysis version).}{fig:soft-DF-BP2}

Specifically, the relay uses Wyner-Ziv coding to compress its estimate $\byBPrelay$, and forwards the resulting signal reliably (i.e., coded) to the destination.
Compression produces a compact representation of the estimate, effectively reducing its block length so that it fits  the capacity of the relay-destination channel. Generally, to achieve the requisite block-length reduction, the compression needs to be {\it lossy}, meaning that it involves distorting the estimate.   This introduces some {\it quantization noise}, which plays a harmful role similar to the channel noise in soft-DF-BP's uncoded communication. To reduce the level of this noise, the  relay and destination apply two techniques.  First, they exploit the statistical dependencies between the components of $\byBPrelay$ to improve the compression rate.
More importantly, they exploit the statistical dependence between $\byBPrelay$ and $\by_\StD$, the source-destination channel output (see Sec.~\ref{sec:BEC_relay_model}).  This vector  serves as {\it side information}, allowing the destination to reconstruct the compressed $\byBPrelay$ with fewer bits obtained from the relay.  The destination uses the reconstructed signal in the same way it used $\by_\RtD$ in soft-DF-BP, to help decode the source codeword.

We begin with a sketch of a simple version of soft-DF-BP2 (see Fig.~\ref{fig:soft-DF-BP2}), which is easy to understand and similar to soft-DF-BP.   Our analysis and numerical results will apply to a different version of
the strategy (henceforth called the {\it analysis version}), which is more involved but more amenable to analysis
(see Sec.~\ref{sec:Analysis_soft_DF_BP2} and Appendix~\ref{apdx:Proof_of_Theorem_Formal}).

Note that unlike soft-DF-BP and soft-IC-BP, our discussion of soft-DF-BP2 does not include a practical algorithm.  Most importantly, we assume that Wyner-Ziv coding is achieved using unstructured, randomly-generated codes, for which practical, low-complexity algorithms are currently unknown.   As noted in Sec.~\ref{sec:Introduction}, however, our main focus in this paper is on a theoretical analysis.

As with soft-DF-BP, we again let $\cC$ denote the LDPC code used by the source and by the BP decoders at the relay and destination.  The strategy also uses another code, denoted $\cC_\textrm{ch}$, explained below.
\begin{algorithm}[Soft-DF-BP2] \label{alg:Soft-DF-BP2}$\:$
\begin{itemize}
\item {\bf Source:}  Select a codeword $\bx_\SRC$ from $\cC$ and transmit it over the channel.
\item {\bf Relay:}
\begin{enumerate}
\item {\it BP:} Apply BP with respect to $\cC$ to compute the estimate of $\bx_\SRC$ from the channel output $\by_\StR$.  The estimate is denoted $\byBPrelay$.
\item {\it Wyner-Ziv:}  Apply a vector quantizer to map $\byBPrelay$ to a distorted version $\hbyBPrelay$, taken from a codebook. Obtain an integer index $I_\RLY$ corresponding to a {\it bin} of $\hbyBPrelay$ (this will be elaborated shortly).
\item {\it Encoder for $\cC_\textrm{ch}$:} Encode $I_\RLY$ using the code $\cC_\textrm{ch}$, to obtain a codeword $\bx_\RLY$, which is transmitted over the channel to the destination.
\end{enumerate}
\item {\bf Destination:}
\begin{enumerate}
\item {\it Decoder for $\cC_\textrm{ch}$:} Decode  $I_\RLY$ from the channel output $\by_\RtD$.
\item {\it Wyner-Ziv Reconstruction:}  Reconstruct $\hbyBPrelay$ from $I_\RLY$, using the output $\by_\StD$ of the source-destination channel as side-information.
\item {\it BP:} Apply BP with respect to $\cC$ to decode the vector $\hbyBPrelay\cdot\by_\StD$ (i.e., provide this vector at the input to the algorithm), where multiplication is defined as in Sec.~\ref{sec:Notation_Erasures}. The output of the algorithm is denoted $\byBPdest$.
\end{enumerate}
\end{itemize}
\end{algorithm}

Following is a brief sketch of Wyner-Ziv coding as used by soft-DF-BP2.  The relay first searches a predefined source codebook $\cC_{\WZ}$ for a codeword $\hbyBPrelay$ that is ``close'' (discussed below) to the estimate $\byBPrelay$.  To conserve bandwidth, rather than send the index of the codeword $\hbyBPrelay$ (within the code $\cC_{\WZ}$), the relay sends the index $I_\RLY$ of its {\it bin}.  That is, the codewords of $\cC_{\WZ}$ are assumed to have been pre-assigned to bins (subsets).  Given the bin number $I_\RLY$, the destination can recover $\hbyBPrelay$ by searching the bin for the source codeword ``near" $\by_\StD$, its own channel output from the source.

In Appendix~\ref{apdx:Proof_of_Theorem_Formal}, we will relate the analysis version  of soft-DF-BP2 (mentioned above) to CF~\cite[Theorem~6]{CoverElGamal},\footnote{The name CF was coined later.} where proximity between vectors is defined in terms of joint-typicality.  We will also follow CF and specify $\cC_\textrm{ch}$, the code used by the relay and destination to communicate $I_\RLY$, as a  randomly generated code.  Similar definitions can be applied to the version above, but are omitted in this paper.
\subsection{Analysis Using Auxiliary Relay Channel}\label{sec:Analysis_soft_DF_BP2}

Our analysis of soft-DF-BP2 follows by applying our results for soft-DF-BP (Sec.~\ref{sec:BEC_relay}).  We define an {\it auxiliary} erasure relay channel model, and prove that under certain conditions, the results of soft-DF-BP over this channel carry over to soft-DF-BP2 over the original, {\it physical} channel.

For this purpose, we model the quantization noise, by which the estimate $\byBPrelay$ and its distorted version $\hbyBPrelay$ are related, as i.i.d. erasure noise.  Formally, we write
\begin{eqnarray}\label{eq:hyBP}
\hYBPrelay{i} = \YBPrelay{i} + \hE_{\BPR,i}\:,
\end{eqnarray}
where $\YBPrelay{i}$, $\hYBPrelay{i}$ and $\hE_{\BPR,i}$ are random variables corresponding to the $i$th components of $\byBPrelay$, $\hbyBPrelay$ and the quantization noise, respectively.
We assume $\hE_{\BPR,i}\sim \eras(\hvarepsilon_\BPR)$ for some $0 \le \hvarepsilon_\BPR \le 1$, and that the components of the noise are mutually independent.

\myfigure{file=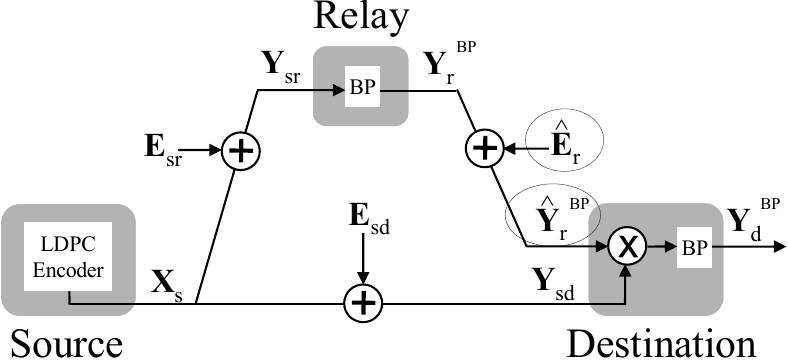, width = 8cm}{Soft-DF-BP over the auxiliary relay channel. This figure is provided for easy reference of our notation, despite its similarity to Fig.~\ref{fig:soft-DF-BP}.  The differences from Fig.~\ref{fig:soft-DF-BP} (encircled) are $\hbYBPrelay$ and $\hbE_\RLY$, which replace $\bY_\RtD$ and $\bE_\RtD$, respectively. }{fig:Stochastic_Relay}

With this model, the stochastic relation between $\byBPrelay$ and $\hbyBPrelay$ is that of a memoryless BEC.   Given an $(\varepsilon_\StD,\varepsilon_\StR,\varepsilon_\RtD)$ erasure relay channel, we define the $(\varepsilon_\StD,\varepsilon_\StR,\hvarepsilon_\BPR)$ auxiliary relay channel (Fig.~\ref{fig:Stochastic_Relay}) as the one obtained by replacing the channel between the relay and destination (in the original channel) with an  EEC($\hvarepsilon_\BPR$) (see Sec.~\ref{sec:Notation_Erasures} for a definition of EEC).  The following theorem relates the performance of soft-DF-BP2 over the physical relay channel, to that of soft-DF-BP over the auxiliary one.

\begin{theorem}\label{theorem:Formal}
Let $(\varepsilon_\StD,\varepsilon_\StR,\varepsilon_\RtD)$ be the parameters of an erasure relay channel as defined in Sec.~\ref{sec:BEC_relay_model}.  Let $\{\cC_n\}_{n=1}^\infty$ be a sequence of LDPC codes with rate $R$, where $n$ is the block length of $\cC_n$, and let $\hvarepsilon_\BPR,\xi \in [0,1]$.  Assume the following conditions hold.
\begin{enumerate}
\item The erasure rate (defined as in~\eqref{eq:erasure_rate}) at the output of soft-DF-BP when applied to the $(\varepsilon_\StD,\varepsilon_\StR,\hvarepsilon_\BPR)$ auxiliary erasure relay, is upper bounded by $\xi$, with a probability that approaches~1 as $n \rightarrow \infty$.
\item The following inequality is satisfied for large enough $n$.
\begin{eqnarray}\label{eq:CF_I_Condition_Block}
\frac{1}{n} I( \bYBPrelay; \hbYBPrelay \given \bY_\StD ) \leq C_\RtD,
\end{eqnarray}
where $C_\RtD$ is the capacity of the relay-destination link, evaluated as in~\eqref{eq:C_EEC}.  The distributions of the various random variables are obtained from the following discussion (see Fig.~\ref{fig:Stochastic_Relay}):  $\hbYBPrelay$ is related to $\bYBPrelay$ via~\eqref{eq:hyBP}.  $\bYBPrelay$ is a deterministic function of the random channel output at the relay $\bY_\StR$, being the output of an application of BP to this vector.  $\bY_\StR$ and $\bY_\StD$ are both obtained from the transmitted codeword $\bX_\SRC$ via the channels from the source to the relay and destination, respectively.  Finally, $\bX_\SRC$ is uniformly distributed within the LDPC code $\cC_n$.
\end{enumerate}
If the above conditions hold, then a rate of $R\cdot (1-h(\xi/R))$ (from source to destination) is achievable using a version of soft-DF-BP2 (the {\it analysis version}), over the $(\varepsilon_\StD,\varepsilon_\StR,\varepsilon_\RtD)$ physical channel.
\end{theorem}
The proof of this theorem, and the above-mentioned analysis version of soft-DF-BP2, rely on the analysis of CF~\cite[Theorem~6]{CoverElGamal} and are provided in Appendix~\ref{apdx:Proof_of_Theorem_Formal}.

The analysis version of soft-DF-BP2 introduces an extra {\it outer} code\footnote{A similar technique was applied in various contexts, e.g., \cite[Theorem~3]{Uri_Lattice_Known_Interference} implicitly relies on a similar derivation.}, which is concatenated with the LDPC code $\cC_n$, and replaces BP decoding at the destination with joint-typicality decoding.  It preserves the main features of Algorithm~\ref{alg:Soft-DF-BP}, namely soft decoding by BP at the relay, and Wyner-Ziv coding.  Typically, we will be interested in negligibly small values of $\xi$ (e.g., $10^{-6}$), and so the term $1-h(\xi/R)$ will be very close to 1.  We conjecture that the theorem can also be extended to the simpler version of soft-DF-BP2 (Algorithm~\ref{alg:Soft-DF-BP2}) but leave  this to future analysis.

Eq.~\eqref{eq:CF_I_Condition_Block} in Theorem~\ref{theorem:Formal} is a condition on the level of quantization noise $\hvarepsilon_\BPR$.  It determines the minimal $\hvarepsilon_\BPR$ required so that the compressed estimate has low enough entropy to enable its communication over the relay-destination channel.  In our analysis of achievable rates (Sec.~\ref{sec:Numerical_Relay} below),  we will choose $\hvarepsilon_\BPR$ as the minimal value still satisfying~\eqref{eq:CF_I_Condition_Block}.


\subsection{Bounds on \normalfont$I(\bYBPrelay; \hbYBPrelay \given \bY_\StD )$}\label{sec:Quantization_Noise}
We now proceed to bound $I( \bYBPrelay; \hbYBPrelay \given \bY_\StD )$ on the left-hand side of of~\eqref{eq:CF_I_Condition_Block}.  As noted above, this value determines the achievable compression rate of  $\bYBPrelay$, as a function of the quantization noise $\hvarepsilon_\BPR$.  In our bound, we will exploit the structure of LDPC codes, and specifically, strong dependence between the components of  $\bYBPrelay$ that result from it.  To demonstrate the significance of this structure,  we begin with a na\"{i}ve bound that does not exploit it.
\begin{remark}\label{remark:ensemble_random}
In our analysis, $I( \bYBPrelay; \hbYBPrelay \given \bY_\StD )$ is actually a random variable.   The distributions of the variables involved in its evaluation (see Sec.~\ref{sec:Analysis_soft_DF_BP2}) are implicitly functions of the code $\cC$, through their dependence on the transmitted codeword $\bX_\SRC$ from $\cC$.  In our analysis, $\cC$ is randomly selected from an ensemble (see Sec.~\ref{sec:LDPC}), and thus, $I( \bYBPrelay; \hbYBPrelay \given \bY_\StD )$ is random too.
\end{remark}
\begin{lemma}[Na\"{i}ve bound]\label{lemma:Naive}
Let $\cC$ be selected at random from a $(\lambda,\rho)$ LDPC
code ensemble\footnote{Note that in Sec.~\ref{sec:LDPC}, we defined the maximal degrees in $\lambda$ and $\rho$ to be finite.  While this is inconsequential for most of the results in this paper, it is essential to the ones of this section.  Namely, their proofs rely (through Lemma~\ref{lemma_DRelayBP}) on~\cite[Theorem~3.107]{Urbanke_Book}, which requires it. } with block length $n$.  Then the following holds for large enough $n$, with probability at least $1 - \exp(-\tau \sqrt{n})$ (the probability being over the random selection of $\cC$).
\begin{eqnarray}
\hspace{-0.7cm}\frac{1}{n}I( \bYBPrelay; \hbYBPrelay \given \bY_\StD ) &\leq& I^+_\textrm{v}(\hvarepsilon_\BPR)  + o(1), \quad\quad \label{eq:Naive_Bound}
\end{eqnarray}
where
\begin{eqnarray*}
I^+_\textrm{v}(\hvarepsilon_\BPR) = h(\deltaRelayBP \circ \hvarepsilon_\BPR)  + (1-\deltaRelayBP \circ \hvarepsilon_\BPR)\cdot\varepsilon_\StD    -   h(\hvarepsilon_\BPR)\cdot(1-\deltaRelayBP),
\end{eqnarray*}
and where  $\deltaRelayBP$ is the expected erasure rate (defined as in~\eqref{eq:erasure_rate}) at the output of the relay's BP decoder (Sec.~\ref{sec:Soft-DF-BP}) as computed by density evolution.\footnote{Standard density evolution for P2P channels~\cite{Luby_Erasure}, \cite{Urbanke_Message_Passing}, \cite{Urbanke_Book}, see Appendix~\ref{apdx:Analysis_DRelayBP} for a brief review.}  $o(1)$ is some function of $n$, dependent on~$\lambda$ and~$\rho$ that approaches zero as $n \rightarrow \infty$.  $\tau> 0$ is a constant similarly dependent on~$\lambda$ and~$\rho$.
\end{lemma}

The proof of the lemma is provided in Appendix~\ref{apdx:Proof_of_Lemma_Naive}.   The lemma is na\"{i}ve in the sense that it ignores statistical dependencies that exist  between the components of the relay estimate $\bYBPrelay$, which result from the structure of LDPC codes.  Specifically, the following observations can be made.

\vspace{0.1cm}
\subsubsection {\bf Dependencies between erasures} In Lemma~\ref{lemma:Naive}, we assumed that the locations (indices) of erased bits in $\bYBPrelay$ resemble the output of a memoryless BEC, and bits are erased or not independent of one another.
 Unlike the output of such a channel, however, the locations are not arbitrary, and can be shown to correspond to a {\it stopping set} of the code $\cC$ (Di~\etal~\cite[Lemma 1.1]{Stopping_Sets}). Typically, the number of stopping sets of a given size is significantly smaller than the number of similar-sized (arbitrary) subsets of $\{1,\ldots,n\}$.  Thus, for the relay to describe to the destination the locations of the erasures in $\bYBPrelay$, it can conserve rate by providing the serial number of a stopping set (given some enumeration of the stopping sets) rather than describing the precise indices.

\subsubsection {\bf Dependencies between non-erasures}  Similarly, in Lemma~\ref{lemma:Naive}, we assumed that the values of non-erased bits at the output of BP at the relay, are statistically independent.  In fact, each bit discovered by BP is equal to the sum of other bits, that were un-erased by the channel, or discovered in previous BP iterations.  A smart compression algorithm spends less rate to communicate those bits, under the assumption that with high probability, the destination will have access to the other bits.

\vspace{0.1cm}

The following theorem exploits these dependencies to obtain tighter bounds.  Unlike Lemma~\ref{lemma:Naive}, the theorem applies to right-regular LDPC codes only (see Sec.~\ref{sec:LDPC}).

\begin{theorem} \label{theorem:I_Bounds}
Let $\cC$ be selected at random from a right-regular LDPC
code ensemble $(\lambda, d)$ with block length $n$.  Then the following holds for large enough $n$, with probability at least $1 - \exp(-\tau \sqrt{n})$.
\begin{eqnarray} \label{eq:I_bound}
\frac{1}{n}I( \bYBPrelay; \hbYBPrelay \given \bY_\StD ) &\leq& I^+(\hvarepsilon_\BPR) + o(1),
\end{eqnarray}
where $\tau$ and $o(1)$ are defined as in Lemma~\ref{lemma:Naive}, and
\begin{eqnarray}\label{eq:I_Plus}
I^+(\hvarepsilon_\BPR) = \textrm{l.d.f.}\left[ I^+_1(\hvarepsilon_\BPR)\right],
\end{eqnarray}
where $ I^+_1(\hvarepsilon_\BPR)$ is provided by equation~\eqref{eq:51} on the following page.
In that equation, $\deltaRelayBP$ is defined as in Lemma~\ref{lemma:Naive}, $f(\cdot)$ is provided by equation~\eqref{eq:f}, $R$ is the design rate~\eqref{eq:design_rate} and $\tlambda_i$ was defined in~\eqref{eq:tlambda}.  The operator $\textrm{l.d.f.}[f_1]$ denotes the largest monotonically decreasing function that is upper bounded by $f_1(\cdot)$.  That is,
\begin{eqnarray*}
f_2 = \textrm{l.d.f.}[f_1] \quad \iff \quad f_2(x) = \inf_{t\leq x} f_1(t)\quad \forall x.
\end{eqnarray*}
\begin{figure*}
\begin{eqnarray}
 I^+_1(\hvarepsilon_\BPR) &=& \varepsilon_\StD(1 -\hvarepsilon_\BPR)\Big[(1-\varepsilon_\StR) +  (\varepsilon_\StR - \deltaRelayBP)\Big(1 - (1-\hvarepsilon_\BPR\cdot \varepsilon_\StD)^{d-1}\Big)\Big]- (1-\deltaRelayBP) \cdot h( \hvarepsilon_\BPR ) + \nonumber \\
&& + \min\Big(h( \deltaRelayBP \circ \hvarepsilon_\BPR),\:f(\deltaRelayBP) + (1 - \deltaRelayBP)\cdot h( \hvarepsilon_\BPR )\Big)\label{eq:51}
\\
f(\alpha) &=&  \max_{\beta \in (0, \gamma)} \left[\log\inf_{x > 0, y > 0}\left(\frac{\prod_i(1 + x y^i)^{\tlambda_i}}{x^\alpha y^\beta}\right)  +
                     \log \inf_{x>0} \left(\frac{\left[(1+x)^d - d\cdot x\right]^{(1-R)}}{x^{\beta}}\right)   -
                    \gamma \cdot h\left( \frac{\beta}{\gamma} \right)\right]\label{eq:f}\\
    \gamma &\defined& \sum_i i\cdot \tlambda_i\nonumber
\end{eqnarray}
\hrulefill
\end{figure*}
\end{theorem}
The proof of the theorem is provided in Appendix~\ref{apdx:Proof_of_Theorem_I_Bounds}.
In Fig.~\ref{fig:QuantizationNoise} we have plotted the bounds of Lemma~\ref{lemma:Naive}
and Theorem~\ref{theorem:I_Bounds} for the erasure relay channel and LDPC code designed for
soft-DF-BP2 in Sec.~\ref{sec:Numerical_Relay} (below).  On this channel, the noise level (erasure probability)
along the relay-destination link is~0.232, and the capacity of this link is $C_\RtD = 0.991$
bits per channel use (evaluated as in~\eqref{eq:C_EEC}).  By Theorem~\ref{theorem:Formal}, we can select $\hvarepsilon_\BPR$ as the
value  where the $I^+(\hvarepsilon_\BPR)$ curve equals $C_\RtD$.
This value, 0.139, is substantially smaller than the original noise level over the link (0.232), demonstrating the effectiveness of Wyner-Ziv coding.

\myfigure{file=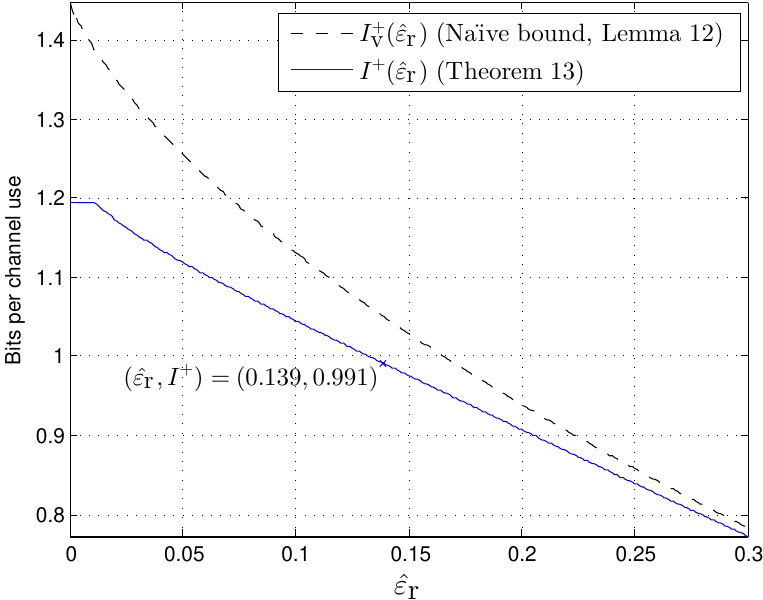, width = 8cm}{Bounds on the achievable compression rate as a function of the quantization noise $\hvarepsilon_\BPR$.  The channel
and LDPC code degree distributions are provided in Sec.~\ref{sec:Numerical_Relay}. Note that the straight line in the vicinity of $\hvarepsilon_\BPR = 0$, in the $I^+(\cdot)$ curve, results from the $\textrm{l.d.f.}[\cdot]$ operation in~\eqref{eq:I_Plus}.}{fig:QuantizationNoise}

\section{Soft-IC-BP}\label{sec:Interference}
\subsection{BIAWGN Interference Channels}\label{sec:BIAWGN_Interference}
We now turn our attention to interference channels, and to communication over them using soft-IC.  Our analysis in this setting focuses on BIAWGN interference channels.  While erasure interference models can be defined (paralleling our above erasure relay model), they typically correspond to {\it strong} interference scenarios, for which the optimal communication strategy is known and achieved by multiuser detection~\cite{StrongInterference}.   In this paper we are interested in the more-challenging {\it weak} interference scenarios.  Such scenarios are characteristic of many wireless settings, where the interfering signal at each destination is weaker than the desired signal from the corresponding source.

Fig.~\ref{fig:BIAWGNInterference} depicts the $(h_1,h_2,\sigma_1,\sigma_2)$ BIAWGN interference channel.  The channel transition probabilities are defined by the following equations.
\begin{eqnarray}
Y_1 &=& X_1 + h_1\cdot X_2 + Z_1 \nonumber\\
Y_2 &=& h_2\cdot X_1 + X_2 + Z_2, \label{eq:Symm_BIAWGN_Interference}
\end{eqnarray}
where $Y_1$ and $Y_2$ are the channel outputs at the two destinations (respectively). $X_1$ and $X_2$ are the transmitted signals.  Unlike typical formulations of AWGN interference channels (e.g.,~\cite{TseOneBit}) we restrict $X_1$ and $X_2$ to $\{\pm1\}$.  $Z_1$ and $Z_2$ are statistically independent zero-mean real-valued Gaussian random variables with variances $\sigma_1^2$ and $\sigma_2^2$, respectively, whose realizations at different time instances are also independent.  $h_1,h_2,\sigma_1$ and $\sigma_2$ are positive constants, known to all nodes.  We confine our attention to weak interference scenarios~\cite{StrongInterference} (as noted above) and restrict $h_1$ and $h_2$ to the interval $(0,1)$.

\myfigure{file=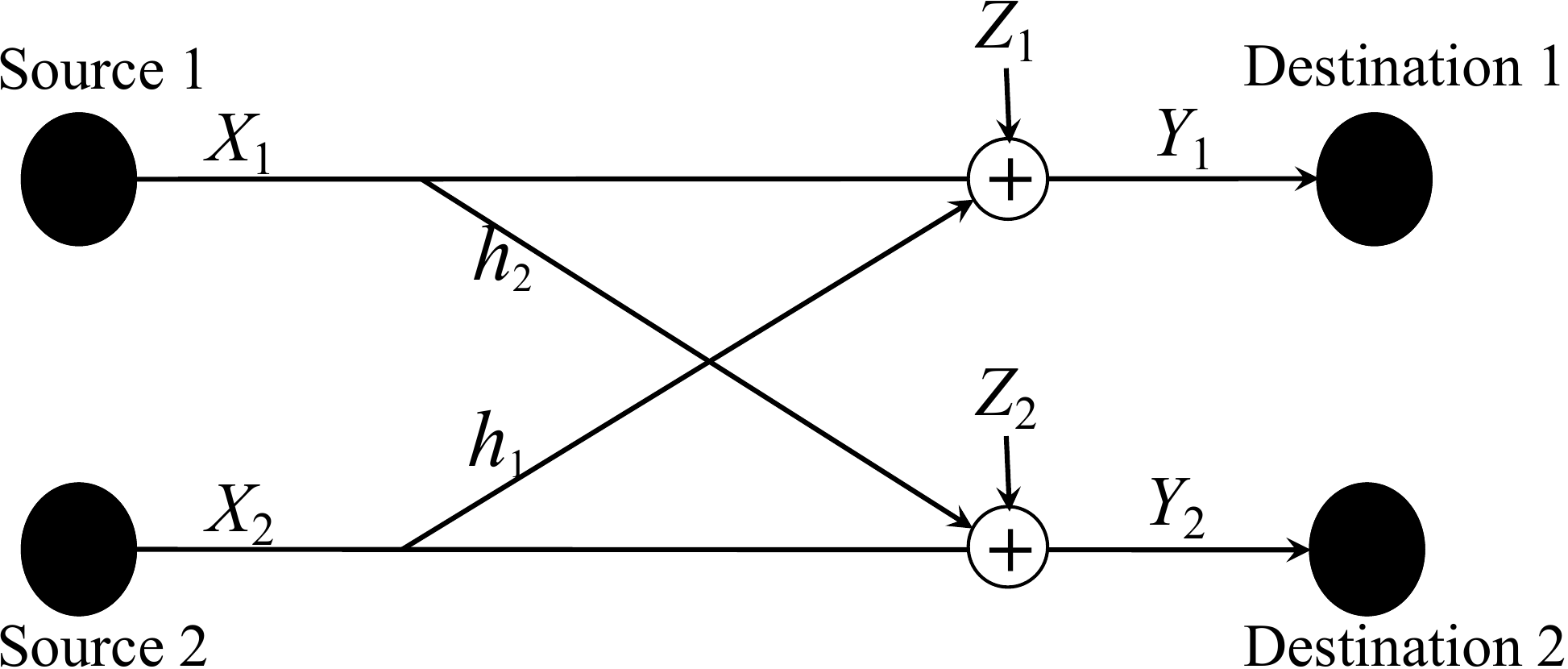, width = 7cm}{The BIAWGN  interference channel.}{fig:BIAWGNInterference}

We define achievable strategies for this channel in the standard way (e.g.,~\cite[Sec.~II]{TseOneBit}).  Theoretic analysis of interference channels typically focuses on the capacity region, i.e., the set of pairs $(R_1, R_2)$ such that communication at rate $R_1$ (resp. $R_2$) is achievable between the first (resp. second) source-destination pair.

In {\it some} of our discussion, we will focus on symmetric rates for symmetric channels, defined as follows.  A $(h,\sigma)$ {\it symmetric} BIAWGN interference channel is a special case where $h_1=h_2=h$ and $\sigma_1=\sigma_2=\sigma$.  An achievable {\it symmetric} rate for this channel is a value $R > 0$ such that the pair $(R,R)$ is achievable.

\begin{remark} The source alphabet of the BIAWGN interference channel, as defined above, is $\{\pm1\}$, while the alphabet of our LDPC code (Sec.~\ref{sec:LDPC}) was defined over $\{0,1\}$.  In our discussion below, we assume some mapping between the two alphabets.\end{remark}
\subsection{Definition and Analysis of Soft-IC-BP}\label{sec:Soft-IC-BP}

Our definition of soft-IC-BP is based on {\it iterative multiuser detection} (iterative-MUD), see e.g., Boutros and Caire~\cite{boutros-caire} and Amraoui~\etal~\cite{AmraouiUrbanke} (and references therein).  The MUD problem is characterized by two (generally, two or more) sources that wish to communicate to one destination.  The MUD destination thus resembles the interference channel destinations, which each receive the output of a combined channel from two sources. Unlike those destinations, however, it is required to decode completely (and not partially) the messages from both sources.
Iterative-MUD (as developed in the above references) is an efficient approach for joint decoding of the two signals.  Like BP, iterative-MUD  is in fact an estimation algorithm (see Sec.~\ref{sec:Introduction}), and approximates bitwise MAP decisions (for both sources).  Typically, the algorithm is applied in settings where both estimation errors can be made negligibly low, essentially amounting to complete decoding of the two signals.

In this paper, we define soft-IC-BP to coincide with iterative-MUD, and
apply it at each of the destinations of our interference channel, to estimate the signals from the two sources (the desired source, and the interfering source).  Unlike the literature on MUD, we are interested in settings where low-error estimation of the two signals is not possible.  Specifically, we tolerate a large error in the estimation of the interference, amounting to partial decoding of that signal.  We continue to require complete decoding (low-error estimation) of the signal from the desired source.

It is convenient to perceive soft-IC-BP as the parallel operation of two decoders, the first estimating the codeword from the desired source, and the second estimating the interference.  The two decoders iteratively exchange information to improve their respective performance.  At each iteration, the information each decoder obtains from the other, assists it in better canceling the signal produced by the other source, in order to better estimate its own signal.  In our work, we assume the two sources use LDPC codes, whose identities will be discussed later.
For a more detailed discussion, see Appendix~\ref{apdx:Overview_Soft_IC_BP}.

Analysis of soft-IC-BP is possible by an application of the density evolution paradigm~\cite[Sec.~IV.A]{boutros-caire} and~\cite[Sec.~5.5]{Urbanke_Book}, similar to the one that was discussed in Sec.~\ref{sec:Simultaneous_DE}, in the context of BP over the P2P BEC.  As in the BEC case, the paradigm includes a numerical algorithm which can be used to approximate the estimation errors (of the desired and interfering signals).  Again, the approximation is asymptotically precise, in the sense that the realized error can be proven to approach it in probability, exponentially with the LDPC code's block length.   For a more detailed discussion, see Appendix~\ref{apdx:Overview_Soft_IC_BP}.

Note that density evolution, as discussed above, produces an (asymptotically precise) approximation of the estimation error of the interfering signal, as well as of the desired signal.  We consider the former a byproduct, and are generally interested in the latter.
\subsection{Limitations of P2P-Optimal Codes}\label{sec:Good-Interference}

We now prove a strong result on P2P-optimal codes over BIAWGN interference channels.  Namely, such codes are incapable of exploiting the power of partial decoding.  Our result applies to {\it any} decoding strategy, including soft-IC and Han-Kobayashi (HK).  For simplicity, we first present our results for symmetric rates over symmetric BIAWGN interference channels.  The extension to the general case is provided in Appendix~\ref{apdx:Arbitrary}.

In this context, our discussion of P2P-optimality focuses on the BIAWGN P2P channel (Sec.~\ref{sec:BIAWGN and BEC}).  Our definition parallels Definition~\ref{def:Good codes} (Sec.~\ref{sec:Limitations_Good_Codes}).
\begin{definition}\label{def:Good codes_AWGN}
Let $\cC = \{\cC_n\}_{n=1}^\infty$ be a sequence of codes of rate $R$.  We say that $\cC$ is {\it P2P-optimal for the BIAWGN channel} if the following holds.
    \begin{eqnarray*}
    \lim_{n \rightarrow \infty} \Pe(\cC_n; \SNR) = 0,\quad\quad \forall \SNR > \SNRstar,
    \end{eqnarray*}
where $\SNRstar$ is the Shannon limit for the BIAWGN channel at rate $R$ and $\Pe(\cC_n; \SNR)$ is the probability of error under maximum-likelihood (ML) decoding, when the code $\cC_n$ is used over a BIAWGN channel with the specified SNR.
\end{definition}

Two communication strategies that do {\it not} involve partial decoding, are multi-user detection (MUD) and single-user detection (SUD).  With MUD, each destination attempts to decode (completely) its respective interference, as well as its desired signal.  With SUD, the destinations do not attempt to decode the interferences at all, instead treating them as random noise vectors, alongside the channel noise.

We now provide achievable rates for both strategies.  The achievability proofs for both rates assume that the transmitters use randomly-generated codes, i.e., codes whose components were selected randomly, independently and with uniform probability from  $\{\pm 1\}$.  Such codes are P2P-optimal for the BIAWGN channel with probability 1 at asymptotically large block lengths. The achievable rates are as follows.
\begin{eqnarray}
R_\MUD &=& \min\Big(  I( X_2; Y_1\: |\: X_1 ),\: \frac{1}{2}I( X_1, X_2;\: Y_1 ) \Big) \quad\quad \label{eq:R_MUD}\\
R_\SUD &=& I( X_1; Y_1) \label{eq:R_SUD},
\end{eqnarray}
where $X_1$ and $X_2$ are uniformly distributed in $\{\pm 1\}$ and are statistically independent.  The joint distribution of $Y_1$ and $X_1,X_2$ is obtained from ~\eqref{eq:Symm_BIAWGN_Interference}.   These two expressions are easily obtained from the analysis of multiple-access channels~\cite[Theorem~14.3.3]{Cover_Book}, and the point-to-point (single-user) channel capacity~\cite[Theorem~8.7.1]{Cover_Book}, respectively.

The above rates correspond to a specific choice of P2P-optimal codes (randomly-generated) and specific decoding strategies at the destinations (SUD and MUD).  We might therefore reasonably expect better performance, by trying other codes and additional strategies. The following theorem proves that if we confine our attention to P2P-optimal codes, this is not possible.

\begin{theorem}\label{theorem:Good_Interference}
Consider communication over a ($h,\sigma$) symmetric BIAWGN interference channel (see Appendix~\ref{apdx:Arbitrary} for the extension to the general case).  Assume the two sources use equal block length codes taken from P2P-optimal code sequences $\{\cC_{1,n}\}_{n=1}^\infty$ and $\{\cC_{2,n}\}_{n=1}^\infty$, respectively, which have rate $R$ (see Sec.~\ref{sec:Limitations_Good_Codes}).  Assume the probabilities of decoding error, under maximum-likelihood decoding, at both destinations, approach zero as
 the block length $n\rightarrow\infty$.  Then the following holds.
\begin{eqnarray}\label{eq:R_Good_Interference}
R \leq \max\Big( R_\MUD, R_\SUD \Big).
\end{eqnarray}
\end{theorem}
The proof of this theorem is provided in Appendix~\ref{apdx:Proof_Theorem_Good_Interference}.  The proof builds on the converse of the capacity theorem of multiple-access channels, see e.g.,~\cite[Sec.~14.3.4]{Cover_Book}.  For example, consider the setting facing Destination 1. Once the destination has decoded the desired codeword, it is able to subtract it.  The remaining signal is equivalent to the output of a point-to-point BIAWGN channel, whose input is the interference $\bX_2$.
If $R$ is lower than the capacity of this channel, then by the P2P-optimality of $\{\cC_{2,n}\}_{n=1}^\infty$, complete decoding of the interference is possible.  Although the destination is not required to decode the interference, the fact that it is {\it able} to do so  enables us to use bounds that apply to   multiple-access scenarios,
leading to the bound $R \leq R_\MUD$ (see Appendix~\ref{apdx:Proof_Theorem_Good_Interference} for the rigorous details).

If $R$ is greater than the capacity of the above point-to-point
BIAWGN channel, then complete decoding of the interference $\bX_2$
is not possible.  However, relying on the P2P-optimality of
$\{\cC_{2,n}\}_{n=1}^\infty$, we are still able to bound its
entropy given the channel output $\bY_1$ and the desired codeword.
We apply this bound in
Appendix~\ref{apdx:Proof_Theorem_Good_Interference} to show that
in this case, $R \leq R_\SUD$ must hold.

\begin{remark} Note that in communication using a code $\cC$, we mean that the source simply maps each message to a codeword of $\cC$, and does not manipulate $\cC$, e.g., by combining it with another code, as in the HK strategy~\cite{HanKobayashi}.
\end{remark}

In Sec.~\ref{sec:Numerical_Interference} below, we provide examples of P2P-{\it suboptimal} codes which are capable of communication at rates that exceed $R_\MUD$ and $R_\SUD$, thus surpassing the performance of P2P-optimal codes.

\section{Numerical Results}\label{sec:Numerical}

The various bounds developed in the previous sections enable an assessment of soft-DF and soft-IC in combined contexts with other strategies for the respective channels, for which the performance is known at asymptotically large block lengths.  We now provide two examples.
\subsection{Erasure Relay}\label{sec:Numerical_Relay}
We considered an erasure relay channel with parameters $\varepsilon_\StR = 0.5$,   $\varepsilon_\StD = 0.85$ and\footnote{The value for $\varepsilon_\RtD$ was selected so that $R_\CF = R_\DF = 0.5$ (see below).} $\varepsilon_\RtD = 0.232$.  We designed applications of soft-DF-BP2 and soft-DF-BP.

As benchmarks for comparison, we used decode-and-forward (DF) and compress-and-forward  (CF) (see e.g.,~\cite{Gerhard_Gaussian_Relay,Junshan,CoverElGamal} as well as the tutorial~\cite{Gerhard_Maric_Cooperative}).  The two strategies do {\it not} involve partial decoding at the relay:  With DF, the relay completely  decodes the source's signal, and with CF it does not decode it at all.  Achievable rates with both strategies are provided by equations~\eqref{eq:R_DF} and~\eqref{eq:R_CF}, on the following page, respectively.  $C_\RtD$ in these expressions is the capacity of the relay-destination link (bits per channel use), and is computed as in~\eqref{eq:C_EEC} (recall that this channel is an EEC). The proofs, which follow~\cite{Primitive_Relay} and~\cite{CoverElGamal}, are provided in Appendix~\ref{apdx:proof_DF_CF}.  Note that these rates assume randomly generated codes (according to a uniform distribution in $\{0,1\}$), which are P2P-optimal for the point-to-point BEC channel.\footnote{More precisely, a sequence of codes generated in this way is P2P-optimal with probability 1.}
\begin{figure*}
\begin{eqnarray}
R_\DF &=& \min\Big{(}1 - \varepsilon_\StR,\: 1-\varepsilon_\StD + C_\RtD\Big{)} \label{eq:R_DF}\\
R_\CF &=& \max_{\hvarepsilon_\BPR} \Big{\{} 1 - (\varepsilon_\StR \circ \hvarepsilon_\BPR) \cdot\varepsilon_\StD \quad :\quad h(\varepsilon_\StR \circ \hvarepsilon_\BPR)  +    (1-\varepsilon_\StR \circ \hvarepsilon_\BPR)\cdot\varepsilon_\StD    -   h(\hvarepsilon_\BPR)\cdot(1-\varepsilon_\StR) \leq C_\RtD\Big{\}}\label{eq:R_CF}
\end{eqnarray}
\hrulefill
\end{figure*}

We designed the following parameters for soft-DF-BP2.
We used {\it differential evolution}~\cite{Differential}, a numerical
optimization algorithm, to design the degree distributions $(\lambda,\rho)$.
We selected  the quantization noise level (see Sec.~\ref{sec:Analysis_soft_DF_BP2})
as the value $\hvarepsilon_\BPR$ which satisfies  $I^+(\hvarepsilon_\BPR) = C_\RtD = 0.991$, where $I^+(\hvarepsilon_\BPR)$ is given by~\eqref{eq:I_Plus}.
We obtained the following values.
\begin{eqnarray*}
&&\hspace{-0.6cm}\lambda_{2,3,4,5,6,7,8,9,10,15,20,50,100} = (0.2301,0.1473,0.01202,\nonumber \\&& 0.02178,0.01656,0.03305,0.02383,0.01614,0.02316,\nonumber \\&&0.1979,0.0001437,0.04566,0.2323),\quad \rho_{10} = 1, \nonumber\\&& \hvarepsilon_\BPR = 0.139.
\end{eqnarray*}

These degree distributions correspond to a design rate (see~\eqref{eq:design_rate})
of $R = 0.506$ (bits per channel use), surpassing both our benchmarks (see above) for this channel, which are $R_\DF = R_\CF = 0.5$.  Note that an upper bound on the capacity of the above relay channel is 0.575, computed using the cutset bound~\cite[Proposition~1]{Primitive_Relay}.  
To assess the strategy's performance with this code, we relied on Theorem~\ref{theorem:Formal}
and the discussion of Sec.~\ref{sec:Simultaneous_DE}, and applied sim-DE to the $(\varepsilon_\StD,\varepsilon_\StR,\hvarepsilon_\BPR)$ auxiliary erasure relay channel model.    The erasure rate at the output of soft-DF-BP
over the auxiliary channel, as computed by sim-DE,
 is negligible (upper bounded by $\xi = 4.35\cdot 10^{-8}$).  The average erasure rate at the output of BP at the relay, is~0.25.  This is less than the noise on the source-relay channel ($\varepsilon_\StR = 0.5$), but greater than zero, indicating that partial decoding was achieved.

We similarly designed degree distributions for soft-DF-BP, again using differential evolution, obtaining the following values.
\begin{eqnarray*}
&&\hspace{-0.6cm}\lambda_{2,3,4,5,6,7,8,9,10,15,20,50,100} = (0.2062,0.1704,6.425\cdot10^{-5},\nonumber \\&& 0.04838,0.05066,0.0236,0.0002077,0.009968,0.01002,\nonumber \\&&0.08731,0.1251,0.05697,0.2111),\quad \rho_{10} = 1.
\end{eqnarray*}

These degree distributions correspond to a design rate
of $R = 0.497$ bits per channel use, which is lower than our above result for soft-DF-BP2 (as expected), and
slightly less than our benchmarks $R_\DF$ and $R_\CF$.
To assess the strategy's performance with this code, we relied on the discussion of Sec.~\ref{sec:Simultaneous_DE}, and applied sim-DE to the $(\varepsilon_\StD,\varepsilon_\StR,\varepsilon_\RtD)$ physical erasure relay channel.
The erasure rate at the output of soft-DF-BP, as computed by sim-DE,
 is upper bounded by $4.71\cdot 10^{-9}$.  The average erasure rate at the output of BP at the relay, is~0.22, again indicating that partial decoding was achieved.

We also experimented with partial-DF (see Sec.~\ref{sec:Introduction}).  Unfortunately, we were not able to design an application of the strategy whose performance exceeds the above rates achieved by DF and CF.  A similar difficulty was reported by~\cite[Sec.~4.2.7]{Gerhard_Maric_Cooperative} in the context of full-duplex AWGN channels.  Further optimization of partial-DF is beyond the scope of this work.
\subsection{BIAWGN Interference}\label{sec:Numerical_Interference}

We considered a symmetric BIAWGN interference channel with parameters $h = 0.839$ and $\sigma = 1.075$ and focused on symmetric achievable rates (Sec.~\ref{sec:BIAWGN_Interference}).   We designed an application of soft-IC-BP.  As benchmarks for comparison, we used $R_\MUD$ and $R_\SUD$, as defined by~\eqref{eq:R_MUD} and~\eqref{eq:R_SUD}.  We also considered the Han-Kobayashi (HK) strategy (see Sec.~\ref{sec:Introduction}).

To design degree distributions for soft-IC-BP codes, we applied
a variation of the hill climbing approach of Richardson~\etal~\cite[Sec. IV.A]{Urbanke_Capacity_Approaching},~\cite[Example 4.139]{Urbanke_Book}.  We obtained the following degree distributions $(\lambda,\rho)$ for codes for both sources.
\begin{eqnarray}\label{eq:LDPC_IC}
&&\hspace{-0.6cm}\lambda_{2,3,10,11,55,56,57} = (0.2949, 0.2036, 0.05943, 0.2399, \nonumber \\&&\quad 0.0001219,0.09542, 0.1065),\quad \rho_6 = 1.
\end{eqnarray}
The design rate~\eqref{eq:design_rate} corresponding to these degree distributions is  0.3243 (bits per channel use). The bit error rate at the output of soft-IC-BP at each destination, for decoding of the desired codeword from the corresponding source (as predicted by density evolution) approaches $4\cdot10^{-6}$ in probability as the block length $n\rightarrow\infty$.  The bit error rate in decoding of the interference approaches 0.062.   This number is less than the bit error rate with bitwise decoding (i.e., when the code structure is not exploited), which equals~0.301, but greater than zero, indicating that partial decoding of the interference was achieved.

The degrees of the check-nodes in codes of the above $(\lambda,\rho)$ code ensemble are bounded (and equal to 6).  By~\cite{DavidUpperBounds}, this implies that such codes are P2P-suboptimal for the BIAWGN channel (in the sense of Definition~\ref{def:Good codes_AWGN}).  Their above-mentioned design rate surpasses benchmarks $R_\MUD = 0.3237$ and $R_\SUD = 0.308$ (bits per channel use), which by Theorem~\ref{theorem:Good_Interference} upper bound the rates achievable with any application of P2P-optimal codes.  Note that an upper bound on the symmetric capacity (maximal achievable symmetric rate) of the above interference channel is 0.378, computed using~\cite[Theorem~2]{Gerhard_UB}. 

The rate of our LDPC codes does {\it not} exceed the best rate achievable by the HK strategy, which as noted in Sec.~\ref{sec:Introduction}, also involves partial decoding.  Using rate-splitting,  HK constructs P2P-suboptimal codes, and is thus not bounded by $R_\MUD$ or $R_\SUD$.
In Appendix~\ref{apdx:HK_Details}, we describe an application of the HK strategy for the above channel which is provably capable of communication at rate 0.333 bits per channel use.  As the parameters in our applications  of both strategies (HK and soft-IC-BP) have not been proven to be optimal, it remains to be seen whether this advantage is fundamental.  In Sec.~\ref{sec:Conclusion} (below) we argue that from a practical perspective, soft-IC-BP enjoys several advantages.

\section{Conclusion}\label{sec:Conclusion}

Our research in this paper has been at the interface of information theory and coding theory.  In our work,  rather than apply ideas from information theory to improve the design of practical codes and strategies, as frequently is the case in similar research over P2P channels,  we have taken the reverse approach.   We developed applications of soft-DF and soft-IC, borrowed from the coding-theoretic literature, and derived bounds on their performance in an information-theoretic context, which focuses on achievable rates at asymptotically long block lengths.

Our interest in this problem was motivated by Kramer~\cite{Gerhard_Turbo_Symp}, who argued that multi-terminal networks offer a much richer problem set  than P2P channels, and thus traditional information-theoretic approaches may not necessarily be optimal for them.  In Theorem~\ref{theorem:Good_Interference}, we have reinforced this intuition, by demonstrating that P2P-optimal codes are often {\it sub}-optimal over BIAWGN interference channels, and are unable to exploit the benefits of partial decoding.  Some additional intuition, which involves the transmitter's degrees of freedom, will be provided later in this section.

In Sec.~\ref{sec:Numerical}, we applied our analysis to compare our strategies with known  information-theoretic ones.  We demonstrated that soft-DF-BP2 can sometimes outperform DF and CF, and soft-IC-BP can outperform MUD and SUD.   We did not obtain a similar result in a comparison with the HK strategy.  Further research could focus on improving our bounds and code design methods.   With soft-DF-BP2, it would be interesting to determine the strategy's gap from the relay channel's capacity (hopefully proving that none exists).  Specifically, our bounds on the affordable quantization noise level can likely be improved, perhaps by introducing memory into the quantization noise model~\eqref{eq:hyBP}.  With soft-IC-BP, it would be interesting to assess the strategy's potential in comparison to the HK strategy.  While we were not able to demonstrate that soft-IC-BP outperforms HK, the perceived gap between them could be an artifact of our suboptimal code design methods.

The two models we have considered (relay and interference), and our solutions for them, share several common features.  In both cases, partial decoding plays an important role.  In both, we applied it by soft estimation, and in both cases  we used LDPC codes' BP algorithm to achieve soft estimation.   With both models, partial decoding is possible at nodes which overhear a communicated signal, but are not the destination of the associated message.   Such nodes exist in many other network scenarios.  Extensions of our methods to them, as well as to additional relay and interference channels, are interesting research directions.

While our main focus in this paper has been theoretical, we now briefly discuss some related practical aspects.

The LDPC codes we designed in Sec.~\ref{sec:Numerical_Interference} for soft-IC-BP offer several practical benefits, in addition to their favorable achievable rates.  Specifically, the low degrees of their check nodes imply low decoding complexity per BP iteration, as well as better error resilience at short block lengths (see e.g.,~\cite[Sec.~II-B]{Urbanke_Capacity_Approaching}).  As noted in Sec.~\ref{sec:Numerical_Interference}, these low degrees, and consequently their benefits, are intrinsically related to the codes' P2P-suboptimality.

For further research, it would be interesting to examine the potential of other known P2P-suboptimal codes.  These include, for example, many algebraic codes (e.g., Reed-Muller and BCH) as well as low constraint-length convolutional codes~\cite[Sec.~5.4]{Viterbi_Omura}.   Many of these codes offer practical benefits similar to those of the above-mentioned LDPC codes.  In light of Theorem~\ref{theorem:Good_Interference}, which highlighted the  differences between coding for multi-terminal and P2P channels, perhaps these codes (and their benefits) can be enjoyed at smaller sacrifices in achievable rates (relative to capacity), in comparison to P2P channels.

Some related intuition can be obtained from the following heuristic discussion, which involves the transmitter's degrees of freedom.  The problems of decoding and estimation, at a destination node, involve opposite requirements from the transmitter.  To reduce the destination's decoding error, the transmitter arguably seeks to exploit all its degrees of freedom (heuristically defined) to maximize the separation of its codewords.  However, to lower the destination's estimation error at rates above capacity, the transmitter may prefer to constrain its signals, to reduce the uncertainty facing the receiver.\footnote{This is best seen by observing that from the perspective of lowering the estimation error over AWGN channels, an optimal code is a degenerate one which maps all source messages to the all-zero vector.}   Over P2P channels, practical benefits often conflict with communication at capacity-approaching rates, because the code structures that enable them (e.g., algebraic or trellis structures) constrain the transmitter's degrees of freedom (see also Forney and Ungerboeck~\cite[Sec.~I]{Forney}).  Over multi-terminal channels, however, such constraints may benefit partial decoding (by estimation) at various network nodes, enabling a smaller gap from capacity.

Practical considerations led Baccelli~\etal~\cite{Shlomo_Rem2} to argue in favor of confining attention to P2P-optimal codes and the SUD and MUD strategies, in research of communication over interference channels.\footnote{They also considered combinations of the strategies, in the context of interference channels with more than two source-destination pairs.}  Their work was motivated in part by practical drawbacks of the HK strategy~\cite[Sec.~VII]{Shlomo_Rem2}, which they considered to be the primary alternative to MUD and SUD.  For example, the structure of the strategy's codes implies a heavy computational burden on the interference channel destinations, which must each decode three codewords:
Two from its own source and one that constitutes a part of the interference.   As the decoding complexity of many algorithms (e.g.,~\cite{boutros-caire}) is exponential in the number of codewords jointly decoded, this gives SUD and MUD a significant advantage.  Soft-IC, however, involves examining just {\it two} codewords; one decoded, and the other (the interference) estimated.   As noted above, we demonstrated that like the HK strategy, soft-IC can often be used to outperform SUD and MUD in terms of achievable rates, making it an attractive alternative to those strategies.

Soft-IC may offer additional benefits.  Theoretical results on the HK strategy involve randomly-generated, unstructured component codes (within the rate-splitting framework), whose decoding requires high-complexity algorithms.  To reduce complexity, these codes could be replaced by lower-complexity ones (e.g., LDPC codes~\cite{Sharifi}), effectively introducing a second form of structure, beyond rate splitting (which facilitates partial decoding).  However, it is likely that soft-IC-BP, whose LDPC codes contribute their structure simultaneously to partial decoding and low-complexity algorithms, enjoys a practical advantage.

One important practical problem involves cases where the transmitter has imperfect knowledge of the channel {\it state}, meaning the received SNRs at the destinations.  Raja~\etal~\cite{Raja} considered this problem, and proposed a HK-based approach that involves rate-splitting between many auxiliary codes (three or more), rather than two.  This gives each destination multiple options from which to choose the amount of partial decoding it wishes to perform, according to the realized channel state.  Equivalently, it enables graceful degradation when the received SNR drops slightly.  By our above discussion, however, this further complicates the decoding scenario at the destinations, which must sometimes jointly decode as many as five (or more) codewords.  We conjecture that LDPC codes, whose estimation errors can be made to change gracefully as a function of SNR (see Fig.~\ref{fig:1}), combined with soft-IC-BP, are better suited for the problem.

\appendices
\section{Details of the Curves in Figure~\ref{fig:1}} \label{apdx:Rigorous_Fig_Bad_Codes}

The MMSE values plotted in Fig.~\ref{fig:1} correspond to the asymptotic normalized MMSE of a {\it sequence} of codes.  More precisely, given a code $\cC_0$, we define,
\begin{eqnarray}\label{eq:mmse}
\mmse( \cC_0; \SNR) \defined \EE \Big[ \: \| \hbX(\bY) - \bX \|^2 \:\Big],
\end{eqnarray}
where $\bX$ is the transmitted codeword, which is uniformly randomly distributed in $\cC_0$, $\bY$ is the channel output at the destination and $\hbX(\bY)$ is the MMSE
estimate of $\bX$ given $\bY$.  Note that we make no distinction between information and parity bits, and estimation of the values of all is performed.

Given a sequence $\cC = \{\cC_n\}_{n=1}^\infty$, we define its normalized MMSE as
\begin{eqnarray*}
\mmse( \cC, \SNR) \defined \lim_{n\rightarrow\infty}  \frac{1}{n}\mmse( \cC_n, \SNR).
\end{eqnarray*}
The curve in Fig.~\ref{fig:1} corresponding to uncoded communications was evaluated as in~\cite[Eq.~(17)]{GuoMutInfMMSE}.   In the curve corresponding to P2P-optimal code-sequences, P2P-optimality is defined as in Definition~\ref{def:Good codes_AWGN}.  In the range  $\SNR < \SNRstar$ where $\SNRstar$ is the Shannon limit for rate 1/2, we have relied on the analysis of~\cite[Eq.~(14)]{PelegExtrinsicGood} to evaluate the curve.  In the range  $\SNR > \SNRstar$, we have relied on an analysis similar to the one in Appendix~\ref{apdx:Proof_Theorem_Good_over_BEC}, with respect to estimation over the BEC, in the range $\varepsilon < \varepsilonstar$.

The LDPC (2,4) curve corresponds to the normalized MMSE of a sequence of codes $\{\cC_n\}_{n=1}^\infty$, where $\cC_n$ was selected at random from the LDPC (2,4) code  ensemble of block length $n$ (see Sec.~\ref{sec:LDPC}).  The bound on the MMSE was explained in our paper~\cite[Sec.~III.A]{BadCodes_Allerton}.  Codes from this ensemble correspond to Tanner graphs whose nodes have very low degrees (2 and 4).  By~\cite[Theorem~3.3]{Gallager_PhD}, \cite{DavidUpperBounds}, \cite{IgalSason}, this implies that the codes are bounded far away from capacity.
\section{Results for Sec.~\ref{sec:BEC_relay}}
\subsection{Proof of Theorem~\ref{theorem:Degrade}}\label{apdx:Proof_of_Lemma_Degrade}

\mmm
We begin with an overview of the proof.   The main idea is best understood by comparing the {\it computation graphs}~\cite[Sec.~3.7.1]{Urbanke_Book} of the messages of soft-DF-BP's destination BP decoder, to those of the corresponding components of sim-BP's message pairs.  A key observation is that the latter graphs can be obtained from the former by ``pruning'' nodes, as explained below.

\begin{figure*}
\centering
\subfigure[Soft-DF-BP's destination BP decoder.]{%
\begin{minipage}[b]{0.5\textwidth}
\centering
\epsfig{file=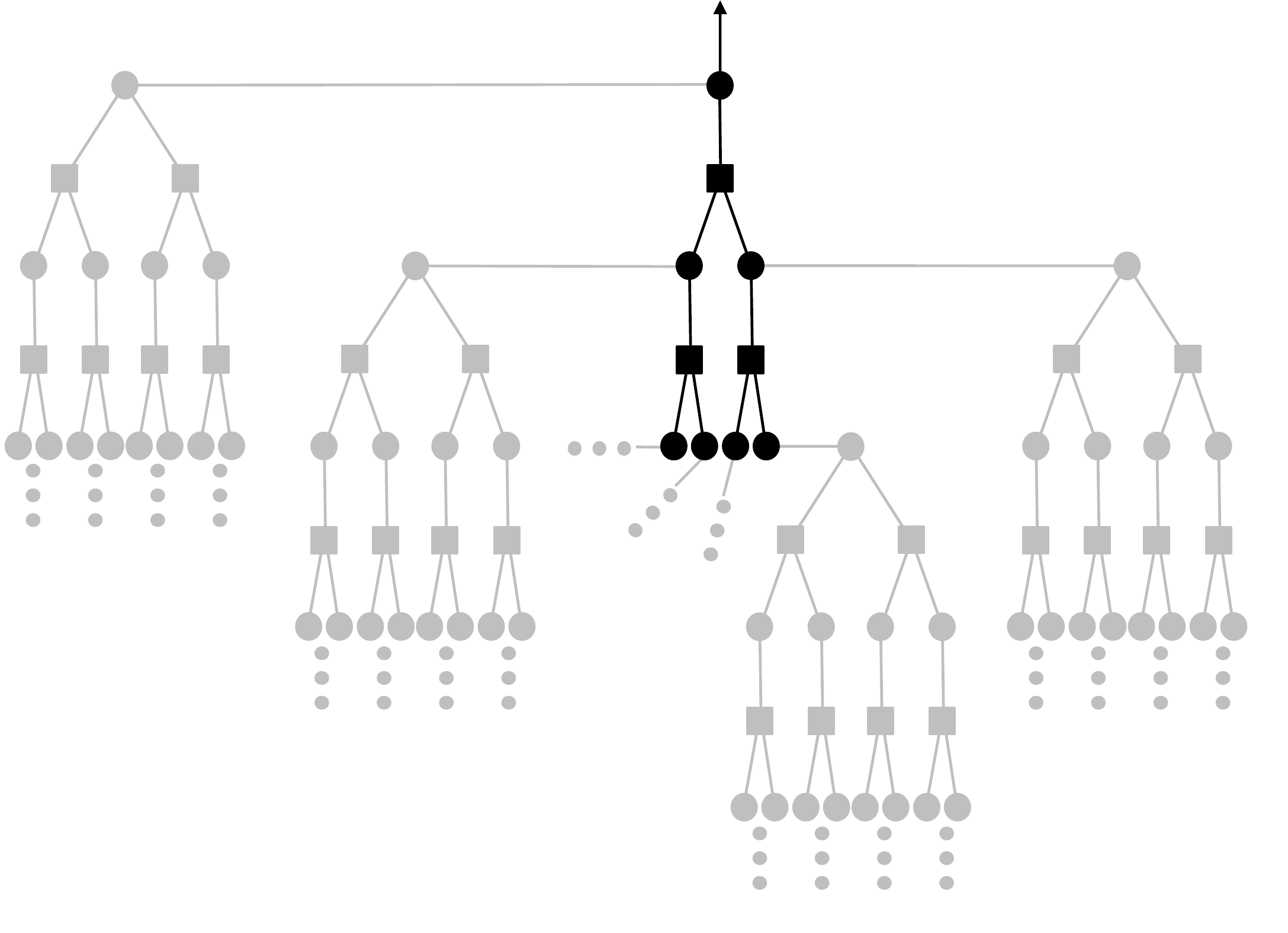, width = 12cm}
\end{minipage}}%
\\
\subfigure[Destination component of sim-BP's message pair.]{%
\begin{minipage}[b]{0.5\textwidth}
\centering
\epsfig{file=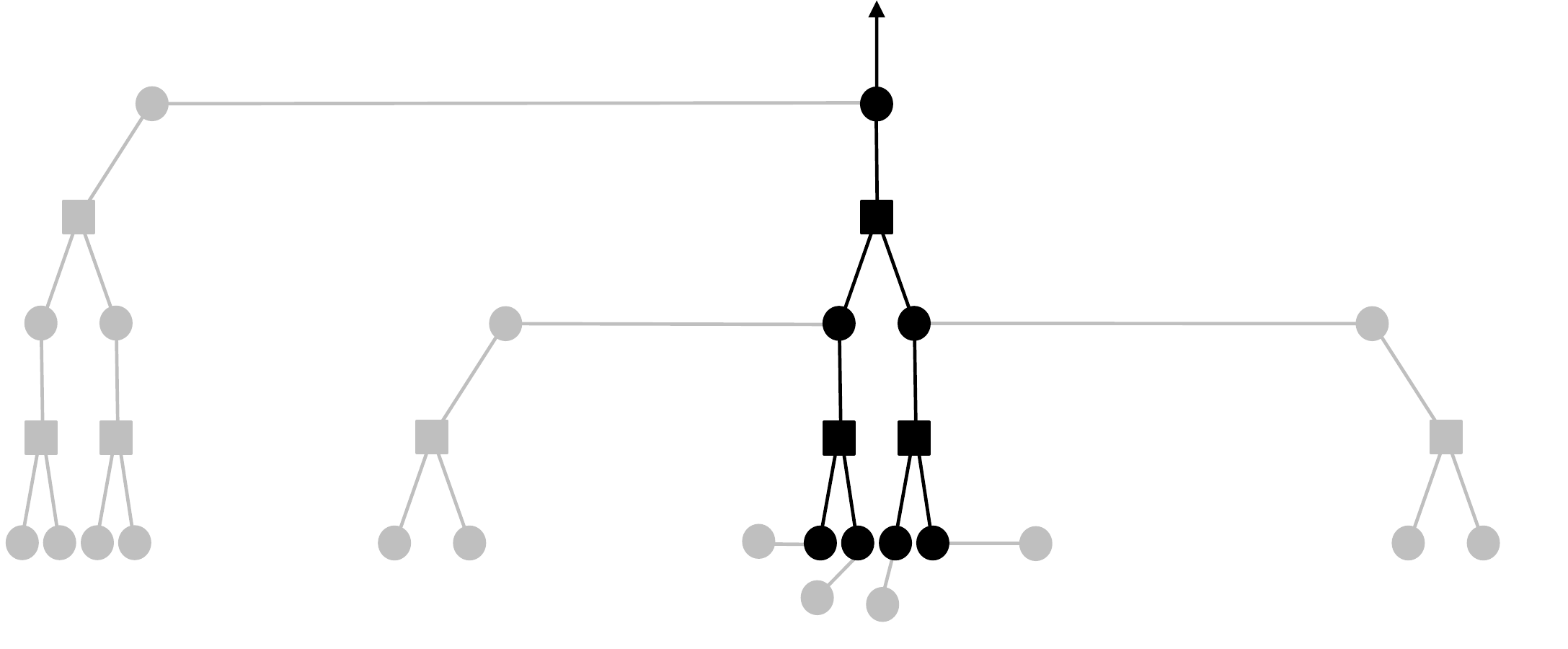, width = 12cm}
\end{minipage}}%
\caption{Computation graphs for a message $\rb{\BPD,2}{i j}$ in an LDPC (2,3) code.  Due to space limitations, many nodes were omitted.  Nodes and edges corresponding to computations at the destination are represented in black, and the corresponding relay components are in grey.  } \label{fig:CompGraphs}
\end{figure*}

Figs.~\ref{fig:CompGraphs}(a) and~\ref{fig:CompGraphs}(b) depict examples of the computation graphs for a variable-to-check message $\rb{\BPD,2}{i j}$ at the second iteration of the above-mentioned decoders (soft-DF-BP's destination decoder and sim-BP), respectively.   A detailed definition of computation graphs is available in~\cite{Urbanke_Book} (we will not require this concept in our rigorous proof later in this section). Each of the graphs tracks the decoding process that produced $\rb{\BPD,2}{i j}$ in its respective decoder.  In each graph, the edge $(i,j)$ along which $\rb{\BPD,2}{i j}$  was sent is drawn on top.  The edges that delivered the messages on which $\rb{\BPD,2}{i j}$ depends, are drawn below.   Each such message, in turn, is a function of additional messages, whose corresponding edges are drawn below, and so forth.  By~\eqref{eq:Variable-to-check_Dest},~\eqref{eq:46},~\eqref{eq:Variable-to-check_Dest_Simultaneous} and~\eqref{eq:hrb}, the value of $\rb{\BPD,2}{i j}$ also depends on values computed at the relay, namely on $\yBPrelay{i}$ (with soft-DF-BP) and $\rb{\BPR,2}{i j}$ (with sim-BP).  It further depends indirectly on additional relay computations, via its reliance on variable-to-check messages of previous decoding iterations.   Consequently, the computation graphs for $\rb{\BPD,2}{i j}$ include nodes that correspond to the relevant computations at the relay.  While the relay computations are associated with the same Tanner graph nodes as those of the destination, in the computation graphs we define distinct nodes to represent them.

A comparison of Figs.~\ref{fig:CompGraphs}(a) and~\ref{fig:CompGraphs}(b) reveals that the latter contains a subset of the nodes and edges of the former.  This follows because the relay messages $\{\rb{\BPR,\ell}{i j}\}$, which sim-BP uses in~\eqref{eq:Variable-to-check_Dest_Simultaneous} and~\eqref{eq:hrb}, are based on less decoding iterations than the final decisions~$\{\yBPrelay{i}\}$, which soft-DF-BP uses in~\eqref{eq:Variable-to-check_Dest} and~\eqref{eq:46}.  They are also based on less check-to-variable messages (produced at nodes $\cN(i)\backslash j$ instead of $\cN(i)$, see~\eqref{eq:Variable-to-check} and~\eqref{eq:Final_Iteration}).  It would thus be natural to expect that the value of $\rb{\BPD,2}{i j}$ with soft-DF-BP to have higher quality than the equivalent sim-BP one.   A similar discussion applies to the other messages of the algorithms, as well as to their output components, $\yBPdest{i}$ and $\yBPdestTag{i}$, respectively, $i=1,\ldots,n$.

We now prove this rigorously.  As noted above, our proof does not directly require an understanding of computation graphs.  We start with  the following definition, which specializes Definition~\ref{def:Degraded} to the case of scalars (vectors of length 1).

\begin{definition}\label{def:Degraded2}
Let $x,y \in \{0,1,\Erasure\}$.  We say that $y$ is {\it degraded} with respect to $x$ if the following two conditions do not hold simultaneously\footnote{This definition is only concerned with whether $x$ and $y$ are erasures.  For example, if $x=1$ and $y=0$, then each is degraded with respect to the other.}:  $y \neq \Erasure$ and $x = \Erasure$.
\end{definition}
With this definition, it is clear that a vector $\by \in \{0,1,\Erasure\}^n$ is degraded with respect to $\bx \in \{0,1,\Erasure\}^n$, in the sense of Definition~\ref{def:Degraded}, if for all $i = 1,\ldots,n$, $y_i$ is degraded with respect to $x_i$.  The proof now relies on the properties of erasure addition and multiplication as defined in~\eqref{eq:Erasure_Addition} and~\eqref{eq:Erasure_Multiplication}, respectively.  Specifically, it is easy to verify that if $x'$ is degraded with respect to $x$ and $y'$ is degraded with respect to $y$, then $x'+y'$ is degraded with respect to $x+y$ and $x'\cdot y'$ is degraded with respect to $x\cdot y$.  Similarly, $x$ is degraded with respect to $x\cdot y$ for all $x,y \in \{0,1,\Erasure\}$.  We proceed with the following lemma.
\begin{lemma}\label{lemma:Degrade_Aux}
Consider an instance of the application of BP (Algorithm~\ref{alg:BP}) over the point-to-point BEC.  Let $\rb{\ell}{i j}$ be a variable-to-check message computed at some intermediate iteration $\ell = 0,\ldots,t-1$, and $\yBP{i}$ the final decision later computed at node $i$.  Then $\rb{\ell}{i j}$ is degraded with respect to $\yBP{i}$.
\end{lemma}

{\it Proof:} We will actually prove a stronger result:  $\rb{\ell-1}{i j}$ is degraded with respect to $\rb{\ell}{i j}$ for all $\ell = 1,\ldots,t-1$, and $\rb{t-1}{i j}$ is degraded with respect to $\yBP{i}$.  The desired result will follow by the obvious transitivity of degradedness.

Our proof follows by induction on the iteration number $\ell$.  We start by comparing $\rb{0}{i j}$ and
$\rb{1}{i j}$.  By~\eqref{eq:Variable-to-check}, $\rb{0}{i j}$ equals the channel output $y_i$ while $\rb{1}{i j}$ is obtained by multiplying $y_i$ with some other components.  The result now follows by the above-mentioned property of erasure multiplication.

We proceed to examine  $\rb{\ell-1}{i j}$ and $\rb{\ell}{i j}$ for $\ell = 2,\ldots,t-1$. By~\eqref{eq:Variable-to-check}, both are functions of check-to-variable messages across the same edges, but at different iterations ($\{\lb{\ell-1}{j'i}\}_{j'\in \cN(i)\backslash j}$ and $\{\lb{\ell}{j'i}\}_{j'\in \cN(i)\backslash j}$, respectively), as well as the same $y_i$.  If we could prove that each message $\lb{\ell-1}{j'i}$ is degraded with respect to the corresponding  $\lb{\ell}{j'i}$, the result would follow by the above-mentioned property of erasure multiplication.  Each such check-to-variable message is computed by~\eqref{eq:Check-to-variable}.  Again, both are functions of variable-to-check messages across the same edges, but at different iterations ($\{  \rb{\ell-2}{i' j'}\}_{i'\in \cN(j')\backslash i}$ and $\{  \rb{\ell-1}{i' j'}\}_{i'\in \cN(j')\backslash i}$, respectively).  By induction, each message $\rb{\ell-2}{i' j'}$ is degraded with respect to $\rb{\ell-1}{i' j'}$ and the result now follows by the above-mentioned property of erasure addition.

The proof of the degradedness of $\rb{t-1}{i j}$ with respect to $\yBP{i}$ follows by similar arguments.  Namely, the expression~\eqref{eq:Final_Iteration} for $\yBP{i}$ differs from the one for $\rb{t-1}{i j}$~\eqref{eq:Variable-to-check}  in that the check-to-variable messages used are of iteration $t$ rather than iteration $t-1$, and also by the fact that $\lb{t}{ji}$ is included in the product.  Degradedness thus follows as in our above discussion, combined with the above-mentioned degradedness of $x$ with respect to its product $x\cdot y$ with any $y$.
{\hfill$\QED$}
\mmm 
We now introduce some notation.  By construction, the components $\rb{\BPR,\ell}{i j}$ and $\lb{\BPR,\ell}{j i}$ of the variable-to-check and check-to-variable message pairs  (respectively) computed by sim-BP  are identical to the messages computed by the relay's BP decoder with soft-DF-BP.  The same does {\it not} hold for the other components of sim-BP and soft-DF-BP's messages at the destination.  We let $\rb{\BPD,\ell}{i j}$ and $\lb{\BPD,\ell}{i j}$ denote the messages computed with soft-DF-BP at the destination, and $\rbt{\BPD,\ell}{i j}$ and $\lbt{\BPD,\ell}{j i}$ denote the corresponding components of sim-BP's message pairs.

The proof proceeds by showing that $\rbt{\BPD,\ell}{i j}$ is degraded with respect to $\rb{\BPD,\ell}{i j}$ for all $\ell, i, j$.  With the above notation, $\rbt{\BPD,\ell}{i j}$ is computed by~\eqref{eq:Variable-to-check_Dest_Simultaneous}, replacing $\lb{\BPD,\ell}{i j}$ with $\lbt{\BPD,\ell}{j i}$.  $\rb{\BPD,\ell}{i j}$ is computed by~\eqref{eq:Variable-to-check_Dest}.
By Lemma~\ref{lemma:Degrade_Aux}, $\rb{\BPR,\ell}{i j}$ is degraded with respect to $\yBPrelay{i}$ for all $\ell,i,j$.  By~\eqref{eq:46} and~\eqref{eq:hrb} and the above-mentioned properties of multiplication, it follows that $\hrb{\BPR,\ell}{i j}$ is degraded with respect to $y_{\RtD,i}$. Each message $\lbt{\BPD,\ell}{j'i},\: j'\in \cN(i)\backslash j$ can be shown to be degraded with respect to $\lb{\BPD,\ell}{j'i}$ by induction, using similar arguments to the ones used in the proof of Lemma~\ref{lemma:Degrade_Aux} above.  The desired degradedness of $\rbt{\BPD,\ell}{i j}$ with respect to $\rb{\BPD,\ell}{i j}$ now follows by the above-mentioned properties of multiplication.

Finally, the proof of degradedness of $\yBPdestTag{i}$ with respect to $\yBPdest{i}$ now follows from the above results by similar arguments and is omitted.
{\hfill$\QED$}

\subsection{Details of Simultaneous Density Evolution (Sec.~\ref{sec:Simultaneous_DE})}\label{apdx:Details_DE}

The description below relies on the discussion of Sec.~\ref{sec:Simultaneous_DE}.  Sim-DE follows the same concepts of density evolution as developed by Richardson and Urbanke~\cite{Urbanke_Message_Passing}.  Its computations follow the expressions for sim-BP.  Like density evolution, it relies on the assumption that the all-zero codeword was transmitted, and thus the distributions $P_\mR^{(\ell)}(x_\BPR, x_\BPD)$ and $P_\mL^{(\ell)}(x_\BPR,x_\BPD)$ that it tracks are confined to the range $\{0,\Erasure\}\times\{0,\Erasure\}$.  The incoming message pairs at each node, on which the computations for the outgoing pairs rely, are assumed to be mutually independent.  At variable nodes, the pairs are also assumed to be independent of the node's channel outputs $(Y_{\StR,i}, Y_{\StD,i},\hE_{\BPR,i})$.  These assumptions are justified by similar arguments to the ones in~\cite{Urbanke_Message_Passing}, relying on the fact that conditioned on the transmission of the all-zero codeword, components $(Y_{\StR,i}, Y_{\StD,i},\hE_{\BPR,i})$ corresponding to different indices $i$, are mutually independent.

\begin{algorithm}[Simultaneous Density Evolution (sim-DE)] \label{alg:Simultaneous_DE}$\:$
\begin{enumerate}
\item {\bf Iterations:} Perform the following steps, alternately.
\begin{itemize}
\item {\em Variable-to-check iteration number $l=0,\ldots,t-1$}: Set $P_\mR^{(\ell)} = \Gamma(\oP_\mR^{(\ell)};\varepsilon_\RtD)$ where
\begin{align} \label{eq:PR}
\oP_\mR^{(\ell)} = \left\{
                 \begin{array}{ll}
                   P_{\varepsilon_\StR}\times P_{\varepsilon_\StD}, & \hbox{$\ell = 0$,} \\
                   \sum_i\lambda_i \cdot \left[\oP_\mR^{(0)}\odot\left(P_\mL^{(\ell)}\right)^{\odot (i-1)}\right], & \hbox{$\ell > 0$,}
                 \end{array}
               \right.
\end{align}
where $P_{\varepsilon}(\cdot)$ is defined for $\varepsilon \in [0,1], x \in \{0,\Erasure\}$ by
\begin{eqnarray*}
P_{\varepsilon}(x) \defined \left\{
                        \begin{array}{ll}
                          \varepsilon, & \hbox{$x = \Erasure$} \\
                          1 - \varepsilon, & \hbox{$x = 0$.}
                        \end{array}
                      \right.
\end{eqnarray*}
$P_{\varepsilon_\StR}\times P_{\varepsilon_\StD}$ is defined as in~\eqref{eq:50} on the following page and the operation $\odot$ is defined by~\eqref{eq:48}.
$P^{\odot i} \defined P \odot P \odot \cdots \odot P$, i.e., the repeated application of the operation $\odot$ a number $i$ times on $P$.
Addition and multiplication by $\lambda_i$ in~\eqref{eq:PR} are performed componentwise (see~\eqref{eq:47}).  Lastly, $\Gamma(\cdot)$ is defined by~\eqref{eq:Gamma}.

\item {\em Check-to-variable iteration number $\ell = 1,\ldots,t$}: $P_\mL^{(\ell)}$
is obtained by
\begin{eqnarray}\label{eq:PL}
P_\mL^{(\ell)} = \sum_j\rho_j\cdot\left[\left(P_\mR^{(\ell-1)}\right)^{\oplus (j-1)}\right],
\end{eqnarray}
where the operation $\oplus$ is defined by~\eqref{eq:49},
and where $P_1$ and $P_2$ are probability functions over $\{0,\Erasure\}^2$.
$P^{\oplus i}$ is defined in the same way as $P^{\odot i}$
\end{itemize}
\item {\bf Stopping Criterion:} Computation stops after a pre-determined number $t$ of iterations.
\item {\bf Final Decisions:} Set $P^{(\Final)} = \Gamma(\oP^{(\Final)};\varepsilon_\RtD)$ where
\begin{eqnarray}\label{eq:PF}
\oP^{(\Final)} = \sum_i\tlambda_i \cdot \left[\oP_\mR^{(0)}\odot\left(P_\mL^{(t)}\right)^{\odot i}\right],
\end{eqnarray}
where $\tlambda_i$ is as defined in Sec.~\ref{sec:LDPC}.
\end{enumerate}
\end{algorithm}

\begin{figure*}
\begin{align}
P = P_{1}\times P_{2} \quad &\Longleftrightarrow
\quad P(x_\BPR,x_\BPD) = P_{1}(x_\BPR) \cdot P_{2}(x_\BPD) &\forall x_\BPR,x_\BPD\in \{0,\Erasure\}\quad\label{eq:50}\\
P = P_1 \odot P_2 \quad &\Longleftrightarrow \quad
P(x_\BPR,x_\BPD) = \sum_{\substack{x^1_\BPR,x^1_\BPD,x^2_\BPR,x^2_\BPD \in \{0,\Erasure\}\\x^1_\BPR \cdot x^2_\BPR = x_\BPR, \: x^1_\BPD \cdot x^2_\BPD = x_\BPD}}P_1(x^1_\BPR,x^1_\BPD)\cdot P_2(x^2_\BPR,x^2_\BPD) &\forall x_\BPR,x_\BPD\in \{0,\Erasure\}\quad\label{eq:48}\\
P = P_1 \oplus P_2 \quad &\Longleftrightarrow \quad P(x_\BPR,x_\BPD) = \sum_{\substack{x^1_\BPR,x^1_\BPD,x^2_\BPR,x^2_\BPD \in \{0,\Erasure\}\\x^1_\BPR + x^2_\BPR = x_\BPR, \: x^1_\BPD + x^2_\BPD = x_\BPD}}P_1(x^1_\BPR,x^1_\BPD)\cdot P_2(x^2_\BPR,x^2_\BPD)&\forall x_\BPR,x_\BPD\in \{0,\Erasure\}\quad\label{eq:49}\\
P = \alpha_1 P_1 + \alpha_2 P_2 \quad &\Longleftrightarrow
\quad P(x_\BPR,x_\BPD) = \alpha_1 P_1(x_\BPR,x_\BPD) + \alpha_2 P_2(x_\BPR,x_\BPD) &\forall x_\BPR,x_\BPD\in \{0,\Erasure\}\quad\label{eq:47}\\
P = \Gamma(\oP; \varepsilon)\quad &\Longleftrightarrow  \quad
P(x_\BPR, x_\BPD) = \sum_{\substack{x,\ox_\BPD\in\{0,\Erasure\},\\ \ox_\BPD\cdot(x_\BPR+x) = x_\BPD}} \oP(x_\BPR, \ox_\BPD)\cdot P_{\varepsilon}(x) &\forall x_\BPR,x_\BPD\in \{0,\Erasure\}\quad\label{eq:Gamma}
\end{align}
\hrulefill
\end{figure*}

Note that above, the computations in a variable-to-check iteration have been simplified by introducing an intermediate step.  Rather than determine $P_\mR^{(\ell)}$ directly, the algorithm first computes an auxiliary value $\oP_\mR^{(\ell)}$, which corresponds to a pair $(\rb{\BPR,\ell}{i,j}, \orb{\BPD,\ell}{i,j})$ where $\orb{\BPD,\ell}{i,j}$ is defined by
\begin{eqnarray*}
\orb{\BPD,\ell}{i j} = \left\{
                      \begin{array}{ll}
                         y_{\StD,i}, & \hbox{$\ell = 0$,} \\
                        y_{\StD,i}\cdot \prod_{j'\in \cN(i)\backslash j}\lb{\BPD,\ell}{j'i}  , & \hbox{$\ell > 0$.}
                      \end{array}
                    \right.
\end{eqnarray*}
That is, $\orb{\BPD,\ell}{i,j}$ coincides with~\eqref{eq:Variable-to-check_Dest_Simultaneous}, except that the multiplication by $\hrb{\BPR,\ell}{i j}$ is omitted.

The weighted sums by $\lambda_i$, $\rho_j$ and $\tlambda_i$ in~\eqref{eq:PR},~\eqref{eq:PL} and~\eqref{eq:PF} respectively, follow from the random construction of the Tanner graph, and are justified by the same arguments as in~\cite[Expression (8)]{Urbanke_Capacity_Approaching}.
\subsection{Proof of Theorem~\ref{theorem:Good_over_BEC}} \label{apdx:Proof_Theorem_Good_over_BEC}

We begin by proving~\eqref{eq:PMAP_Good} for $\varepsilon > \varepsilonstar$.  In this range, we wish to prove that $P_\MAP(\cC_n;\:\varepsilon) =\varepsilon + o(1)$. We begin with the following equalities.
\begin{eqnarray}
&&\hspace{-2cm}\int_{\varepsilonstar}^1 \left(\frac{1}{\varepsilon}\right)\cdot P_\MAP(\cC_n;\:\varepsilon)\: d\varepsilon =\nonumber \\ && \quad \addabove{=}{a} \quad \frac{1}{n}I(\cC_n;\:\varepsilonstar) - \frac{1}{n}I(\cC_n;\:1) \nonumber \\
&& \quad \addabove{=}{b} \quad \Big{(}1 - \varepsilonstar +o(1) \Big{)} \quad - \quad 0 \nonumber \\
&& \quad \:= \quad \int_{\varepsilonstar}^1 \left(\frac{1}{\varepsilon}\right)\cdot \varepsilon\: d\varepsilon \quad + \quad o(1). \label{eq:86}
\end{eqnarray}
In (a), we have defined for an arbitrary code $\cC$ and $\varepsilon \in [0,1]$
\begin{eqnarray}\label{eq:87}
I(\cC;\:\varepsilon) \defined I(\bX;\:\bY),
\end{eqnarray}
 where $\bX$ is uniformly distributed within the codewords of $\cC$ and $\bY$ is randomly related to $\bX$ via the transition probabilities of a $\BEC(\varepsilon)$.

 The equality in (a) follows by Lemma~\ref{lemma:1} (Appendix~\ref{apdx:Lemma_1} below).  In (b)
we have used Fano's inequality (e.g.,~\cite[Sec.~8.9]{Cover_Book}) and the P2P-optimality of the sequence $\cC_n$, to argue that $
\lim_{n \rightarrow \infty} (1/n) \cdot I(\cC_n;\:\varepsilonstar) = R = 1-\varepsilonstar$.  We have also used $I(\cC_n;\:1) = 0$ (which can be verified straightforwardly).

We would now like to use~\eqref{eq:86} to argue that $P_\MAP(\cC_n;\:\varepsilon) =\varepsilon + o(1)$.  To do so, we apply two additional observations.  First, $P_\MAP(\cC_n;\varepsilon) \leq \varepsilon$, which follows because the MAP decoder outputs no more erasures than it obtains via the channel output.  Second, $P_\MAP(\cC_n;\varepsilon)$ is non-decreasing as a function of $\varepsilon$.  This holds because if $\varepsilon_2 > \varepsilon_1$, then $\BEC(\varepsilon_2)$ is stochastically degraded with respect to $\BEC(\varepsilon_1)$.

Formally, let $\xi > 0$ and assume that $P_\MAP(\cC_n;\varepsilono) < \varepsilono - \xi$ for some $\varepsilono > \varepsilonstar + \xi$.
We proceed with the string of equations ending with~\eqref{eq:88} on the following page.  In (a), we have applied our above-discussed observations, first $P_\MAP(\cC_n;\varepsilon) \leq \varepsilon$.  Second, by our assumption, $P_\MAP(\cC_n;\varepsilono) < \varepsilono - \xi$ and thus, by the monotonicity of of $P_\MAP(\cC_n;\varepsilon)$, we must also have $P_\MAP(\cC_n;\varepsilon) < \varepsilono - \xi$ for $\varepsilon < \varepsilono$.  (b) follows by a straightforward evaluation of the integrals.  Finally, the content of the brackets is strictly positive for all $\xi > 0$. This follows from $\ln(\varepsilono/(\varepsilono - \xi)) < \xi/(\varepsilono - \xi)$, which holds by the well-known inequality $\ln(1+x) < x$ for all $x \neq 0, x > -1$.  Thus, for large enough $n$, $\xi >0$ implies a violation of~\eqref{eq:86}. This concludes the proof of~\eqref{eq:PMAP_Good} for $\varepsilon > \varepsilonstar$
\begin{figure*}
\begin{eqnarray}
\int_{\varepsilonstar}^1 \left(\frac{1}{\varepsilon}\right)\cdot P_\MAP(\cC_n;\:\varepsilon)\: d\varepsilon  &\addabove{\leq}{a}& \int_{[\varepsilonstar, \varepsilono - \xi]\cup[\varepsilono,1]} \left(\frac{1}{\varepsilon}\right)\cdot \varepsilon\: d\varepsilon \quad+\quad\int_{[\varepsilono - \xi,\varepsilono]} \left(\frac{1}{\varepsilon}\right)\cdot (\varepsilono - \xi)\: d\varepsilon \nonumber\\
&\addabove{=}{b}& \int_{\varepsilonstar}^1 \left(\frac{1}{\varepsilon}\right)\cdot \varepsilon\: d\varepsilon \quad - \quad \left[\xi - (\varepsilono - \xi)\ln\left(\frac{\varepsilono}{\varepsilono - \xi}\right)\right]\label{eq:88}
\end{eqnarray}
\hrulefill
\end{figure*}

We now turn to prove~\eqref{eq:PMAP_Good} in the range  $\varepsilon < \varepsilonstar$.  The proof is obtained straightforwardly by examining the output of ML decoding.  An ML decoder can be perceived as a suboptimal bitwise estimator which is not allowed to output erasures.  The bit error rate (normalized by $n$) at the output of ML decoding cannot exceed the word error rate (denoted $\Pe(\cC_n; \varepsilon)$ in Definition~\ref{def:Good codes}), because the worst-case number of bit errors in a decoded codeword cannot exceed $n$.  By the P2P-optimality of $\{\cC_n\}$, $\Pe(\cC_n; \varepsilon)$ must approach zero for $\varepsilon < \varepsilonstar$.  The bit error rate with optimal estimation equals half the erasure rate at the output of bitwise MAP estimation as defined in Sec.~\ref{sec:Limitations_Good_Codes} (the optimal bitwise estimator makes a uniform random decision in $\{0,1\}$ whenever the MAP estimator of Sec.~\ref{sec:Limitations_Good_Codes} outputs an erasure).  This bit error cannot exceed $\Pe(\cC_n; \varepsilon)$, and thus must approach zero as well.
{\hfill$\QED$}
\subsection{Relation between \normalfont$I(\cC;\:\varepsilon)$ and \normalfont$P_\MAP(\cC;\:\varepsilon)$ in the Proof of Theorem~\ref{theorem:Good_over_BEC}}\label{apdx:Lemma_1}
The following lemma parallels~\cite[Expression~(1)]{GuoMutInfMMSE}.  The lemma extends results from~\cite{PalomarMutInf}.
\begin{lemma}\label{lemma:1}
The following holds for any linear code $\cC$ and $\varepsilon \in [0,1]$.
\begin{eqnarray}
\frac{d}{d \varepsilon} I(\cC;\:\varepsilon) = n\cdot \left(-\frac{1}{\varepsilon}\right)\cdot P_\MAP(\cC;\:\varepsilon),\label{eq:22}
\end{eqnarray}
where $I(\cC;\:\varepsilon)$ is defined as in~\eqref{eq:87}.
\end{lemma}

{\it Proof:}
We begin with the following identity, which follows from~\cite[Expression (8)]{PalomarMutInf} (similar expressions are available in \cite[Theorem~1]{CyrilGeneralizedArea} and~\cite[Theorem~1]{AlexeiEXIT}).
\begin{eqnarray}\label{eq:20}
&&\frac{d}{d \varepsilon} I(\cC;\:\varepsilon) = \nonumber \\&&\quad = \sum_{i=1}^{n}\EE\left[\frac{\partial \ln P^{(\varepsilon)}_{Y_i|X_i}(Y_i|X_i)}{\partial \varepsilon}\log P^{(\varepsilon)}_{X_i|\bY}(X_i|\bY)\right].\nonumber \\
\end{eqnarray}
Recall from Sec.~\ref{sec:Notation} that $\ln$ denotes the natural logarithm, and $\log$ denotes the base-2 logarithm.  The expectation is over both $\bX$ and $\bY$.  $P^{(\varepsilon)}_{Y_i|X_i}(y|x)$ and $P^{(\varepsilon)}_{X_i|\bY}(x|\by)$ denote the conditional probability functions corresponding to $\bX$ and $\bY$, where the superscript $\varepsilon$ denotes the BEC erasure probability.

Rewriting~\eqref{eq:20} we obtain
\begin{eqnarray}\label{eq:21}
\frac{d}{d \varepsilon} I(\cC; \:\varepsilon) &=& \EE_\bY\left\{\sum_{i=1}^{n}\EE_{X_i}\left[\frac{\partial \ln P^{(\varepsilon)}_{Y_i|X_i}(Y_i|X_i)}{\partial \varepsilon}\times \nonumber\right.\right.\\&&\quad\quad\quad\quad\quad\quad\quad\quad\left.\left. \times\log P^{(\varepsilon)}_{X_i|\bY}(X_i|\bY)\right]\right\},\nonumber\\
\end{eqnarray}
where the first expectation is over $\bY$ and the second over $X_i$.

Let $\hX_i(\by)$ denote the MAP decoder output corresponding to a channel output vector $\by$ and index $i$.  As mentioned in Sec.~\ref{sec:Limitations_Good_Codes}, this output is obtained by mapping of the {\it a posteriori} probability $P^{(\varepsilon)}_{X_i|\bY}(1|\by)$ to the set $\{0,1,\Erasure\}$.
\begin{eqnarray*}
\hX_i(\by) = \left\{
               \begin{array}{ll}
                 x, & \hbox{$P^{(\varepsilon)}_{X_i|\bY}(1|\by) = x \in \{0,1\}$,} \\
                 \Erasure, & \hbox{$P^{(\varepsilon)}_{X_i|\bY}(1|\by) = 1/2$,}
               \end{array}
             \right.
\end{eqnarray*}
where we have relied on the fact that since $\cC$ is linear, $P^{(\varepsilon)}_{X_i|\bY}(1|\by)$ is guaranteed to be in the set $\{0,1,1/2\}$~\cite[Sec.~3.2.1]{Urbanke_Book}.

If $\hX_i(\by) = x\in \{0,1\}$, the transmitted $X_i$, conditioned on $\bY=\by$, equals $x$ with probability 1, and thus $\log P^{(\varepsilon)}_{X_i|\bY}(X_i|\by) = 0$ with probability 1.  $\ln P^{(\varepsilon)}_{Y_i|X_i}(Y_i=y_i|X_i=x)$ equals $\ln \varepsilon$ if $y_i = \Erasure$ and $\ln (1-\varepsilon)$ otherwise, and so its derivative is finite.   If $\hX_i(\by) = \Erasure$ we have $P^{(\varepsilon)}_{X_i|\bY}(x_i|\by) = 1/2$ for $x_i \in \{0,1\}$.  Furthermore, $\by$ must clearly satisfy $y_i = \Erasure$ (or else $\hX_i(\by) = \Erasure$ cannot hold) and thus $P^{(\varepsilon)}_{Y_i|X_i}(y_i \given x_i) = \varepsilon$ for $x_i\in\{0,1\}$.  We can now rewrite~\eqref{eq:21} as
\begin{eqnarray*}
\frac{d}{d \varepsilon} I(\cC;\:\varepsilon) &=& \EE_\bY\left\{\sum_{\hX_i(\bY) = \Erasure}\EE_{X_i}\left[\frac{\partial \ln \varepsilon}{\partial \varepsilon}\log (1/2)\right]\right\}\nonumber\\&=&
\EE_\bY\left\{\sum_{\hX_i(\bY) = \Erasure}\left(-\frac{1}{\varepsilon}\right)\right\}\nonumber\\
&=&
\left(-\frac{1}{\varepsilon}\right)\cdot \EE_\bY\left(\left|\left\{i:\hX_i(\bY) = \Erasure\right\}\right|\right).
\end{eqnarray*}
The desired~\eqref{eq:22} now follows by the definition of $P_\MAP(\cC;\:\varepsilon)$.

{\hfill$\QED$}
\section{Proof of Theorem~\ref{theorem:Formal}}\label{apdx:Proof_of_Theorem_Formal}

For simplicity of notation, we focus on a given $n$, and let $\cC$ denote an LDPC code from the sequence $\{\cC_n\}$, dropping the index $n$.  We begin by quoting~\cite[Proposition 3]{Primitive_Relay}, which specializes~\cite[Theorem 6]{CoverElGamal} (we have changed the notation).

\myfigure{file=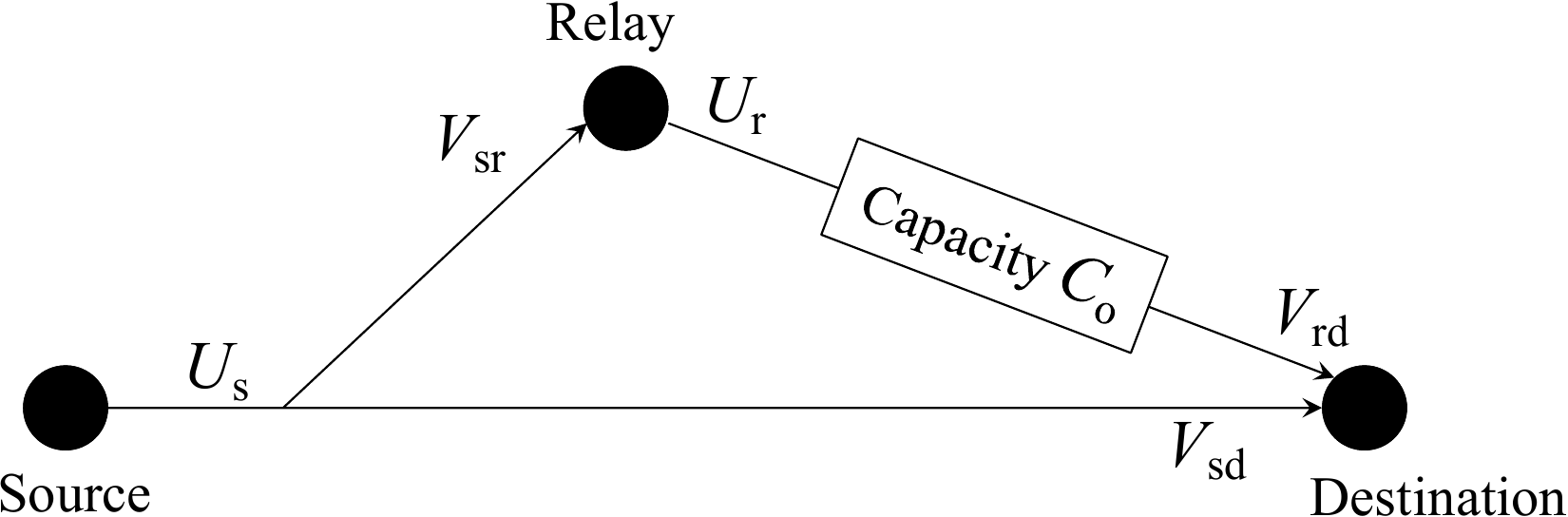, width = 8cm}{Channel model for Proposition~\ref{proposition:CF}.  Like the channel of Sec.~\ref{sec:BEC_relay}, the channels to the destination are decoupled.  $U_\SRC$ denotes the channel input at the source, and $V_\StR$ and $V_\StD$ denote the channel outputs at the relay and the destination, respectively.  $C_\textrm{o}$ is the capacity of the relay-destination channel (a more precise characterization of this channel is inconsequential).}{fig:Primitive_Relay}
\begin{proposition} (\cite[Proposition~3]{Primitive_Relay}) \label{proposition:CF}
Consider a memoryless relay channel as depicted in Fig.~\ref{fig:Primitive_Relay}.  The following rate is achievable using CF.
\begin{eqnarray*}
R_{\CF} = \max
\{I(U_\SRC; \hV_\StR, V_\StD) : I(V_\StR; \hV_\StR\given V_\StD ) \le C_\textrm{o}\}.
\end{eqnarray*}
$\hV_\StR$ is an auxiliary variable which is statistically dependent on $V_\StR$, and the maximum is over distributions $p(u_\SRC)$ of $U_\SRC$ and conditional distributions $p(\hv_\StR \given  v_\StR)$ for $\hV_\StR$ given $V_\StR$.
\end{proposition}

We now apply this proposition to prove Theorem~\ref{theorem:Formal}.  Our technique involves focusing on the {\it virtual} channel, obtained from the setting of soft-DF-BP2 (Algorithm~\ref{alg:Soft-DF-BP2}) by encapsulating LDPC encoding at the source, and soft decoding at the relay, into the channel model.  With the notation of Fig.~\ref{fig:Primitive_Relay} we have: $U_\SRC = \bX_\SRC$ where $\bX_\SRC$ is the transmitted LDPC codeword.  That is, the input alphabet of this channel is the LDPC code $\cC$.   We define $V_\StR = \bYBPrelay$ where $\bYBPrelay$ is the soft relay estimate, and $V_\StD = \bY_\StD$ where $\bY_\StD$ is the relay channel output.  Finally, $C_\textrm{o} = nC_\RtD$, because each use of the virtual channel corresponds to $n$ uses of the
original channel.

We define the analysis version of soft-DF-BP2 (mentioned in Theorem~\ref{theorem:Formal}) to coincide with CF, as defined by~\cite[Theorem 6]{CoverElGamal}, over the above virtual channel.    We define $U_\SRC$ (see Proposition~\ref{proposition:CF}) to be uniformly distributed in the LDPC code $\cC$, and $\hV_\StR = \hbYBPrelay$, where the components of $\hbYBPrelay$ are defined as in~\eqref{eq:hyBP}.  Recalling our above assignment $V_\StR = \bYBPrelay$, this specifies the joint distribution of $\hV_\StR$ and $V_\StR$.  While these might not be  the maximizing distributions, they guarantee achievable rates, which are our focus.

We now briefly sketch the main components of the strategy, borrowed from the proof of~\cite[Theorem 6]{CoverElGamal} and applied to our setting.  For a complete discussion, the reader is referred to~\cite{CoverElGamal}.  The source uses a code $\cC_\CF$ which is defined over the virtual source alphabet $\cC$ and has block length $n_\CF$ (measured in symbols of the virtual channel).  Equivalently, it uses a code $\cCstar$ which is the concatenation of $\cC_\CF$ and $\cC$.  It transmits a codeword $\bu_\SRC = (\bx_{\SRC,1},\ldots,\bx_{\SRC,n_\CF})$ from $\cCstar$.  The relay obtains a vector $(\by_{\StR,1},\ldots,\by_{\StR,n_\CF})$ and applies BP estimation independently to each received subvector $\by_{\StR,i}$, $i = 1,\ldots,n_\CF$, obtaining a vector $\bv_\StR= (\byBP_{\RLY,1},\ldots,\byBP_{\RLY,n_\CF})$.  It then searches a codebook $\cC_\WZ$ (with block length $n_\CF\cdot n$) for a codeword $\hbv_\StR$ which is ``close'' to $\bv_\StR$ (to be elaborated shortly).  It records its bin number $I_\RLY$ (discussed below), and transmits it using a channel code $\cC_\textrm{ch}$ to the destination.  The destination decodes $I_\RLY'$ from the output $\bv_\RtD$ of the relay-destination channel.  It searches the corresponding bin for a codeword $\hbv_\StR'$ which is ``close'' to the output $\bv_\StD$ of the source-destination channel.  Finally, it seeks a codeword $\bu_\SRC'$ from $\cCstar$ which is simultaneously ``close'' to both $\hbv_\StR'$ and $\bv_\StD$.

Following~\cite[Theorem 6]{CoverElGamal}, the various CF codes ($\cC_\CF$, $\cC_\WZ$ and $\cC_\textrm{ch}$) are generated randomly, and $\cC_\WZ$ is randomly partitioned into equal-sized bins.   Proximity between vectors is defined in terms of {\it typicality} (see~\cite[Definition~1]{CoverElGamal}).   For example, the codeword $\hbv_\StR$ sought by the relay must be such  that the pair  $(\bv_\StR,\hbv_\StR)$ is typical\footnote{In~\cite{CoverElGamal}, the relay and destination enforce joint typicality also with the relay's transmitted codeword.  In our setting, due to the decoupling of the channels to the destination, this can be omitted.} to the above-defined distribution of $(V_\StR,\hV_\StR)$.    We also assume that the destination applies typicality decoding to $\cCstar$, rather than BP decoding on codewords of $\cC$.  For additional details of the various components, see~\cite{CoverElGamal}.\footnote{Note that the distribution of the $V_\StR$, its joint distribution with $\hV_\StR$, as well as several other distributions involved, are complex to describe.  A precise characterization of these distributions would be required should we choose to implement the strategy, but is not necessary to prove the achievability of rates, which is our focus.}

Proposition~\ref{proposition:CF}  guarantees that if~\eqref{eq:CF_I_Condition_Block} holds (Condition~2 of the theorem), we can select the various codes such that the rate
\begin{eqnarray} \label{eq:44}
R_\CF = \frac{1}{n} I(U_\SRC; \hV_\StR,  V_\StD )= \frac{1}{n} I(\bX_\SRC; \hbYBPrelay,  \bY_\StD )
\end{eqnarray}
is achievable, where normalization by $n$ is required because we are measuring rate in bits per use of the original erasure relay channel.
Finally, to evaluate~\eqref{eq:44}, we apply the analysis of soft-DF-BP over the auxiliary channel.  Relying on Condition~1 of the theorem, the following inequality can now be shown to hold
\begin{eqnarray}\label{eq:31}
\frac{1}{n} I(\bX_\SRC; \hbYBPrelay, \bY_\StD ) > R\cdot\Big(1 - h(\xi/R)\Big) + o(1),
\end{eqnarray}
where $R$ is the rate of $\cC$ and $o(1)$ is a term that approaches zero as $n\rightarrow\infty$.
The proof of~\eqref{eq:31} relies on concepts similar to the proof of the joint source-channel coding theorem (see e.g.,~\cite[Sec.~10.5]{MacKay_Book}) and is omitted.  The argument $\xi/R$ to the entropy function is an upper bound on the fraction of erroneous information bits ($\xi$ bounds the fraction of codebits), as assumed in the bound of~\cite{MacKay_Book}.
{\hfill$\QED$}
\section{Proof of Lemma~\ref{lemma:Naive}}\label{apdx:Proof_of_Lemma_Naive}

For a quick reference of the main notations used in this proof, see Fig.~\ref{fig:Stochastic_Relay}.

\begin{remark}\label{remark:1}
Recall from Remark~\ref{remark:ensemble_random} that our probability space is conditioned on the code $\cC$, which is randomly selected from an ensemble.  For simplicity of notation, we drop the conditioning on $\cC$ from our expressions.  However, all information measures (mutual information and entropies, including $I( \bYBPrelay; \hbYBPrelay \given \bY_\StD )$) and all probabilities, are dependent on it.  Thus, as $\cC$ is randomly selected, they are random variables themselves.
\end{remark}
We begin by writing
\begin{eqnarray}
 && \hspace{-0.5cm}I( \bYBPrelay; \hbYBPrelay \given \bY_\StD ) \nonumber\\
 && \: = H( \hbYBPrelay \given \bY_\StD ) - H( \hbYBPrelay \given \bYBPrelay, \bY_\StD ) \nonumber \\
&&\addabove{=}{a} H( \hbYBPrelay,\hbEBPrelay \given \bY_\StD )- H( \hbYBPrelay \given \bYBPrelay, \bY_\StD )\nonumber\\
&&\: =  H( \hbYBPrelay\given \hbEBPrelay, \bY_\StD ) + H( \hbEBPrelay \given \bY_\StD )- H( \hbYBPrelay \given \bYBPrelay, \bY_\StD )\nonumber\\
&& \addabove{=}{b} H( \hbYBPrelay\given \hbEBPrelay, \bY_\StD ) + H( \hbEBPrelay )- H( \hbYBPrelay \given \bYBPrelay, \bY_\StD ).\nonumber \\  \label{eq:11}
\end{eqnarray}
In (a), we have defined $\hbEBPrelay = I_\Erasure(\hbYBPrelay)$ as the erasure indicator vector of $\hbYBPrelay$ (see~\eqref{eq:erasure_indicator}).  To justify (b), we argue that $\hbEBPrelay$ is independent of $\bY_\StD$.  To see this,
first observe that by the above definitions and~\eqref{eq:hyBP}, the following holds for $i = 1,\ldots,n$
 \begin{eqnarray}\label{eq:hEBPrelay_is_sum}
 \hEBPrelay{i} = \EBPrelay{i} + \hE_{\BPR,i}\:,
 \end{eqnarray}
where $\EBPrelay{i}$ is a component of $\bEBPrelay$, defined as $\bEBPrelay = I_\Erasure(\bYBPrelay)$, and $\hE_{\BPR,i}$ is simply the quantization noise (see Sec.~\ref{sec:Analysis_soft_DF_BP2}).  By this definition, $\bEBPrelay$ specifies the set of erased indices at the output of the relay's BP, and is thus a function of $\bE_\StR$, the erasure noise on the source-relay channel. Both $\bE_\StR$ and $\hbE_\BPR$ are independent of $\bY_\StD$, and thus so is $\hbEBPrelay$, proving equality (b).

Our proof proceeds by bounding the three terms on the right-hand side of~\eqref{eq:11}.  We begin with the first term.
\begin{eqnarray}
&&\hspace{-0.5cm}H( \hbYBPrelay\given \hbEBPrelay, \bY_\StD ) \nonumber\\
&&\quad \:\leq\: \sum_{i=1}^n H( \hYBPrelay{i}\given \hbEBPrelay, \bY_\StD )\nonumber\\
&&\quad \:\leq\: \sum_{i=1}^n H( \hYBPrelay{i}\given \hEBPrelay{i}, Y_{\StD,i} )\nonumber\\
&&\quad \addabove{=}{a} \sum_{i=1}^n  H( X_{\SRC,i}\given \hEBPrelay{i}=0, Y_{\StD,i} ) \cdot\Pr[\hEBPrelay{i}=0]\nonumber\\
&&\quad \addabove{=}{b} \sum_{i=1}^n  H( X_{\SRC,i}\given \hEBPrelay{i}=0, Y_{\StD,i} = \Erasure) \cdot\Pr[Y_{\StD,i} = \Erasure]\times\nonumber\\
&&\qquad\qquad \times \Pr[\hEBPrelay{i}=0]\nonumber\\
&&\quad \addabove{\leq}{c} \sum_{i=1}^n  1 \cdot\varepsilon_\StD\cdot (1-\DeltaRelayBPInd{i}\circ \hvarepsilon_\BPR)\nonumber\\
&&\quad \:=\: n\left\{\varepsilon_\StD\cdot \left[1-\left(\frac{1}{n}\sum_{i=1}^n\DeltaRelayBPInd{i}\right)\circ \hvarepsilon_\BPR\right]\right\}\nonumber\\
&&\quad \addabove{=}{d} n\Big\{\varepsilon_\StD\cdot [1-\deltaRelayBP\circ \hvarepsilon_\BPR] + o(1)\Big\}. \label{eq:77}
\end{eqnarray}
In (a), we have relied on the equality  $\hbYBPrelay = \bX_\SRC + \hbEBPrelay$, which holds by arguments similar to~\eqref{eq:error_vector} and by~\eqref{eq:hyBP}.  With this equality, if $\hEBPrelay{i}=\Erasure$ then $\hYBPrelay{i}=\Erasure$ with probability 1 and so $H( \hYBPrelay{i}\given \hEBPrelay{i}=\Erasure, Y_{\StD,i} ) = 0$.  If $\hEBPrelay{i}=0$ then
$\hYBPrelay{i}=X_{\SRC,i}$.
In (b), we have observed that if $Y_{\StD,i} = x$ where $x\in \{0,1\}$, then by the transition probabilities of the BEC, $X_{\SRC,i}=x$ with probability 1, and so $H( X_{\SRC,i}\given \hEBPrelay{i}=0, Y_{\StD,i} = x)=0$.  In (c), we have observed that since $X_{\SRC,i}$ is defined over $\{0,1\}$, $H( X_{\SRC,i}\given \hEBPrelay{i}=0, Y_{\StD,i} = \Erasure) \leq 1$.  We have also evaluated $\Pr[Y_{\StD,i} = \Erasure] = \varepsilon_\StD$.  Finally, we have defined  $\DeltaRelayBPInd{i} = \Pr[\EBPrelay{i}=\Erasure]$.  By~\eqref{eq:hEBPrelay_is_sum} and the fact of $\hE_{\BPR,i}$ being distributed as $\eras(\hvarepsilon_\BPR)$ (see Sec.~\ref{sec:Analysis_soft_DF_BP2}), we have $\Pr[\hEBPrelay{i}=\Erasure] = \DeltaRelayBPInd{i} \circ \hvarepsilon_\BPR$ (invoking~\eqref{eq:circ}).  Observe that each $\DeltaRelayBPInd{i}$ is in fact a function of the code $\cC$, and therefore by Remark~\ref{remark:1}, it is a random variable.  In (d), we have relied on the following derivation.
\begin{eqnarray}
\frac{1}{n}\sum_{i=1}^n\DeltaRelayBPInd{i} &\addabove{=}{a}& \frac{1}{n}\sum_{i=1}^n \Pr[\EBPrelay{i} = \Erasure] \nonumber \\
&\addabove{=}{b}&  \EE\left[ \frac{1}{n}\sum_{i=1}^n \chi_{\EBPrelay{i} = \Erasure} \right]\nonumber\\
&\addabove{=}{c}&  \EE\left[ \DRelayBP \right] \nonumber\\
&\addabove{=}{d}& \deltaRelayBP  + o(1). \label{eq:37}
\end{eqnarray}
In (a), we have invoked the definition of $\DeltaRelayBPInd{i}$.  In (b), $\chi_{\EBPrelay{i} = \Erasure}$ is an indicator random variable, which equals 1 if $\EBPrelay{i} = \Erasure$.  The expectation is over the channel transitions, but the code $\cC$ is assumed to be fixed. In (c), $\DRelayBP \defined  P(\Erasure\given \bYBPrelay)$ is the realized erasure rate (defined as in~\eqref{eq:erasure_rate}). In (d) we have relied on~\eqref{eq:10} (see Appendix~\ref{apdx:Analysis_DRelayBP} below).  This bound holds for large enough $n$ with probability at least $1-\exp(-\tau \sqrt{n})$ for $\tau > 0$, thus complying with the conditions of Lemma~\ref{lemma:Naive}.

We now turn to bound the second term on the right-hand side of~\eqref{eq:11}.
\begin{eqnarray}
H( \hbEBPrelay ) &\leq& \sum_{i=1}^n H( \hEBPrelay{i} ) \nonumber \\
&\addabove{=}{a}& \sum_{i=1}^n h( \DeltaRelayBPInd{i} \circ \hvarepsilon_\BPR ) \nonumber\\
&\addabove{\leq}{b}&  n \cdot h\left( \frac{1}{n}\sum_{i=1}^n\DeltaRelayBPInd{i} \circ \hvarepsilon_\BPR \right) \nonumber \\
&\addabove{=}{c}& n \cdot \Big{[} h( \deltaRelayBP \circ \hvarepsilon_\BPR) + o(1) \Big{]}.\label{eq:76}
\end{eqnarray}
In (a), we have invoked $\Pr[\hEBPrelay{i}=\Erasure] = \DeltaRelayBPInd{i} \circ \hvarepsilon_\BPR$, which was justified above.  In (b), we have applied Jensen's inequality, relying on the concavity of the entropy function.   In (c), we have relied on~\eqref{eq:37} and invoked the continuity of $h(\cdot)$.

We now turn to evaluate the last term on the right-hand side of~\eqref{eq:11}.
\begin{eqnarray}
&&\hspace{-1.5cm}H( \hbYBPrelay \given \bYBPrelay, \bY_\StD ) \nonumber \\
&&\addabove{=}{a} H( \hbYBPrelay \given \bYBPrelay)\nonumber\\
&&\addabove{=}{b} \sum_{i=1}^n H( \hYBPrelay{i} \given \bYBPrelay, \hYBPrelay{1},\ldots,\hYBPrelay{i-1}) \nonumber\\
&&\addabove{=}{c} \sum_{i=1}^n H( \hYBPrelay{i} \given \YBPrelay{i})\nonumber\\
&&\addabove{=}{d} \sum_{i=1}^n h( \hvarepsilon_\BPR )(1-\DeltaRelayBPInd{i})\nonumber\\
&&\:=\: n\cdot\left[h( \hvarepsilon_\BPR )\left(1-\frac{1}{n}\sum_{i=1}^n\DeltaRelayBPInd{i}\right)\right]\nonumber\\
&&\addabove{=}{e} n \cdot \Big[  h( \hvarepsilon_\BPR )\cdot(1-\deltaRelayBP) + o(1)\Big].\label{eq:15}
\end{eqnarray}
In (a), we have relied on the fact that the three random vectors on both sides of the equation form a Markov chain: $\bY_\StD \leftrightarrow \bYBPrelay \leftrightarrow \hbYBPrelay$.  In (b), we have simply applied the chain rule for entropy.  In (c), we have relied on the fact that by~\eqref{eq:hyBP}, the random variables on the previous line form the following Markov chain:  $\hbYBPrelaySub{\sim i} \leftrightarrow \bYBPrelaySub{\sim i} \leftrightarrow \YBPrelay{i} \leftrightarrow \hYBPrelay{i}$, where $\hbYBPrelaySub{\sim i}$ is defined by,
$\hbYBPrelaySub{\sim i} \defined (\hYBPrelaySub{1},\ldots,\hYBPrelaySub{i-1},\hYBPrelaySub{i+1},\ldots,\hYBPrelaySub{n})$, and $\bYBPrelaySub{\sim i}$ is similarly defined.  In (d), we have relied on the observation that if $\YBPrelay{i} = \Erasure$, then $\hYBPrelay{i} = \Erasure$ with probability 1 and thus $H( \hYBPrelay{i} \given \YBPrelay{i} = \Erasure) = 0$, and if $\YBPrelay{i} = x \in \{0,1\}$ then $\hYBPrelay{i} = x$ with probability $1-\hvarepsilon_\BPR$ and  $\hYBPrelay{i} = \Erasure$ with probability $\hvarepsilon_\BPR$.  By definition, $\EBPrelay{i}=\Erasure$ if and only if $\YBPrelay{i}=\Erasure$ and so $\Pr[\YBPrelay{i}=\Erasure] = \DeltaRelayBPInd{i}$, where  $\DeltaRelayBPInd{i}$ is defined as above.  In (e), we have applied~\eqref{eq:37}.

Finally, combining~\eqref{eq:11},~\eqref{eq:77},~\eqref{eq:76} and~\eqref{eq:15}, we obtain our desired~\eqref{eq:Naive_Bound}.
{\hfill$\QED$}
\subsection{Analysis of \normalfont$\DRelayBP$ and \normalfont$\deltaRelayBP$}\label{apdx:Analysis_DRelayBP}

In our above proof, we defined $\DRelayBP =  P(\Erasure\given \bYBPrelay)$, the erasure rate (defined as in~\eqref{eq:erasure_rate}) at the output of BP at the relay.  In Lemma~\ref{lemma:Naive}, we defined $\deltaRelayBP$ to be its asymptotic mean, as computed by density evolution.  We use the following expressions to evaluate $\deltaRelayBP$, following~\cite[Chapter~3]{Urbanke_Book}.
\begin{eqnarray*}
\deltaRelayBP = \varepsilon_\StR \tlambda(1 - \rho(1-\xRelayBP)),
\end{eqnarray*}
where $\xRelayBP$ is the largest fixed point of the function $F(x) = \varepsilon_\StR \lambda(1 - \rho(1-x))$ in the range $0 \leq x \leq \varepsilon_\StR$,
$\lambda(x) = \sum_{i}\lambda_i x^{i-1}$, $\rho(x) = \sum_{j}\rho_j x^{j-1}$ and $\tlambda(x) = \sum_{i}\tlambda_i x^{i}$ where $\tlambda_i$ was defined by~\eqref{eq:tlambda}.

The following lemma, which is based on~\cite[Theorem~3.107]{Urbanke_Book}, relates $\DRelayBP$ and $\deltaRelayBP$.
\begin{lemma}\label{lemma_DRelayBP}
Let $\cC$ and $\deltaRelayBP$ be defined as in Lemma~\ref{lemma:Naive}, and let $\DRelayBP$ be defined as above.  Then the following two inequalities hold for large enough $n$ with probability at least $1-\exp(-\tau \sqrt{n})$ (the probability being over the random selection of $\cC$, see Remark~\ref{remark:1}).
\begin{eqnarray}
\hspace{-0.2cm}\Pr\Big[ | \DRelayBP - \deltaRelayBP| > a_1 n^{-1/6} \Big] &\le& a_2n^{1/6}\e^{-\tau\sqrt{n}}\label{eq:36}\\
\Big|\:\EE [\DRelayBP] - \deltaRelayBP\:\Big| &\le& a_1 n^{-1/6} + a_2n^{1/6}\e^{-\tau\sqrt{n}}, \nonumber \\ \label{eq:10}
\end{eqnarray}
where $\tau,a_1,a_2 > 0$ are some constants, dependent on~$\lambda$ and~$\rho$.
\end{lemma}
{\it Proof:}
In~\cite{Urbanke_Book}, the authors present bounds which are valid for a probability space that includes the random selection of $\cC$, as well as the random channel transitions.  Our proof amounts simply to applying their results to our setting, where the probabilities are conditioned on $\cC$.  We let $Q[\cdot]$ denote probabilities as in their setting, and $P(\cC)$ denote the left-hand side of~\eqref{eq:36}.  With this notation
\begin{eqnarray*}
P(\cC) = Q\Big[| \DRelayBP - \deltaRelayBP| > a_1 n^{-1/6}   \given \textrm{The code $\cC$ is used}\Big].
\end{eqnarray*}
By our above discussion, $P(\cC)$ is a random variable, which depends on the randomly selected $\cC$.

The following result is a direct application of~\cite[Theorem~3.107]{Urbanke_Book}\footnote{The results of~\cite[Theorem~1]{Luby_Irregular} and~\cite[Theorem~2]{Urbanke_Message_Passing} are unsuitable to our setting, because they assume that the number of BP iterations is limited by a predefined maximum.}
\begin{eqnarray}
Q\Big[ | \DRelayBP - \deltaRelayBP| > a_1 n^{-1/6} \Big] &\le& a_2n^{1/6}\e^{-2\tau\sqrt{n}},\label{eq:32}
\end{eqnarray}
for appropriately selected constants.  Recall that in Sec.~\ref{sec:LDPC}, we defined the maximal degrees in $\lambda$ and $\rho$ to be finite, and this is required by the conditions of~\cite[Theorem~3.107]{Urbanke_Book}.  We now use Markov's inequality to bound the probability that $P(\cC)$ is too large.
\begin{eqnarray*}
Q\Big[ P(\cC) > a_2n^{1/6}\e^{-\tau\sqrt{n}} \Big] &\addabove{\leq}{a}&  \frac{\EE\left[P(\cC)\right]}{a_2n^{1/6}\e^{-\tau\sqrt{n}}}\\
&\addabove{=}{b}& \frac{Q\Big[ | \DRelayBP - \deltaRelayBP| > a_1 n^{-1/6} \Big]}{a_2n^{1/6}\e^{-\tau\sqrt{n}}}\\
&\addabove{\leq}{c}&  \e^{-\tau\sqrt{n}}.
\end{eqnarray*}
Above, (a) follows by Markov's inequality.  The expectation on the right-hand side is over the random selection of the code $\cC$. (b) follows by the law of total probability and the definition of $P(\cC)$ and (c) follows by~\eqref{eq:32}.~\eqref{eq:36} now follows.  \eqref{eq:10}~also follows by the observation that $\DRelayBP$ is confined to $[0,1]$.
{\hfill$\QED$}

\section{Proof of Theorem~\ref{theorem:I_Bounds}}\label{apdx:Proof_of_Theorem_I_Bounds}

Our proof is based on the proof of Lemma~\ref{lemma:Naive} (cf.~Appendix~\ref{apdx:Proof_of_Lemma_Naive}).  Once again we use~\eqref{eq:11} as a starting point, and bound its terms.  As noted in Sec.~\ref{sec:Quantization_Noise}, we improve upon the bound of Lemma~\ref{lemma:Naive} by exploiting dependencies between the components of $\bYBPrelay$, the output of BP at the relay with soft-DF-BP2.  In Appendix~\ref{sec:Bound_H_hbYBPrelay} below we will exploit dependencies between the values of non-erased bits  (see Sec.~\ref{sec:Quantization_Noise}) to tighten the bound on the first term in~\eqref{eq:11} ($H( \hbYBPrelay\given \hbEBPrelay, \bY_\StD )$).  In Appendix~\ref{apdx:IPlus2} we will exploit dependencies between erased bits at the output of BP, to produce a bound on the second term ($H( \hbEBPrelay )$), which is sometimes tighter than~\eqref{eq:76}.  As the new bound is not always tighter than~\eqref{eq:76}, we have applied the minimum operation in~\eqref{eq:51}.  Finally, the l.d.f. operation in~\eqref{eq:I_Plus} will be justified in Appendix~\ref{apdx:Justification_ldf}.

\subsection{Upper Bound on \normalfont$H( \hbYBPrelay\given \hbEBPrelay, \bY_\StD )$}\label{sec:Bound_H_hbYBPrelay}

We begin with the string of equations ending with~\eqref{eq:12} on the following page.
In (a), $\bE_\StR$ is the noise along the source-relay link (see~\eqref{eq:Relay_Erasure_Noise}).  In this equation, we have relied on the Markov chain relation between the random variables, $\bY_\StD \leftrightarrow \bX_\SRC\leftrightarrow \hbYBPrelay  \leftrightarrow \hbEBPrelay \leftrightarrow \bE_\StR$.  In (b), we have applied the definition of conditional entropy:  The expectation is over the variable $\bcErelay$, which is defined to be distributed identically as $\bE_\StR$.  In (c), we have applied the chain rule for entropy.  (d) will be discussed shortly.

For simplicity, rather than consider BP (Algorithm~\ref{alg:BP}), we examine the following algorithm, called the {\it peeling decoder}~\cite[Sec.~3.19]{Urbanke_Book}, originally due to Luby~\etal~\cite[Algorithm~1]{Luby_Erasure}.
In~\cite[Sec.~3.22]{Urbanke_Book} the algorithm was shown to be equivalent to BP, in the sense that it yields precisely the same output as BP.\footnote{Note that with both algorithms, we let the decoding iterations continue without restriction, until they can no longer provide a benefit.   This is required for the equivalence between the algorithms to hold. }  Therefore, we may assume without loss of generality that it is the one applied by the relay of soft-DF-BP2.   Like BP, this algorithm is iterative, and relies on the Tanner graph representation of the LDPC code. However, it is not a message-passing algorithm.
    \begin{algorithm}[Peeling Decoder] \label{alg:Simplified_BP}$\:$
    \begin{enumerate}
    \item {\bf Initialization:} Set the value of each variable node to the channel output.
    \item {\bf Iterations:} Each iteration involves examining all check nodes one-by-one according to some predefined order.  At each check node, if the values at {\it all but one} of the adjacent variables are known (not erased), set the remaining unknown variable node to the modulo-2 sum of the others.\footnote{In~\cite[Algorithm~1]{Luby_Erasure} and~\cite[Sec.~3.22]{Urbanke_Book}, the check node and its adjacent edges are removed from the graph.  For simplicity of exposition, we omit this part of the algorithm.  It is straightforward to see that this has no effect on the final outcome.}
  \item {\bf Stopping criterion:} Stop iterating as soon as the values of no new variables are discovered.
        \end{enumerate}
    \end{algorithm}

We assume, without loss of generality, that the components of $\hbYBPrelay$ are ordered in the order that they would have been discovered by Algorithm~\ref{alg:Simplified_BP} had the channel erasures corresponded to $\bcErelay$. That is, the first components are the ones revealed at the initialization step of the algorithm (i.e., not erased by the channel).  They are followed by the components that the algorithm revealed at iteration 1, in the order that they were discovered, and so forth.  Components that were not revealed at any iteration are ordered last.

In~\myeqref{eq:12}{d}, we have separated the sums of components that were revealed by the channel, components that were revealed at the various iterations of the algorithm, and components that remained unknown at the algorithm's end.  We let $\cD_\StR \defined P(\Erasure\given \bcErelay)$ (see~\eqref{eq:erasure_rate}), and $\cDRelayBP \defined P(\Erasure\given \bcErelayBP)$ where $\bcErelayBP$ denotes the output of BP when its input is $\bcErelay$.  It is distributed as
$\DRelayBP$ (see Appendix~\ref{apdx:Proof_of_Lemma_Naive}).
\begin{figure*}
\begin{eqnarray}
H( \hbYBPrelay\given \hbEBPrelay, \bY_\StD ) &\addabove{=}{a}& H( \hbYBPrelay\given \hbEBPrelay, \bY_\StD, \bE_\StR )\nonumber\\
&\addabove{=}{b}& \EE_{\bcErelay} \left[ H( \hbYBPrelay\given \hbEBPrelay, \bY_\StD, \bE_\StR=\bcErelay ) \right] \nonumber\\
&\addabove{=}{c}& \EE_{\bcErelay} \left[\sum_{i=1}^{n} H( \hYBPrelay{i}\given \hYBPrelay{1},\ldots,\hYBPrelay{i-1},\hbEBPrelay, \bY_\StD, \bE_\StR=\bcErelay ) \right]\nonumber\\
&\addabove{=}{d}& \EE_{\bcErelay} \left[\sum_{i=1}^{(1-\cD_\StR) n} H( \hYBPrelay{i}\given \hYBPrelay{1},\ldots,\hYBPrelay{i-1},\hbEBPrelay, \bY_\StD, \bE_\StR=\bcErelay ) +\right. \nonumber\\&&\quad\quad\quad\quad\quad+ \sum_{i=(1-\cD_\StR) n + 1}^{(1-\cDRelayBP) n} H( \hYBPrelay{i}\given \hYBPrelay{1},\ldots,\hYBPrelay{i-1},\hbEBPrelay, \bY_\StD, \bE_\StR=\bcErelay )+ \nonumber\\
&&\quad\quad\quad\quad\quad\left. +\sum_{i=(1-\cDRelayBP) n + 1}^{ n} H( \hYBPrelay{i}\given \hYBPrelay{1},\ldots,\hYBPrelay{i-1},\hbEBPrelay, \bY_\StD, \bE_\StR=\bcErelay )\right]\label{eq:12}
\end{eqnarray}
\hrulefill
\end{figure*}

We now examine the three sums on the right-hand side of~\eqref{eq:12}.  The desired exploitation of the dependencies between the bits discovered by BP will take place in the second sum, which we will examine last.  The third sum is easily evaluated to equal zero.  This is because $\hYBPrelay{i} = \Erasure$ with probability 1 for all components of the sum, which follows from~\eqref{eq:hyBP} because $\YBPrelay{i} = \Erasure$ at these components.  Turning to the components of the first sum, let $i \in \{1,\ldots,(1-\cD_\StR) n\}$.
\begin{eqnarray}
&&\hspace{-0.5cm}H( \hYBPrelay{i}\given \hYBPrelay{1},\ldots,\hYBPrelay{i-1},\hbEBPrelay, \bY_\StD, \bE_\StR=\bcErelay ) \nonumber\\
&&\addabove{\leq}{a} H( \hYBPrelay{i}\given \hEBPrelay{i}, Y_{\StD,i}, \bE_\StR=\bcErelay ) \nonumber\\
&&\addabove{=}{b} \sum_{x \in \{0,\Erasure\}}H( \hYBPrelay{i}\given \hEBPrelay{i} = x, Y_{\StD,i},\bE_\StR=\bcErelay  ) \times \nonumber\\
&&\quad\quad\quad\quad\quad\times\Pr[\hEBPrelay{i} = x\given \bE_\StR=\bcErelay] \nonumber\\
&&\addabove{=}{c} H( \hYBPrelay{i}\given \hEBPrelay{i} = 0, Y_{\StD,i},\bE_\StR=\bcErelay  )\cdot (1 -\hvarepsilon_\BPR)\nonumber\\
&&\addabove{=}{d} H( X_{\SRC,i}\given Y_{\StD,i} )\cdot (1 -\hvarepsilon_\BPR)\nonumber\\
&&\addabove{\leq}{e} \varepsilon_\StD(1 -\hvarepsilon_\BPR). \label{eq:13}
\end{eqnarray}
In (a), we have reduced the conditions on the entropy to obtain an upper bound on its value.  In (b), we have applied the definition of conditional entropy.  In (c), we have applied $H( \hYBPrelay{i}\given \hEBPrelay{i} = \Erasure, Y_{\StD,i},\bE_\StR=\bcErelay ) = 0$.  This holds because by construction of $\hbEBPrelay$ as the erasure indicator vector of $\hbYBPrelay$, if  $\hEBPrelay{i} = \Erasure$ then $\hYBPrelay{i} = \Erasure$ with probability~1.  We have also evaluated  $\Pr[\hEBPrelay{i} = 0 \given \bE_\StR=\bcErelay] = \Pr[ \hE_{\BPR,i} = 0] = (1-\hvarepsilon_\BPR)$.  This holds because we are currently examining $i \in \{1,\ldots,(1-\cD_\StR) n\}$, for which $\cE_{\StR,i} = 0$ by definition, implying $\EBPrelay{i} = E_{\StR,i} = 0$, and because by~\eqref{eq:hEBPrelay_is_sum}, given $\EBPrelay{i} = 0$, we have $\hEBPrelay{i}  = \hE_{\BPR,i}$.  In (d), we have relied on the fact that if $\hEBPrelay{i} = 0$ then $\hYBPrelay{i} = X_{\SRC,i}$, where $X_{\SRC,i}$ is the transmitted signal from the source at time $i$.  This holds by similar arguments to those of~\myeqref{eq:77}{a}.  We have also applied  the independence  between the variables $(X_{\SRC,i},Y_{\StD,i})$ and  $(\hEBPrelay{i},\bE_\StR)$, which holds because the latter pair is a  function of the noise over the source-relay channel.  In (e), we have applied $H( X_{\SRC,i}\given Y_{\StD,i} = \Erasure) = H(X_{\SRC,i}) \leq 1$ and $H( X_{\SRC,i}\given Y_{\StD,i} = 0) = H( X_{\SRC,i}\given Y_{\StD,i} = 1)= 0$.  The latter equality holds by the transition probabilities of the BEC.

We now turn to the components of the second sum in~\eqref{eq:12}.  Let $i \in \{(1-\cD_\StR) n+1, \ldots, (1-\cDRelayBP) n\}$.
\begin{eqnarray*}
&&\hspace{-0.8cm}H( \hYBPrelay{i}\given \hYBPrelay{1},\ldots,\hYBPrelay{i-1},\hbEBPrelay, \bY_\StD, \bE_\StR=\bcErelay ) \\ &&\addabove{\leq}{a} H( \hYBPrelay{i}\given \hYBPrelay{j_{i,1}},\ldots,\hYBPrelay{j_{i,d-1}}, \hEBPrelay{i}, \\
&& \hspace{2.5cm}Y_{\StD,j_{i,1}}, \ldots,Y_{\StD,j_{i,d-1}}, Y_{\StD,i}, \bE_\StR=\bcErelay ) \\
&&\addabove{\leq}{b} H( \hYBPrelay{i}\given \oYBPrelay{i}, \hEBPrelay{i}, Y_{\StD,i}, \bE_\StR=\bcErelay ) \\
&&\addabove{=}{c} H( \hYBPrelay{i}\given \oYBPrelay{i}, \hEBPrelay{i} = 0, Y_{\StD,i}, \bE_\StR=\bcErelay )\cdot (1 -\hvarepsilon_\BPR) \\
&&\addabove{=}{d} H( X_{\SRC,i}\given \oYBPrelay{i},Y_{\StD,i}, \bE_\StR=\bcErelay  )\cdot (1 -\hvarepsilon_\BPR)\\
&&\addabove{\leq}{e} \varepsilon_\StD\Big(1 - (1-\hvarepsilon_\BPR\cdot \varepsilon_\StD)^{d-1}\Big)(1 -\hvarepsilon_\BPR).
\end{eqnarray*}

The analysis follows in lines similar to the derivation leading to~\eqref{eq:13}, and we will elaborate only on the differences.  Recall that each component at indices $i \in \{(1-\cD_\StR) n+1, \ldots, (1-\cDRelayBP) n\}$ was discovered in the application of Algorithm~\ref{alg:Simplified_BP}.  Let $j_{i,1},\ldots,j_{i,d-1}$ be the indices of the other variable nodes that were connected to the check node by which index $i$ was discovered.  By the nature of Algorithm~\ref{alg:Simplified_BP}, these indices necessarily correspond to bits that were discovered previously by the algorithm.  Thus, since we have assumed that the indices $i=1,\ldots,n$ are arranged by the order in which components of $\bYBPrelay$ were discovered by Algorithm~\ref{alg:Simplified_BP}, we have $\{j_{i,1},\ldots,j_{i,d-1} \} \subseteq \{1,\ldots,i-1\}$.  (a) now follows.

In (b), we have defined
\begin{eqnarray}\label{eq:1000}
\oYBPrelay{i} = (\hYBPrelay{j_{i,1}}\cdot Y_{\StD,j_{i,1}})+\cdots+(\hYBPrelay{j_{i,d-1}}\cdot Y_{\StD,j_{i,d-1}}),
\end{eqnarray}
 where addition  and multiplication are defined as in Sec.~\ref{sec:Notation_Erasures}.   Equality (c) follows in similar lines to~\myeqref{eq:13}{c}, except that the justification for $\EBPrelay{i} = 0$ now follows from the fact that, as noted above, we are now focusing on indices $i$ that correspond to codebits that were discovered by BP, and thus $\YBPrelay{i} \neq \Erasure$.  Equality (d) follows in similar lines as~\myeqref{eq:13}{d}.  $\oYBPrelay{i}$ is not independent of $\bE_\StR$, and thus we have not removed the conditioning on $\bE_\StR=\bcErelay$ from the equation.  However,
$\oYBPrelay{i}$ and  $\hEBPrelay{i}$ are independent,  because  by our above definitions, conditioned on $\bE_\StR=\bcErelay$ and recalling~\eqref{eq:hyBP}, $\oYBPrelay{i}$ is a function of $\bX_\SRC$, of  $\hE_{\BPR,j_{i,1}},\ldots,\hE_{\BPR,j_{i,d-1}}$ and of $Y_{\StD,j_{i,1}},\ldots,Y_{\StD,j_{i,d-1}}$, while $\hEBPrelay{i} = \hE_{\BPR,i}$.

Equality (e) parallels~\myeqref{eq:13}{e}.  Specifically, $H( X_{\SRC,i}\given \oYBPrelay{i}= \Erasure,Y_{\StD,i}= \Erasure, \bE_\StR=\bcErelay  ) \leq 1$ and $H( X_{\SRC,i}\given (\oYBPrelay{i}\neq \Erasure\: \textrm{or}\: Y_{\StD,i} \neq \Erasure), \bE_\StR=\bcErelay  ) = 0$.  The latter follows because by construction~\eqref{eq:1000}, whenever $\oYBPrelay{i} \neq \Erasure$ we have  $\oYBPrelay{i} = X_{\SRC,i}$.  We have also evaluated $\Pr[\oYBPrelay{i} = \Erasure] = 1 - (1-\hvarepsilon_\BPR\cdot \varepsilon_\StD)^{d-1}$.  This holds by the fact that  $\oYBPrelay{i}$ is erased if any of $(\hYBPrelay{j_{i,1}}\cdot Y_{\StD,j_{i,1}}),\ldots,(\hYBPrelay{j_{i,d-1}}\cdot Y_{\StD,j_{i,d-1}})$ are erased, and conditioned on $\bE_\StR=\bcErelay$, these variables are independent and  erased with probability $\hvarepsilon_\BPR \cdot \varepsilon_\StD$ (this follows in part by arguments as in our above analysis of $\Pr[\hEBPrelay{i} = 0 \given \bE_\StR=\bcErelay]$).  Finally, the event $\oYBPrelay{i} = \Erasure$ is independent of the event $Y_{\StD,i}= \Erasure$ because conditioned on $\bE_\StR=\bcErelay$, the former event is determined by the values of $\hE_{\BPR,j_{i,1}},\ldots,\hE_{\BPR,j_{i,d-1}}$ and of $E_{\StD,j_{i,1}},\ldots,E_{\StD,j_{i,d-1}}$ while the latter is determined by $E_{\StD,i}$.

We now return to~\eqref{eq:12}.  Relying on the above discussion, we have the string of equations ending with~\eqref{eq:14} on the following page.
\begin{figure*}
\begin{eqnarray}
H( \hbYBPrelay\given \hbEBPrelay, \bY_\StD ) &\leq& \EE_{\bcErelay} \Big[n(1 -\cD_\StR) \cdot \varepsilon_\StD(1 -\hvarepsilon_\BPR) + n(\cD_\StR - \cDRelayBP)\varepsilon_\StD\Big(1 - (1-\hvarepsilon_\BPR\cdot \varepsilon_\StD)^{d-1}\Big)(1 -\hvarepsilon_\BPR)\Big] \nonumber\\
&\addabove{=}{a}& n \Big[(1-\varepsilon_\StR)\cdot \varepsilon_\StD(1 -\hvarepsilon_\BPR) + (\varepsilon_\StR - \deltaRelayBP)\varepsilon_\StD\Big(1 - (1-\hvarepsilon_\BPR\cdot \varepsilon_\StD)^{d-1}\Big)(1 -\hvarepsilon_\BPR) + o(1)\Big] \nonumber\\
&=& n\cdot \varepsilon_\StD(1 -\hvarepsilon_\BPR)\Big[(1-\varepsilon_\StR) +  (\varepsilon_\StR - \deltaRelayBP)\Big(1 - (1-\hvarepsilon_\BPR\cdot \varepsilon_\StD)^{d-1}\Big) + o(1)\Big] \label{eq:14}
\end{eqnarray}
\hrulefill
\end{figure*}
In (a), as $\cD_\StR$ is the erasure rate of the noise vector $\bcErelay$ on the source-relay link, its expected value is clearly $\varepsilon_\StR$.  We have also relied on~\eqref{eq:10} (Appendix~\ref{apdx:Analysis_DRelayBP} above) to express the expected values of $\cDRelayBP$, recalling that it is identically distributed as $\DRelayBP$.

This bound~\eqref{eq:14} concludes our analysis of $H( \hbYBPrelay\given \hbEBPrelay, \bY_\StD )$.

{\hfill$\QED$}
\subsection{Upper Bound on \normalfont$H( \hbEBPrelay )$}\label{apdx:IPlus2}
In this section we exploit dependencies  between erased bits at the output of BP.  An outline of the main idea behind the proof was provided in Sec.~\ref{sec:Quantization_Noise}.  We begin as follows.
\begin{eqnarray}
H( \hbEBPrelay ) &\leq& H( \bEBPrelay, \hbEBPrelay) \nonumber\\
&=& H( \bEBPrelay ) +  H(\hbEBPrelay\given \bEBPrelay),\label{eq:16}
\end{eqnarray}
where $\bEBPrelay$ is as defined in Appendix~\ref{apdx:Proof_of_Lemma_Naive}, i.e., $\bEBPrelay = I_\Erasure(\bYBPrelay)$.  Focusing on the second term on the right-hand side of~\eqref{eq:16} we obtain
\begin{eqnarray}
H(\hbEBPrelay\given \bEBPrelay)
&\leq& \sum_{i=1}^n H(\hEBPrelay{i}\given \EBPrelay{i}) \nonumber\\ &\addabove{=}{a}& \sum_{i=1}^n h(\hvarepsilon_\BPR)(1-\DeltaRelayBPInd{i})\nonumber\\
&\addabove{=}{b}& n\Big[h(\hvarepsilon_\BPR)(1-\deltaRelayBP) + o(1)\Big].\label{eq:18}
\end{eqnarray}
In (a), we have observed that if $\EBPrelay{i} = \Erasure$, then by~\eqref{eq:hEBPrelay_is_sum}, $\hEBPrelay{i} = \Erasure$ with probability 1 and thus $H(\hEBPrelay{i}\given \EBPrelay{i}=\Erasure) = 0$.  If $\EBPrelay{i} = 0$ then $\hEBPrelay{i} = \hE_{\BPR,i}$  and thus $H(\hEBPrelay{i}\given \EBPrelay{i}=0) = h(\hvarepsilon_\BPR)$.  We have also defined $\DeltaRelayBPInd{i}$ to equal $\Pr[\EBPrelay{i}=\Erasure]$ as in Appendix~\ref{apdx:Proof_of_Lemma_Naive}.  Finally, (b) follows in the same lines as in our derivation of~\eqref{eq:15} in Appendix~\ref{apdx:Proof_of_Lemma_Naive}.

We now turn to the first term in~\eqref{eq:16}, see the string of equations ending with~\eqref{eq:40} on the following page.

In (a),  we have again defined $\DRelayBP \defined P(\Erasure\given \bYBPrelay)$ (see~\eqref{eq:erasure_rate}).  In (b), we have relied on the fact that $\DRelayBP$ is confined to the set $\{0, 1/n, 2/n, \ldots,1\}$, which contains $n+1$ elements, thus $H( \DRelayBP ) \leq \log(n+1) = n\cdot o(1)$.

\begin{figure*}
\begin{eqnarray}
H( \bEBPrelay )
&\addabove{\leq}{a}& H( \bEBPrelay, \DRelayBP ) \nonumber\\
&=& H( \bEBPrelay \given \DRelayBP ) + H( \DRelayBP ) \nonumber \\
&\addabove{\leq}{b}& H( \bEBPrelay \given \DRelayBP ) + n\cdot o(1) \nonumber \\
&\addabove{\leq}{c}&\EE \Big[\:\log N_\cC(\DRelayBP n) \:\Big]+ n\cdot o(1) \nonumber \\
&\addabove{\leq}{d}& \EE \Big[\: n\cdot f(\DRelayBP) \:\Big]+ n\cdot o(1) \nonumber \\
&\addabove{=}{e}& \EE\left[\: n\cdot f(\DRelayBP) \:\Big{|}\: |\DRelayBP - \deltaRelayBP| \leq a_1n^{-1/6}\right]\cdot \Pr\left[|\DRelayBP - \deltaRelayBP| \leq a_1n^{-1/6}\right] + \nonumber\\
&&  \EE\left[\: n\cdot f(\DRelayBP) \:\Big{|}\: |\DRelayBP - \deltaRelayBP| > a_1n^{-1/6}\right]\cdot  \Pr\left[|\DRelayBP - \deltaRelayBP| > a_1n^{-1/6}\right] + n\cdot o(1) \nonumber \\
&\addabove{\leq}{f}& \EE\left[\: n\cdot f(\DRelayBP) \:\Big{|}\: |\DRelayBP - \deltaRelayBP| \leq a_1n^{-1/6}\right]\cdot 1 \:+ \: n\cdot a_2n^{1/6}\e^{-\tau\sqrt{n}} \: + \:n\cdot o(1) \nonumber \\
&\addabove{\leq}{g}& n\cdot f(\deltaRelayBP) + n\cdot o(1)
\label{eq:40}
\end{eqnarray}
\hrulefill
\end{figure*}
In (c), we have made the following key observation.  The vector $\bEBPrelay$ specifies the set of bits that remained undecoded (erased) at the output of BP.  Di~\etal~\cite[Lemma~1.1]{Stopping_Sets} proved that these bits correspond to a {\it stopping set} of the code $\cC$ (see~\cite{Stopping_Sets} for its definition).  The set of indices implied by  $\bEBPrelay$ is thus a stopping set of size $\DRelayBP n$.  For $s \in \{0,1,\ldots,n\}$, let $N_\cC(s)$ denote the number of stopping sets of size $s$, of the code $\cC$.   We thus have
\begin{eqnarray*}
H( \bEBPrelay \given \DRelayBP = \alpha  ) \leq \log  N_\cC(\alpha n).
\end{eqnarray*}
In (d), we have applied the following results by Burshtein and Miller~\cite[Theorem 9]{DavidGadiStopping} and Orlitsky~\etal~\cite[Theorem 5]{Orlitsky_Stopping}.  They examined $\EE[ N_\cC(\alpha n) ]$ where the expectation is over all codes $\cC$ in the $(\lambda, d)$ LDPC code ensemble.  They obtained the following bound, for all $\alpha = k/n,\quad  k = 0,\ldots,n$.
\begin{eqnarray}\label{eq:39}
 \frac{1}{n}\log \EE\Big{[}N_\cC(\alpha n)\Big{]} \leq f(\alpha) + o(1),
\end{eqnarray}
where $f(\cdot)$ is given by~\eqref{eq:f} and the term $o(1)$ is independent of $\alpha$.  A few minor remarks are deferred to Remark~\ref{remark:2} (Appendix~\ref{apdx:Remarks} below).  In Remark~\ref{remark:3} (same appendix below), we applied~\eqref{eq:39} to obtain a bound that holds with high probability for an individual code $\cC$ (as in our setting), rather than the expected value over all codes.

In (e), $a_1$ is defined as in Lemma~\ref{lemma_DRelayBP}  (Appendix~\ref{apdx:Analysis_DRelayBP} above).  In (f), we have relied on Remark~\ref{remark:4} (Appendix~\ref{apdx:Remarks}) to argue that $f(\alpha) \leq 1$, and  we have also applied~\eqref{eq:36} (Lemma~\ref{lemma_DRelayBP}).
Finally, (g) follows by the continuity of $f(\cdot)$, which holds by~\cite[Corollary~6]{Orlitsky_Stopping}.

Combining~\eqref{eq:16},~\eqref{eq:18} and~\eqref{eq:40} we obtain our desired bound on  $H( \hbEBPrelay )$.
\begin{eqnarray}
H( \hbEBPrelay ) \leq n\Big[ f(\deltaRelayBP) + (1-\deltaRelayBP)\cdot h(\hvarepsilon_\BPR) + o(1) \Big].\label{eq:78}
\end{eqnarray}
{\hfill$\QED$}
\subsection{Some Remarks Regarding \normalfont$f(\cdot)$} \label{apdx:Remarks}

\begin{remark}\label{remark:2}
Our expression~\eqref{eq:f} for $f(\cdot)$ is a slight variation of~\cite[Theorem~5]{Orlitsky_Stopping} ($\gamma(\cdot)$ in their notation).  In~\cite{Orlitsky_Stopping}, expressions for the minimizers $x$ and $y$ of the various minimizations (denoted $x_0$ and $y_0$) are provided, and the expression for $f(\cdot)$ is provided as a function of them.   The range of the maximization of $\beta$ is also different from the one we used in~\eqref{eq:f}.  An examination of their proof shows that these differences do not affect the final outcome.

We now discuss~\eqref{eq:39} (most importantly, with the $o(1)$ term being independent of $\alpha$).  To justify its validity, we argue that in~\cite[Theorem~5]{Orlitsky_Stopping}, adding the term $(1/n)\cdot \log n$ to the right-hand side of the equation, produces an upper bound on $(1/n)\cdot \log \EE [ N_C(\alpha n) ]$ for all $n$.  To see this, observe that in~\cite[Lemmas~3 and~4]{Orlitsky_Stopping} each limit may be replaced by a supremum over all $n$. This holds by replacing the asymptotic saddle-point analysis in the lemmas' proofs with an upper bound as in~\cite[Eq.~(6)]{DavidGadiStopping}.  In~\cite[Eq.~(11)]{Orlitsky_Stopping}, where these lemmas were applied, we may discard the limit, replace the sum by a supremum, and add a compensation term $(1/n)\cdot \log n$, to obtain a bound on $(1/n)\cdot \log \EE [ N_C(\alpha n) ]$ rather than an evaluation of its limit.  The desired result will then follow as in the proof of~\cite[Theorem~5]{Orlitsky_Stopping}.
\end{remark}
\begin{remark} \label{remark:3}
In~\eqref{eq:39}, the expectation is over all codes $\cC$ in our ensemble.  In our analysis, however, we are interested in the probability that an individual code $\cC$ has $\log N_\cC(\alpha n)$ that greatly exceeds $f(\alpha)$.  As in the proof of Lemma~\ref{lemma_DRelayBP} (Appendix~\ref{apdx:Analysis_DRelayBP}), we apply Markov's inequality to bound this probability.  For fixed $n$ we let $\hf(\alpha; n)$ denote the left-hand side of~\eqref{eq:39}.  We now derive
\begin{eqnarray*}
&&\Pr\left[ N_\cC(\alpha n) > \e^{n(\hf(\alpha; n) + n^{-1/2})} \right] \leq \frac{\EE [N_\cC(\alpha n)]}{\e^{n(\hf(\alpha; n) + n^{-1/2})}}
\\&&\qquad=  \frac{\e^{n\hf(\alpha; n)}}{\e^{n(\hf(\alpha; n) + n^{-1/2})}} \:= \: \e^{-n^{1/2}},
\end{eqnarray*}
where the probability is over the random selection of a code $\cC$ from the $(\lambda, d)$ ensemble.   By a union bound argument we obtain
\begin{eqnarray*}
&&\hspace{-0.6cm}\Pr\left[ \exists \alpha \in \{0,1/n,2/n,\ldots,1\} \: : \:  N_\cC(\alpha n) > \e^{n(\hf(\alpha; n) + n^{-1/2})} \right] \\&&\quad\quad\leq (n+1) \cdot \e^{-n^{1/2}}.
\end{eqnarray*}
By these results, for large enough $n$, with probability at least $1 -  \exp(-\tau\sqrt{n})$ (for a $\tau > 0$, as required by Theorem~\ref{theorem:I_Bounds}'s conditions), a randomly selected code $\cC$ satisfies for all $\alpha \in \{0,1/n,2/n,\ldots,1\}$
\begin{eqnarray*}
\frac{1}{n} \log N_\cC(\alpha n) &\addabove{\leq}{a}& f(\alpha; n) + n^{-1/2}\nonumber\\
&\addabove{\leq}{b}& f(\alpha) + o(1),
\end{eqnarray*}
where (a) follows by our above discussion, and (b) follows by~\eqref{eq:39}, recalling that we have defined $f(\alpha; n)$ to be the left-hand side of that equation.
\end{remark}
\begin{remark}\label{remark:4} To show that $f(\alpha) \leq 1$, we observe that by~\cite[Theorem 9]{DavidGadiStopping} and~\cite[Theorem 5]{Orlitsky_Stopping},
\begin{eqnarray*}
f(\alpha) = \lim_{n\rightarrow\infty}  \frac{1}{n}\log \EE\Big{[}N_\cC(\alpha n)\Big{]}
\addabove{\leq}{a} \lim_{n\rightarrow\infty} \frac{1}{n}\log \EE\Big{[} 2^n\Big{]},
\end{eqnarray*}
where in (a) we relied on the fact that $N_\cC(\alpha n)$, being the number of stopping sets of size $\alpha n$, cannot exceed the total number of subsets of $\{1,2,\ldots,n\}$.
\end{remark}
\subsection{Justification of the l.d.f. Operator}\label{apdx:Justification_ldf}

The operator l.d.f. in~\eqref{eq:I_Plus} is easily justified by the fact that $I( \bYBPrelay; \hbYBPrelay \given \bY_\StD )$ {\it must} be non-ascending as a function of $\hvarepsilon_\BPR$.  To see this, let $\hvarepsilon_\BPR' < \hvarepsilon_\BPR''$ and let $\hbYBPrelayTag$ and $\hbYBPrelayDoubleTag$ denote random vectors obtained as in~\eqref{eq:hyBP}, replacing $\hE_{\BPR,i}$ with components $\hE_{\BPR,i}'$ and $\hE_{\BPR,i}''$ which are distributed as $\eras(\hvarepsilon_\BPR')$ and $\eras(\hvarepsilon_\BPR'')$, respectively.   We now seek to prove that $I( \bYBPrelay; \hbYBPrelayTag \given \bY_\StD ) \ge I( \bYBPrelay; \hbYBPrelayDoubleTag \given \bY_\StD )$.

The random variables $\hbYBPrelayTag$ and $\hbYBPrelayDoubleTag$ can equivalently be modeled as having been obtained from $\bYBPrelay$ via an EEC($\hvarepsilon_\BPR'$) and an EEC($\hvarepsilon_\BPR''$), respectively.  We now argue that EEC($\hvarepsilon_\BPR''$) is stochastically degraded with respect to EEC($\hvarepsilon_\BPR'$).  This follows because an EEC($\hvarepsilon_\BPR''$) can be modeled as a concatenation of an EEC($\hvarepsilon_\BPR'$) and an EEC($\varepsilon$), where $\varepsilon$ satisfies $\hvarepsilon_\BPR' \circ \varepsilon = \hvarepsilon_\BPR''$ (see~\eqref{eq:circ}), i.e., $\varepsilon = (\hvarepsilon_\BPR''- \hvarepsilon_\BPR')/(1 - \hvarepsilon_\BPR')$.  Thus, $\hbYBPrelayDoubleTag$ can be modeled as having been obtained from $\hbYBPrelayTag$ via a memoryless EEC($\varepsilon$), and the desired result follows by the data processing inequality.
{\hfill$\QED$}
\section{Overview of Soft-IC-BP (Sec.~\ref{sec:Soft-IC-BP})}\label{apdx:Overview_Soft_IC_BP}
We now provide a general overview of soft-IC-BP and its analysis methods.  A complete discussion is available in the literature on iterative-MUD, e.g.,~\cite{boutros-caire} and~\cite{Declercq}.  Our description is intended to point out specific features of our implementation, which we have assumed in our discussion of numerical results in Sec.~\ref{sec:Numerical_Interference}.

Soft-IC-BP progresses through the exchange of messages between the nodes of a {\it factor graph}~\cite{FactorGraph} (see Fig.~\ref{fig:FactorGraph}), which represents the communication setting at the destination.  This graph contains the Tanner graphs of the LDPC codes (see Sec.~\ref{sec:LDPC}) at the desired and interfering sources, as well as $n$ additional factor nodes, which we call {\it channel nodes}.  Each channel node corresponds to the received signal at one time instance.  It is linked to one variable node from each Tanner graph, each corresponding to a transmitted bit from one of the two sources at the given time instance.  Decoding includes standard LDPC decoding iterations (see e.g.,~\cite{Urbanke_Message_Passing}) as well as variable-to-channel and channel-to-variable iterations, which implement an exchange of information between the two Tanner graphs.  For precise details regarding the computation of the messages see~\cite[Sec.~2]{Declercq}.\footnote{In~\cite[Fig.~1]{Declercq}, the authors have included three additional nodes for each channel observation, one of them called a ``state-check'' node.  In our work, we have eliminated them and retained only the factor node (our {\it channel node}) connecting the two variable nodes.  The messages to and from this node remain unchanged, as in~\cite[Eq.~(3)]{Declercq}.}

\myfigure{file=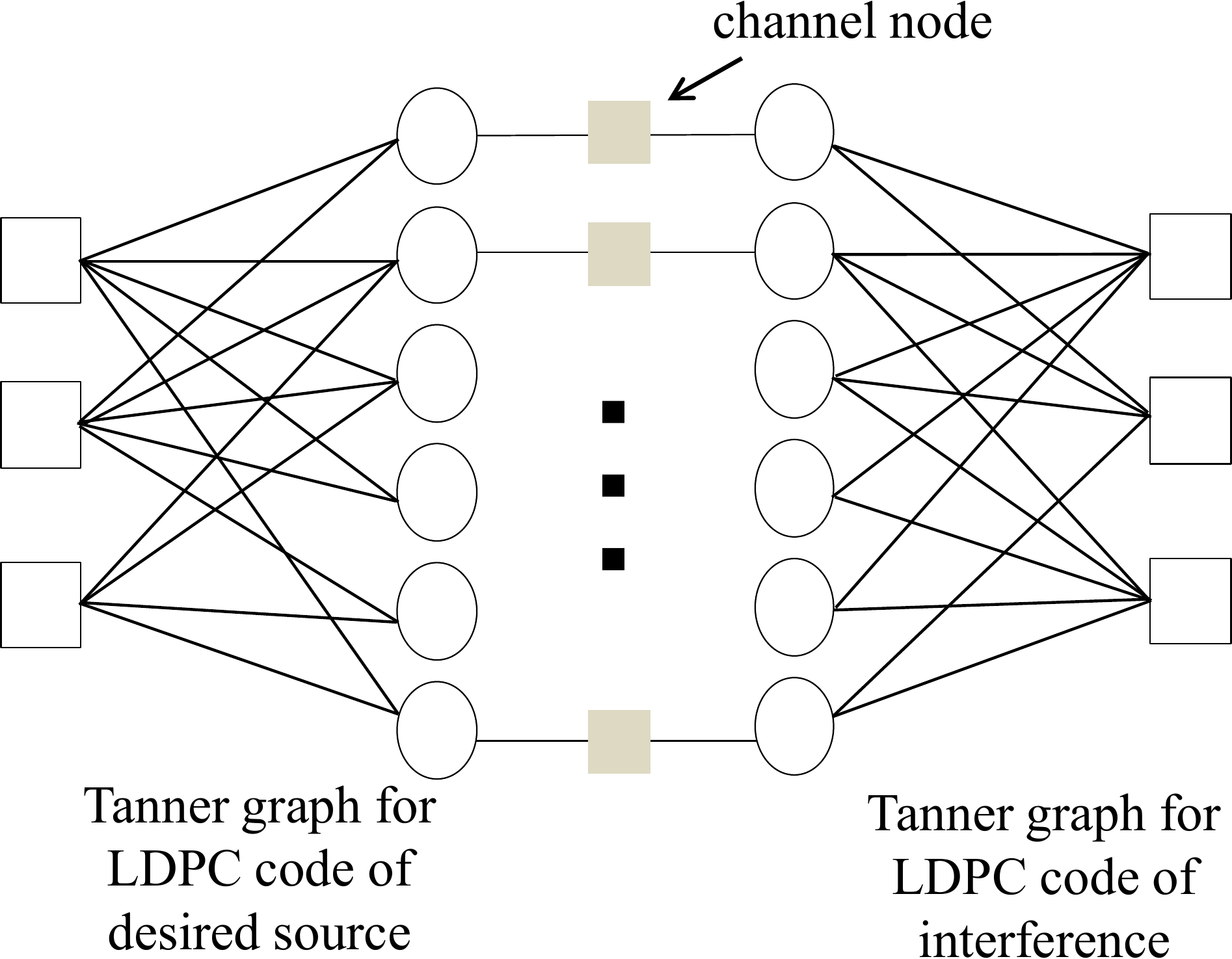, width = 7.5cm}{An example of the factor graph for an application of soft-IC.}{fig:FactorGraph}
In this paper, we have adopted a number of attributes of the design of~\cite{Declercq}.  Namely, we have assumed that the LDPC codes used by the two sources have the same block lengths and degree distributions $(\lambda,\rho)$.  Under this assumption, the number of nodes of any given degree within the Tanner graphs of the codes is the same.  We further assumed that the nodes are arranged so that the two variable nodes that are linked to each channel node have the same degree.  In~\cite[Sec.~4]{Declercq} this is known as the {\it no-interleaver hypothesis}. Lastly, we assumed  {\it parallel scheduling}.  This means that decoding iterations at both Tanner graphs are computed in parallel.

A detailed discussion of density evolution for iterative-MUD (equivalently, soft-IC-BP), is available in~\cite[Sec.~IV.A]{boutros-caire} and~\cite[Sec.~5.5]{Urbanke_Book}.   Like sim-DE (Sec.~\ref{sec:Simultaneous_DE}), density evolution for soft-IC-BP tracks the distributions of messages exchanged at the various iterations of the algorithm.  In addition to variable-to-check and check-to-variable LDPC code BP iterations, the algorithm tracks the distributions of variable-to-channel and channel-to-variable messages.  A distinction is made between the messages exchanged in the Tanner graph of the desired codeword, and the messages in the graph of the interference, whose distributions are expected to be different.  Unlike sim-BP over the BEC, the messages of soft-IC-BP are taken from a large, continuous alphabet (the real number field), and so their distributions as tracked by density evolution are defined over this alphabet (more precisely, a fine grid over the real-number field as in~\cite{Urbanke_Message_Passing}).

Like~\cite{Urbanke_Book}, our computation of the evolution of distributions through variable-to-channel and channel-to-variable iterations is performed precisely, rather than by Monte-Carlo simulations as in~\cite{boutros-caire}.  Unlike~\cite[Sec.~IV.A]{boutros-caire} and~\cite[Example~5.34]{Urbanke_Book}, our reliance on the above-mentioned no-interleaver hypothesis implies that the degrees of the variable node linked to each channel node are {\it not} independent (in fact they are equal).  In our implementation of density evolution, we account for this by considering the variable-to-channel, channel-to-variable and following variable-to-check iteration as a combined single iteration.

A {\it concentration theorem} exists~\cite[Proposition~1]{boutros-caire}, which asserts that the realized bit error rates of the desired and interference codewords approach density evolution's prediction in probability, exponentially in the block length $n$, as $n\rightarrow \infty$.


\section{Proof of Theorem~\ref{theorem:Good_Interference}}\label{apdx:Proof_Theorem_Good_Interference}

In our analysis, we focus on the first source-destination pair.  An outline of the proof was provided in Sec.~\ref{sec:Good-Interference}.  We let $X^*_1$ and $X^*_2$ denote scalar random variables that are distributed as in the discussion following~\eqref{eq:R_MUD} and~\eqref{eq:R_SUD}.  That is, both are uniformly distributed in $\{\pm1\}$.  We also let $Y^*_1$ be a random variable that is related to them via the channel transition equation~\eqref{eq:Symm_BIAWGN_Interference}.

We distinguish between two cases, $R < I(X_2^*; Y^*_1\given X_1^*)$ and $R \geq I(X_2^*; Y^*_1\given X_1^*)$.  In the first case, which is discussed in Appendix~\ref{apdx:R_lt_I} below, we will prove that $R \leq R_\MUD$.  In the second case, which is discussed in Appendix~\ref{apdx:R_geq_I}, we will prove $R \leq R_\SUD$.  The desired~\eqref{eq:R_Good_Interference} thus follows.
\subsection{Analysis in the Range $R < I(X_2^*; Y^*_1\given X_1^*)$}\label{apdx:R_lt_I}
Our proof begins in lines similar to the proof of the converse of the capacity of the multiple-access channel,~\cite[Sec.~14.3.4]{Cover_Book}.
\begin{eqnarray}
n\cdot 2R &\addabove{=}{a}& n\Big(R_{1,n} + R_{2,n} + o(1)\Big) \nonumber\\
&\addabove{=}{b}& H(W_1,W_2) + n\cdot o(1)\nonumber\\
&=& I(W_1,W_2; \bY_1) + H(W_1,W_2\given \bY_1) + n\cdot o(1) \nonumber\\
&=& I(W_1,W_2; \bY_1) + H(W_1\given \bY_1) + \nonumber\\
&& \quad\quad + H(W_2\given W_1, \bY_1) + n\cdot o(1)\nonumber\\
&\addabove{=}{c}& I(W_1,W_2; \bY_1) + n\cdot o(1) + n\cdot o(1)  + n\cdot o(1)\nonumber\\
&\addabove{\leq}{d}& \sum_{i=1}^n I( X_{1i}, X_{2i}; Y_{1i} )  + n\cdot o(1)\nonumber\\
&\addabove{\leq}{e}& nI( X^*_1, X^*_2; Y^*_1 ) +  n\cdot o(1).\label{eq:79}
\end{eqnarray}
In (a), $R_{1,n}$ and $R_{2,n}$ are the rates of the codes $\cC_{1,n}$ and $\cC_{2,n}$, respectively, and the equality holds by the definition of $R$ as being the rate of the code sequences $\{\cC_{1,n}\}_{n=1}^\infty$ and $\{\cC_{2,n}\}_{n=1}^\infty$.  In (b), $W_1$ and $W_2$ are the messages that were transmitted by Sources 1 and~2, respectively, defined as in~\cite[Sec.~14.3.4]{Cover_Book}.  The equality holds because the two messages are statistically independent, and uniformly distributed in $\{1,\ldots,2^{nR_{1,n}}\}$ and~$\{1,\ldots,2^{nR_{2,n}}\}$, respectively.  In (c), $H(W_1\given \bY_1) = n\cdot o(1)$ holds by Fano's inequality~\cite[Theorem~2.11.1]{Cover_Book}, relying on the fact that the probability of error in the decoding of $W_1$, by the conditions of Theorem~\ref{theorem:Good_Interference}, approaches zero as $n\rightarrow\infty$.  The justification for $H(W_2\given W_1,\bY_1) = n\cdot o(1)$ will be provided shortly.
(d) follows by the same arguments as in~\cite[Eq.~(14.116)]{Cover_Book}.  Finally, (e) follows by our discussion in Appendix~\ref{apdx:Analysis_I_X1_X2_Y} below.

By~\eqref{eq:79}, recalling that we are now focusing our attention to the range $R < I(X_2^*; Y^*_1\given X_1^*)$, we obtain by~\eqref{eq:R_MUD} (recalling our above definitions of $X^*_1,X^*_2$ and~$Y^*_1$), $R \leq R_\MUD$.

We now prove $H(W_2\given W_1,\bY_1) = n\cdot o(1)$ in the above equation (c).  Consider the scenario facing a decoder of $W_2$ (at Destination~1), recalling the channel equation~\eqref{eq:Symm_BIAWGN_Interference}.   Note that the destination is not required to decode $W_2$ and thus our discussion of this decoder is for analysis purposes only. As noted in Sec.~\ref{sec:Good-Interference}, given $W_1$, the decoder is able to eliminate $X_1$, and is thus faced with a point-to-point BIAWGN communication scenario, with $\SNR = h^2/\sigma^2$.  The capacity of this channel is clearly $C(\SNR) = I(X_2^*; Y^*_1\given X_1^*)$.  By the fact that $R <  I(X_2^*; Y^*_1\given X_1^*)$ (we are currently focusing on such $R$), we have that $\SNR > \SNRstar$, where $\SNRstar$ is the Shannon limit for rate $R$.  By the P2P-optimality of $\{\cC_{2,n}\}_{n=1}^\infty$, recalling Definition~\ref{def:Good codes_AWGN}, we obtain that the probability of error, under maximum-likelihood decoding, of $W_2$ given $\bY_1$ and $W_1$, must approach zero as $n \rightarrow \infty$.  Thus, by Fano's inequality (as in our analysis of $H(W_1\given \bY_1)$), we obtain $H(W_2\given W_1,\bY_1) = n\cdot o(1)$ as desired.
\subsection{Analysis in the Range $R \geq I(X_2^*; Y^*_1\given X_1^*)$}\label{apdx:R_geq_I}
Our analysis begins as in Appendix~\ref{apdx:R_lt_I}.
\begin{eqnarray}
&&\hspace{-0.5cm}n\cdot 2R \addabove{=}{a} I(W_1,W_2; \bY_1) + H(W_1\given \bY_1) + \nonumber\\
&&\quad\quad\quad + H(W_2\given W_1, \bY_1) + n\cdot o(1)\nonumber\\
&&\addabove{\leq}{b} I(W_1,W_2; \bY_1) + n\cdot o(1) + \nonumber\\
&&\quad\quad\quad + n\Big( R - I(X^*_2; Y^*_1\given X^*_1) + o(1)\Big) + n\cdot o(1)\nonumber\\
&&\addabove{\leq}{c} \sum_{i=1}^n I( X_{1i}, X_{2i}; Y_{1i} )  + nR - n I(X^*_2; Y^*_1\given X^*_1) + \nonumber\\
&&\quad\quad\quad +n\cdot o(1)\nonumber\\
&&\addabove{\leq}{d} nI( X^*_1, X^*_2; Y^*_1 ) + nR - n I(X^*_2; Y^*_1\given X^*_1) + n\cdot o(1)\nonumber\\
&&\addabove{=}{e} nI(X^*_1; Y^*_1) + nR + n\cdot o(1)\nonumber\\
&&\addabove{=}{f} nR_\SUD + nR + n\cdot o(1).\label{eq:80}
\end{eqnarray}
(a) follows as in Appendix~\ref{apdx:R_lt_I}.   In (b), $H(W_1\given \bY_1) = n\cdot o(1)$ follows again as in Appendix~\ref{apdx:R_lt_I}, and $H(W_2\given W_1,\bY_1) \leq n( R - I(X^*_2; Y^*_1\given X^*_1) + o(1))$ will be justified shortly.  (c) and (d) follow as in Appendix~\ref{apdx:R_lt_I}.  (e) follows by the chain rule for mutual information~\cite[Theorem~2.5.2]{Cover_Book}.  Finally, (f) follows by~\eqref{eq:R_SUD}, recalling our above definitions of $X^*_1$ and $Y_1^*$.

Subtracting $nR$ from both sides of the above inequality, dividing by $n$ and taking $n$ to infinity, we obtain $R \leq R_\SUD$ as desired.

We now prove $H(W_2\given W_1,\bY_1) \leq n( R - I(X^*_2; Y^*_1\given X^*_1) + o(1))$, as required in inequality (b) above.
\begin{eqnarray}
&&\hspace{-1cm}H( W_2 \given W_1, \bY_1) \nonumber\\
&&\quad\addabove{\leq}{a} H( \bX_2 \given W_1, \bY_1) + n\cdot o(1)\nonumber\\
&&\quad\:=\: H(\bX_2\given W_1) - I( \bX_2; \bY_1\given W_1) + n\cdot o(1)\nonumber\\
&&\quad\addabove{\leq}{b} n\Big(R + o(1)\Big) - I( \bX_2; \bY_1\given W_1) + n\cdot o(1)\nonumber\\
&&\quad\addabove{=}{c} nR  - I( \bX_2; \btY_2) + n\cdot o(1)\nonumber\\
&&\quad\addabove{\leq}{d} nR - n\Big( C(\SNR) + o(1) \Big) + n\cdot o(1)\nonumber\\
&&\quad\addabove{=}{e} nR - n I( X^*_2; Y^*_1 \given X^*_1)  + n\cdot o(1).\label{eq:81}
\end{eqnarray}
(a) will be proven in Appendix~\ref{apdx:Analysis_H_W2}, below.  (b) follows by the fact that the cardinality of the range of the random vector $\bX_2$ cannot exceed that of $W_2$, which is $2^{nR_{2,n}}$, and $R_{2,n} = R + o(1)$.  In (c), we have defined $\btY_2 \defined \bY_1 - \bX_1(W_1)$, where $\bX_1(W_1)$ is the codeword corresponding to $W_1$.  As noted in Sec.~\ref{sec:Good-Interference} and Appendix~\ref{apdx:R_lt_I}, by~\eqref{eq:Symm_BIAWGN_Interference}, the channel from $\bX_2$ to $\btY_2$ is a BIAWGN channel with $\SNR = h^2/\sigma^2$.  The capacity of this channel is clearly $C(\SNR) = I(X_2^*; Y^*_1\given X_1^*)$.  As we have confined our attention to $R \geq I(X_2^*; Y^*_1\given X_1^*)$, we have $\SNR \leq \SNRstar$ where $\SNRstar$ is again the Shannon limit for rate $R$.  The justification for (d) will be provided shortly. Finally, in (e) we have simply rewritten $C(\SNR) = I(X_2^*; Y^*_1\given X_1^*)$.

To justify (d), we argue that $I( \bX_2; \btY_2) \geq n(C(\SNR) + o(1))$.  Had the $\SNR$ satisfied $\SNR > \SNRstar$, this would have held trivially by the P2P-optimality of code sequence $\{\cC_{2,n}\}_{n=1}^\infty$ and Fano's inequality.  However, as mentioned above, we are now interested in the range $\SNR \leq \SNRstar$.  Our justification in this range of $\SNR$ is based on the following discussion.  We define, for any code $\cC$, $I(\cC; \SNR) \defined I(\bX;\bY)$ where $\bX$ is uniformly distributed within $\cC$ and $\bY$ is related to it via the transitions of a BIAWGN channel, as in~\eqref{eq:BIAWGN}.  With this definition, by our above discussion $I(\bX_2; \btY_2) = I(\cC_{2,n},\SNR)$.  We let $C(\SNRstar)$ denote the capacity of a BIAWGN channel with an SNR of $\SNRstar$.  We now have
\begin{eqnarray}
&&\hspace{-0.8cm}nC(\SNRstar) \nonumber\\
&&\hspace{-0.4cm}\addabove{=}{a} I(\cC_{2,n}; \SNRstar) + n\cdot o(1)\nonumber\\
&&\hspace{-0.4cm} \addabove{=}{b} I(\cC_{2,n}; \SNR) + \int_{\SNR}^{\SNRstar} \frac{1}{2}\mmse(\cC_{2,n}; \snr)d\snr + n\cdot o(1)\nonumber\\\
&&\hspace{-0.4cm} \addabove{\leq}{c} I(\cC_{2,n}; \SNR) + \int_{\SNR}^{\SNRstar} \frac{1}{2}n\cdot\mmse(\textrm{bitwise}; \snr)d\snr +\nonumber\\
&&\quad + n\cdot o(1)\nonumber\\
&&\hspace{-0.4cm} \addabove{=}{d} I(\cC_{2,n}; \SNR) + n\cdot\Big( C(\SNRstar) - C(\SNR) \Big) + n\cdot o(1).\nonumber\\
\label{eq:82}
\end{eqnarray}
(a) follows by similar arguments to~\myeqref{eq:86}{b}, relying on Fano's inequality and the P2P-optimality of $\{\cC_{2,n}\}_{n=1}^\infty$.  In (b), the $\mmse$ function is defined as in~\eqref{eq:mmse} (Appendix~\ref{apdx:Rigorous_Fig_Bad_Codes}).   The equality follows from the relation between mutual information and the MMSE, see~\cite[Eq.~(1)]{GuoMutInfMMSE}.  In (c), $\mmse(\textrm{bitwise}; \snr)$ denotes the MMSE in the estimation of a symbol $X$ which is uniformly distributed in $\{\pm1\}$, from $Y$, which is related to $X$ via a BIAWGN channel with noise variance $1/\snr$.  In such an estimation, the decoder does not have the benefit of the code structure to draw upon, and so the estimation error clearly increases in comparison to the estimation of a given bit in $\cC_{2,n}$.  In (d), we have relied on the fact that the derivative of the function $C(\SNR)$ with respect to $\SNR$ is $1/2 \cdot \mmse(\textrm{bitwise}; \snr)$.  This follows from the discussion of~\cite[Sec.~II.A]{GuoMutInfMMSE}.\footnote{Specifically, \cite[Eq.~(17)]{GuoMutInfMMSE}
corresponds to $\mmse(\textrm{bitwise}; \snr)$ and \cite[Eq.~(18)]{GuoMutInfMMSE} corresponds to $C(\SNR)$.}  Finally, recalling $I(\bX_2; \btY_2) = I(\cC_{2,n},\SNR)$, we have our desired result.
{\hfill$\QED$}

\subsection{Analysis of $I( X_{1i}, X_{2i}; Y_{1i} )$}\label{apdx:Analysis_I_X1_X2_Y}
We now prove the inequalities~\myeqref{eq:79}{e} and~\myeqref{eq:80}{d}.  Our proof relies on the properties of P2P-optimal codes for the BIAWGN channel.  Specifically, we show that the marginal distributions of the individual code symbols $X_{1i}$ and $X_{2i}$, $i=1,\ldots,n$, cannot stray too far from the uniform distribution over $\{\pm1\}$, which is the capacity-achieving input distribution over the BIAWGN channel~\cite[Theorem~4.5.1]{Gallager_book}.  Our proof is a variation of a similar result by~\cite[Theorem~4]{ShlomoVerduEmpiricalGood}.
\begin{eqnarray}
\sum_{i=1}^n I( X_{1i}, X_{2i}; Y_{1i} ) &\addabove{=}{a}& \sum_{i=1}^n i_2\Big( p_{1i} \times p_{2i}  \Big)\nonumber\\
&\addabove{\leq}{b}& n\cdot  i_2\Big(\frac{1}{n} \sum_{i=1}^n(p_{1i} \times p_{2i})  \Big)\nonumber\\
 &\addabove{=}{c}& n\cdot i_2\Big( p^* \times p^*  + o(1) \Big) \nonumber\\
 &\addabove{=}{d}& n\cdot \left[i_2\Big( p^* \times p^* \Big) + o(1)\right] \nonumber\\
 &\addabove{=}{e}& n\cdot I( X^*_1, X^*_2; Y^*_1 )  + n\cdot o(1). \nonumber\\ \label{eq:85}
\end{eqnarray}
In (a), we have made the following definitions.  $i_2(\cdot)$ is a function whose argument is a probability function $p(x_1,x_2)$ where $(x_1,x_2) \in \{\pm1\}^2$. Its value is $I( X_1, X_2; Y_1 )$ where $(X_1,X_2)$ are distributed as $p(x_1,x_2)$ and $Y_1$ is related to them via the transitions of the interference channel,~\eqref{eq:Symm_BIAWGN_Interference}.  $p_{1i}$ is a probability function over $x_1 \in \{\pm1\}$, corresponding to the distribution of $X_{1i}$.  $p_{2i}$ is similarly defined, corresponding to $X_{2i}$.  $p_{1i}\times p_{2i}$ is defined by
\begin{eqnarray*}
p = p_{1i}\times p_{2i}\quad \Rightarrow \quad p(x_1,x_2) = p_{1i}(x_1)\cdot p_{2i}(x_2)\\ \forall (x_1,x_2) \in \{\pm1\}^2.
\end{eqnarray*}
The independence between $X_{1i}$ and $X_{2i}$, implied by equality (a), follows from the independence between the messages $W_1$ and $W_2$, as in~\cite[Eq.~(14.122)]{Cover_Book}.

Inequality (b) follows by Jensen's inequality and the concavity of the mutual information as a function of the marginals of its distributions,~\cite[Theorem~2.7.4]{Cover_Book}.   In (c), we have defined $p^*$ to be the distribution of $X^*_1$ (and of $X_2^*$).  The justification for this equality will be provided shortly.  (d) follows by the continuity of $i_2(\cdot)$, and (e) follows by the definition of $i_2(\cdot)$.

To prove equality (c) above, we consider communication over a point-to-point BIAWGN channel with an SNR equal to the Shannon limit for rate $R$, $\SNRstar$.  We let $i_1\left( p \right)$ denote $I(X;\hY)$, where $X$ takes the value $1$ with probability $p$ and $-1$ with probability $1-p$.  $\hY$ is related to $X$ via the transitions of the above-mentioned BIAWGN channel, see~\eqref{eq:BIAWGN}.
\begin{eqnarray*}
n\cdot i_1\left( \frac{1}{2} \right) &\addabove{\leq}{a}& I( \bX_1; \hbY_1 ) + n\cdot o(1)\\
&\addabove{\leq}{b}& \sum_{i=1}^n I( X_{1i}; \hY_{1i} ) + n\cdot o(1)\\
&\addabove{=}{c}& \sum_{i=1}^n i_1( \pi_{1i} ) + n\cdot o(1)\\
&\addabove{=}{d}& \sum_{i=1}^n i_1( \hpi_{1i} ) + n\cdot o(1)\\
&\addabove{\leq}{e}& n \cdot i_1\left( \frac{1}{n}\sum_{i=1}^n \hpi_{1i} \right) + n\cdot o(1)\\
&\addabove{\leq}{f}& n \cdot i_1\left( \frac{1}{2} \right) + n\cdot o(1).
\end{eqnarray*}
In (a), $\hbY_1$ corresponds to the output of the above-mentioned BIAWGN channel, when provided with $\bX_1$ as its input.  We have relied on the fact that as the capacity-achieving input distribution for the BIAWGN channel corresponds to $p = 1/2$, its capacity is $i_1(1/2)$.  The inequality now follows by the same arguments as the ones used to justify~\myeqref{eq:82}{a}, relying on the P2P-optimality of $\{\cC_{1,n}\}_{n=1}^\infty$.  (b) follows as in~\cite[Eq.~(8.104)]{Cover_Book}, relying on the memorylessness of the BIAWGN channel.  In (c), we have defined $\pi_{1i}$ to be the probability that $X_{1i}$ is equal to 1.  In (d), we have defined $\hpi_{1i} = \min(\pi_{1i}, 1-\pi_{1i} )$ and the equality follows by the obvious symmetry of $i_1(\cdot)$.  In (e), we have again applied Jensen's inequality and~\cite[Theorem~2.7.4]{Cover_Book}.  In (f), we have relied on the fact that the maximum of $i_1(\cdot)$ is achieved at $1/2$, as this corresponds to the capacity-achieving input distribution of the BIAWGN channel.

We now have
\begin{eqnarray*}
\lim_{n \rightarrow \infty } i_1\left(\frac{1}{n}\sum_{i=1}^n \hpi_{1i}\right) = i_1\left(\frac{1}{2}\right).
\end{eqnarray*}
The function $i_1(\cdot)$ achieves its maximum uniquely at $p = 1/2$.  We thus obtain,
\begin{eqnarray}\label{eq:84}
\lim_{n \rightarrow \infty }\frac{1}{n}\sum_{i=1}^n \hpi_{1i} = \frac{1}{2}
\end{eqnarray}
We now define
\begin{eqnarray}\label{eq:83}
f_1(n) \defined \frac{1}{2} - \frac{1}{n}\sum_{i=1}^n \hpi_{1i}\:,
\end{eqnarray}
and,
\begin{eqnarray*}
I_1(n) = \left\{ i\ :\ \hpi_{1i} \geq  \frac{1}{2} - \sqrt{f_1(n)} \right\}.
\end{eqnarray*}
By a simple argument, relying on~\eqref{eq:83} and the fact that $\hpi_{1i} \leq 1/2$ for all $i$, we have
\begin{eqnarray*}
\left|I_1(n)^\mathrm{c}\right| \leq \sqrt{f_1(n)}\cdot n,
\end{eqnarray*}
where $I_1(n)^\mathrm{c}$ denotes the complement set of $I_1(n)$.  Note that $I_1(n)$ satisfies
\begin{eqnarray*}
&&i \in I_1(n) \Rightarrow \\
&&\quad \frac{1}{2} - \sqrt{f_1(n)}\leq p_{1i}(x) \leq  \frac{1}{2} + \sqrt{f_1(n)} \qquad \forall x \in \{\pm 1\},
\end{eqnarray*}
where $p_{1i}(\cdot)$ is as defined above.  We similarly define $f_2(n)$ and $I_2(n)$.  Eq.~\myeqref{eq:85}{c} now follows by the observation that $f_1(n)$ and $f_2(n)$ approach zero as $n\rightarrow\infty$, and by the above definition of $p^*$.
{\hfill$\QED$}
\subsection{Analysis of $H( W_2 \given W_1, \bY_1)$}\label{apdx:Analysis_H_W2}

We now justify inequality~\myeqref{eq:81}{a}.  Namely, we have
\begin{eqnarray*}
&&\hspace{-1cm}H( W_2 \given W_1, \bY_1) \leq H( W_2, \bX_2 \given W_1, \bY_1)\\
&&\quad \:=\: H( \bX_2 \given W_1, \bY_1)+ H( W_2 \given \bX_2, W_1, \bY_1)\\
&&\quad \addabove{=}{a} H( \bX_2 \given W_1, \bY_1)+ H( W_2 \given \bX_2)\\
&&\quad \addabove{\leq}{b} H( \bX_2 \given W_1, \bY_1)+ H( W_2 \given \hbY)\\
&&\quad \addabove{=}{c} H( \bX_2 \given W_1, \bY_1) + n\cdot o(1).
\end{eqnarray*}
(a) follows by the Markov chain relation $W_2 \leftrightarrow \bX_2 \leftrightarrow (W_1, \bY_1)$.  In (b), we have defined $\hbY$ to be the output of a BIAWGN channel~\eqref{eq:BIAWGN}, whose  SNR is greater  than $\SNRstar$, which is provided with the input $\bX_2$.  The inequality follows by the data processing inequality, using the Markov chain relation $W_2 \leftrightarrow \bX_2 \leftrightarrow \hbY$.   By the P2P-optimality of $\{\cC_{2,n}\}_{n=1}^\infty$, recalling Definition~\ref{def:Good codes_AWGN}, the probability of error, when decoding $W_2$ from $\hbY$, must approach zero as $n\rightarrow\infty$. Equality (c) now follows, using Fano's inequality.\footnote{Note that if the mapping from $W_2$ to $\bX_2$ is injective, then $H( W_2 \given \bX_2) = 0 = n\cdot o(1)$ holds trivially.  In a random construction of a P2P-optimal code, however, this is not guaranteed, requiring steps (b) and (c) in the above derivation.}.
{\hfill$\QED$}
\section{Extension of Theorem~\ref{theorem:Good_Interference} to Arbitrary Rates over Arbitrary BIAWGN Interference Channels}\label{apdx:Arbitrary}

We now extend the discussion of Sec.~\ref{sec:Good-Interference}, and specifically Theorem~\ref{theorem:Good_Interference}, from symmetric rates over symmetric BIAWGN interference channels, to a general setting.  We continue to use the same Definition~\ref{def:Good codes_AWGN} of P2P-optimality as in Sec.~\ref{sec:Good-Interference}.  We begin by focusing on Destination~1.  Using randomly-generated P2P-optimal codes, the following inequalities ensure that reliable decoding is possible at Destination~1 using MUD~\cite[Theorem~14.3.3]{Cover_Book}.
\begin{eqnarray}
R_1 + R_2 &<& I(X_1, X_2;\: Y_1), \nonumber \\
R_1 &<& I(X_1; Y_1 \given X_2), \nonumber \\
R_2 &<& I(X_2; Y_1 \given X_1), \label{eq:MUD_Arbitrary}
\end{eqnarray}
where $R_1$ and $R_2$ are the rates of sources 1 and 2, and $X_1, X_2, Y_1$ are distributed as in Sec.~\ref{sec:Good-Interference}.  A condition for reliable decoding using SUD is obtained as in~\eqref{eq:R_SUD}.
\begin{eqnarray}
R_1 &<& I(X_1 ; Y_1) \label{eq:SUD_Arbitrary}.
\end{eqnarray}
The following theorem now extends Theorem~\ref{theorem:Good_Interference}.

\begin{theorem}\label{theorem:Good_Interference2}
Consider communication over a ($h_1,h_2,\sigma_1,\sigma_2$) BIAWGN interference channel.  Assume the two sources use equal block length codes taken from P2P-optimal code sequences $\{\cC_{1,n}\}_{n=1}^\infty$ and $\{\cC_{2,n}\}_{n=1}^\infty$, respectively, which have rates $R_1$ and $R_2$, respectively.  Assume the probabilities of decoding error, under maximum-likelihood decoding, at both destinations, approach zero as the block length $n\rightarrow\infty$.  Then the following conditions must hold.
\begin{enumerate}
\item {\it Destination~1} \label{enum:Good_Interference2:1} Either inequalities~\eqref{eq:MUD_Arbitrary} or else~\eqref{eq:SUD_Arbitrary} must hold.
\item {\it Destination~2}: The same as Condition~\ref{enum:Good_Interference2:1}, exchanging $X_1,Y_1,R_1$ and $X_2,Y_2,R_2$.
\end{enumerate}
\end{theorem}
{\it Proof:}
The proof follows in the lines of the proof of Theorem~\ref{theorem:Good_Interference} (Appendix~\ref{apdx:R_lt_I}).  We again, without loss of generality, focus on Destination~1.  We distinguish between two ranges, $R_2 < I(X_2^*; Y^*_1\given X_1^*)$ and $R_2 \geq I(X_2^*; Y^*_1\given X_1^*)$.  Once again, we let $X_1^*,X_2^*, Y^*_1$ denote random variables distributed as $X_1,X_2, Y_1$ in~\eqref{eq:MUD_Arbitrary} or else~\eqref{eq:SUD_Arbitrary}, to avoid confusion with $X_{1i},X_{2i}, Y_{1i}$, which we let denote the components of the transmitted and received signals with the above codes $\cC_{1,n}$ and $\cC_{2,n}$.

In the first range of $R_2$, we seek to prove that inequalities~\eqref{eq:MUD_Arbitrary} hold. $R_2 < I(X^*_2; Y^*_1 \given X^*_1)$ holds by the definition of this range.  The proof of $R_1 + R_2 <I(X^{*}_1, X^{*}_2;\: Y^{*}_1)$ follows by the similar arguments to~\eqref{eq:79}.  Finally, we prove

\begin{eqnarray*}
nR_1 &\addabove{\le}{a}& \sum_{i=1}^n I( X_{1i}; Y_{1i} \given X_{2i})  + n\cdot o(1) \\
&\addabove{=}{b}& n\cdot I(X^*_1; Y^*_1 \given X^*_2) + n\cdot o(1).
\end{eqnarray*}
(a) follows by the same lines as in the converse of the capacity of the multiple-access channel~\cite[Eq.~(14.105)]{Cover_Book}.  (b) follows by arguments similar to those of Appendix~\ref{apdx:Analysis_I_X1_X2_Y} above.

In the second range, ~\eqref{eq:SUD_Arbitrary} holds by the similar arguments to~\eqref{eq:80}.
{\hfill$\QED$}
\section{Results for Sec.~\ref{sec:Numerical}}
\subsection{Proof of the Achievable Rates for DF and CF}\label{apdx:proof_DF_CF}

By~\cite[Proposition~2]{Primitive_Relay} (which specializes~\cite{CoverElGamal}) any rate $R$ satisfying the following inequality is achievable with DF.
\begin{eqnarray*}
R \leq \min\Big{(}I(X_\SRC; Y_\StR),\: I(X_\SRC; Y_\StD ) + C_\RtD\Big{)},
\end{eqnarray*}
where the distribution $P_{X_\SRC}(x_\SRC)$ of $X_\SRC$ is a parameter that can be optimized, and the distributions of the rest of the variables are derived from the channel transitions (see Sec.~\ref{sec:BEC_relay_model}).  The maximum achievable rate $R_\DF$ can easily be shown to equal~\eqref{eq:R_DF} and is obtained by a uniform distribution in $\{0,1\}$.

Turning to CF, the following achievable rates are proved by~\cite[Proposition~3]{Primitive_Relay} (which again specializes~\cite{CoverElGamal}), which was provided in Proposition~\ref{proposition:CF} (Appendix~\ref{apdx:Proof_of_Theorem_Formal}).
\begin{eqnarray} \label{eq:R_CF_BEC_Raw}
R_{\CF} = \max
\{I(X_\SRC; \hY_\StR, Y_\StD)\: :\: I(Y_\StR; \hY_\StR\given Y_\StD ) \le C_\RtD\},
\end{eqnarray}
where we have substituted  $X_\SRC, \hY_\StR, Y_\StD, Y_\StR, C_\RtD$ for  $U_\SRC, \hV_\StR, V_\StD, V_\StR,  C_\textrm{o}$ of Proposition~\ref{proposition:CF}.
As in that proposition, the distributions $P_{X_\SRC}(x_\SRC)$ and $P_{\hY_\StR\given Y_\StR}(\hy_\StD\given y_\StR)$ are parameters that can be optimized.
Evaluation of the optimal choices, however, is beyond the scope of our work.  In this paper, we confine ourselves to $X_\SRC$ which is uniformly distributed in $\{0,1\}$ and $\hY_\StR$ which is distributed as
\begin{eqnarray*}
\hY_\StR = Y_\StR + \hE_\BPR\:,
\end{eqnarray*}
where $\hE_\BPR \sim \eras(\hvarepsilon_\BPR)$ for some $\hvarepsilon_\BPR \in [0,1]$ and is independent of $Y_\StR$.  This choice was guided by ease of analysis.\footnote{Similar motivation guided the choice of auxiliary variables in~\cite[Sec.~VII.A]{Gerhard_Gaussian_Relay}, in the context of CF over the Gaussian relay channel.}  We have made a similar choice in the analysis of soft-DF-BP2 (Sec.~\ref{sec:Analysis_soft_DF_BP2}) and so the comparison will be fair.  With this choice, \eqref{eq:R_CF_BEC_Raw} coincides with~\eqref{eq:R_CF}.

\subsection{Details of the Application of HK in Sec.~\ref{sec:Numerical_Interference}}\label{apdx:HK_Details}

An overview of the HK strategy was provided in Sec.~\ref{sec:Introduction}.  Our discussion below follows the results and notation of~\cite[Sec.~III]{HanKobayashi}.  An application for the strategy involves constructing codes (here denoted $\cX_1$ and $\cX_2$) that are each obtained by combining two auxiliary codes, $\cU_i$ and $\cW_i$, $i = 1,2$, with rates $S_i$ and $T_i$, respectively.  Destination 1, for example, decodes the codewords $\bu_1 \in \cU_1$  and $\bw_1 \in \cW_1$, produced at its corresponding source, as well as $\bw_2 \in \cW_2$, amounting to a partial decoding of $\cX_2$.

Following~\cite{HanKobayashi}, we assume that the codes $\cU_i$ and $\cW_i$ are generated randomly, by independent selection of the components of their codewords according to the distributions of random variables
$U_i$ and $W_i$.  We define $U_i \sim \Bernoulli(0.055)$ and $W_i \sim \Bernoulli(1/2)$.  We also define $X_i = \BPSK(U_i \xor W_i)$, $i=1,2$, where $\xor$ denotes modulo-2 addition and the function $\BPSK$
maps the digits $\{0, 1\}$ to $\{1, -1\}$.  This means that the codewords of $\cX_i$ are similarly obtained by applying the above operation componentwise to pairs of codewords $(\bu_i,\bw_i)$ from $\cU_i$ and $\cW_i$.
An evaluation of the rate implied by~\cite[Theorem~3.1]{HanKobayashi} gives 0.333 bits per channel use.  More precisely, this figure is obtained by maximizing $S_1+T_1 = S_2+T_2$ (we restricted $S_1=S_2$ and $T_1=T_2$), as defined there, subject to~\cite[Eqs.~(3.2)--(3.15)]{HanKobayashi}.  The maximizing choices were $S_i = 0.101$ bits and $T_i = 0.231$ bits per channel use, respectively.

Note that our use of binary rather than real-valued random variables as in applications of the HK strategy for the AWGN interference channel (e.g.,~\cite{TseOneBit}), as well as modulo-2 addition, follow from the setting of our channel (Sec.~\ref{sec:BIAWGN_Interference}) which includes a binary input alphabet.

Consider sequences of codes $\{\cX_{i,n}\}_{n=1}^\infty$, $i=1,2$ where $\cX_{i,n}$ has block length $n$, constructed as described above.
 By Theorem~\ref{theorem:Good_Interference}, relying on the fact that the rate of the code sequences (0.333 bits per channel use) exceeds both $R_\MUD$ and $R_\SUD$ (see Sec.~\ref{sec:Numerical_Interference}), these code sequences are P2P-suboptimal for the BIAWGN channel (Definition~\ref{def:Good codes_AWGN}).  This can also be verified independently by observing that the decoding of $\cX_{i,n}$ over a P2P BIAWGN channel is equivalent to a multiple-access decoding problem (see e.g.,~\cite[Sec.~14.3]{Cover_Book}), where one destination must decode component codes $\cU_{i,n}$ and $\cW_{i,n}$ sent by two virtual users.  Using methods similar to~\cite[Eqs.~(14.99),~(14.111)]{Cover_Book} it is possible to show that the code sequence requires an SNR of at least 0.7684 for reliable communication, exceeding the Shannon limit for rate 0.333 bits per channel use, which is $\SNRstar = 0.5941$.  Thus, by Definition~\ref{def:Good codes_AWGN}, the code sequence is P2P-suboptimal.

\section*{Acknowledgements}
We are very much indebted to the associate editor, Pascal Vontobel, whose devoted review produced invaluable remarks and suggestions which greatly enhanced the quality of this work.

We also greatly appreciate the work of  the anonymous reviewers, who did a thorough and diligent job.  Suggestions by Rudiger Urbanke regarding the design of degree distributions for LDPC codes, and a discussion with Sergio \Verdu,\ are gratefully acknowledged.  Remarks by David Burshtein on a slide presentation of the material, and by Yuval Kochman on an initial draft of the paper, are very much appreciated.
        \bibliographystyle{IEEEtranS}
\providecommand{\noopsort}[1]{}


\end{document}